\documentclass[preprint,12pt]{elsarticle}
\usepackage{siunitx}
\usepackage{hyperref}
\usepackage{physics}
\usepackage[numbers]{natbib}
\usepackage{setspace}
\usepackage{amsmath}
\usepackage{amssymb}
\usepackage{graphics, subcaption}
\usepackage{caption}
\usepackage{pdfpages} 
\usepackage{lipsum}
\usepackage[percent]{overpic}
\usepackage{enumitem}
\usepackage{float}
\journal{Acta Materialia}

\begin{document}
\doublespacing
\title {Stability, Evolution and Switching of Ferroelectric Domain Structures in
Lead-free BaZr$_{0.2}$Ti$_{0.8}$O$_3$ - Ba$_{0.7}$Ca$_{0.3}$TiO$_3$ System: 
Thermodynamic Analysis and Phase-field Simulations}                      
\author[label1]{Soumya Bandyopadhyay}%

\author[label1]{Tushar Jogi\corref{correspondingauthor}}
\ead{ms14resch11003@iith.ac.in}

\author[label1]{Ranjith Ramadurai}%

\author[label1]{Saswata Bhattacharya\corref{correspondingauthor}}
\ead{saswata@msme.iith.ac.in}

\address[label1]{Department of Materials Science and 
Metallurgical Engineering, Indian Institute of Technology 
Hyderabad, Sangareddy- 502285, India}

\cortext[correspondingauthor]{Corresponding authors}


\begin{abstract}
\begin{singlespacing}
Enhanced room-temperature electromechanical coupling in the lead-free ferroelectric system 
$(1-x)$BaZr$_{0.2}$Ti$_{0.8}$O$_{3}$ - $x$Ba$_{0.7}$Ca$_{0.3}$TiO$_{3}$ (abbreviated as BZCT) at $x=0.5$ is attributed to the existence of a morphotropic phase region (MPR) containing an intermediate orthorhombic ($O$) phase between terminal rhombohedral ($R$) BZT and tetragonal ($T$) BCT phases. 
However, there is ambiguity regarding morphotropic phase transition in BZCT at room temperature -  while some experiments suggest a single 
$O$ phase within the MPR, others indicate  coexistence of three polar phases ($T+R+O$). 
Therefore, to understand thermodynamic stability of polar phases and its relation to  electromechanical switching during morphotropic phase transition in BZCT, 
we develop a Landau potential based on the theory of polar anisotropy. 
Since intrinsic electrostrictive anisotropy changes as a function of electromechanical processing, we establish a correlation between the parameters of our potential and the coefficients of electrostriction. 
We also conducted phase-field simulations based on this potential to demonstrate changes in domain configuration from single-phase $O$ to three-phase $T+R+O$ at the equimolar composition with the increase in electrostrictive anisotropy. 
Diffusionless phase diagrams and the corresponding piezoelectric coefficients obtained from our model compare well with the experimental findings. 
Increase in electrostrictive anisotropy increases the degeneracy of the free energy at ambient temperature and pressure leading to decreasing polar anisotropy, although there is an accompanying increase in the electromechanical anisotropy manifested by an increase in the difference  
between effective longitudinal and transverse piezo-coefficients, $d_{33}$ and $d_{31}$. 
Additionally, application of mechanical constraint (clamping) shows a change in phase stability from orthorhombic ($O$) (in stress-free condition) to tetragonal ($T$) (in clamped condition) with a lower effective piezoresponse for the latter.

\end{singlespacing}
\end{abstract}


\maketitle
\section{Introduction}




Due to lead toxicity concerns,  
one of the key challenges in the field of oxide-based electronics is the development of environment friendly lead-free ferroelectric materials that can replace the high-performance lead-based counterparts, such as lead zirconate titanate (PZT) and lead magnesium niobate - lead titanate (PMN-PT) systems~\cite{saito2004lead,shrout2007lead,rodel2009perspective,zhang2007lead, panda2015pzt,he2020advances}. 
Although several lead-free ferroelectric systems have been identified in the last decade, attaining a room-temperature piezoresponse 
superior to the best available lead-based system (abbr. PZT) 
is still a challenge~\cite{doshida2007miniature,tou2009properties,doshida2013investigation}.

There is renewed interest in barium titanate (abbr. BT), the first discovered perovskite ferroelectric with perfect cubic perovskite structure (AB$O_3$ with point group $m\bar{3}m$) above \SI{120}{\degreeCelsius} which transforms to tetragonal $4mm$ symmetry at room temperature~\cite{shrout2007lead,rodel2009perspective,bechmann1956elastic}. 
However, the piezoresponse of undoped bulk BT 
at room temperature, characterized by electromechanical coupling coefficients d$_{33}$ and d$_{31}$, is far lower than that of PZT. Therefore, there are continuing efforts to improve the 
electromechanical coupling efficiency or piezoresponse
of BT at room temperature using a combination of doping, nanostructuring, and strain-tuning~\cite{schlom2007strain,buscaglia2020size}.

Yu et al.~partially substituted Ti$^{4+}$ ions with zirconium to a 
maximum of 30\% and observed highest piezoresponse d$_{33}=$\SI{230}{\pico\coulomb\per\newton} for 5\% Zr substituted BT~\cite{yu2002piezoelectric}. Tian et al.~reported an increase in d$_{33}$ to \SI{305}{\pico\coulomb\per\newton} with partial replacement of Ti with 5\% hafnium~\cite{tian2007preparation}. Recently, 
Kalyani et al.~compared the effects of doping of BT with zirconium, hafnium and tin~\cite{kalyani2014orthorhombic} and reported substantial 
increase in piezoresponse with a  maximum d$_{33}=$\SI{425}{\pico\coulomb\per\newton} 
when BT is doped with 2\% tin. With 2\% hafnium or 2\% zirconium, there was reduction in d$_{33}$. Also, there have been attempts to increase piezoresponse in 
BT by microstructural engineering that includes refinement of grain size 
in polycrystalline BT, dimensional reduction in the 
form of thin films/nanowires with a view to confining 
polar order within small volumes; see the comprehensive 
review~\cite{buscaglia2020size} by Buscaglia and Randall and the references therein.
Huan et al.~obtained a maximum d$_{33}=$\SI{519}{\pico\coulomb\per\newton} for a polycrystalline BT film of \SI{20}{\nano\meter} thickness containing columnar grains 
with an average surface grain size of \SI{1}{\micro\meter}~\cite{huan2014grain}. 
Additionally, there has been extensive research on the application of  
strain engineering through epitaxial growth of single/multi-layer thin films/superlattices on a variety of strain-compatible substrates to enhance the room temperature electromechanical properties of BT~\cite{schlom2007strain}.
Choi et al.~reported an enhancement in ferroelectric properties 
(increase in Curie temperature T$_{c}$ beyond \SI{120}{\degreeCelsius} with a subsequent increase in the remanent polarization P$_{s}$) by tuning epitaxial strain and thickness of film using molecular beam epitaxy~\cite{choi2004enhancement}. 
However, they reported a decrease in d$_{33}$. 
Kim et al.~reported a maximum piezoresponse d$_{33}=$\SI{54}{\pico\coulomb\per\newton} for epitaxially grown BT on Pt and LSCO/Pt electrodes using pulsed laser deposition~\cite{kim2005study}. 
Jo et al.~also reported a maximum d$_{33}=$ \SI{54}{\pico\coulomb\per\newton} using an epitaxially grown 
BaTiO$_{3}$/CaTiO$_{3}$ superlattice with eighty periods of two BaTiO$_3$ and four CaTiO$_{3}$ repeating units ~\cite{jo2010piezoelectricity}.
Since the effective piezo-coefficient d$_{33}$ of doped, nanostructured, and strain-engineered BT remained less than \SI{550}{\pico\coulomb\per\newton}, 
there is a focus on developing lead-free ferroelectric alloys 
with a view to mimicking unique thermodynamic characteristics 
(such as morphotropic behavior) of lead-based solid solutions~\cite{saito2004lead,shrout2007lead,rodel2009perspective,liu2009large,acosta2014relationship,brandt2014mechanical,bao2010modified}.

One of the first attempts in developing a lead-free BT-based solid solution is alloying of BT with varying amounts of calcium titanate (CaTiO$_3$, abbr. CT). The solubility limit of CT to produce a stable tetragonal ferroelectric phase at room temperature is 34\%. However, even at maximum solubility there was no appreciable improvement in piezoresponse for this solid solution due to increase 
in leakage current~\cite{varatharajan2000ferroelectric, victor2003normal,fu2010invariant}. Alloying of BT 
with strontium titanate (SrTiO$_3$, abbr. ST) or 
barium zirconate (BaZrO$_3$, abbr. BZ) has yielded better piezoresponse (around \SI{400}{\pico\coulomb\per\newton})~\cite{khassaf2014strain,buscaglia2014average}. 
However, these could not surpass the piezoresponse of PZT at room temperature.
Recently, a new BT-based alloy with 
zirconium-doped BT and calcium-doped BT as the two components (chemical formula: 
${(1-x)}\textrm{Ba}(\textrm{Zr}_{0.2}\textrm{Ti}_{0.8}\textrm{O}_{3}  )-x(\textrm{Ba}_{0.7}\textrm{Ca}_{0.3})\textrm{TiO}_{3}\;(0 \leq x \leq 1))$, abbr. BZCT)
has emerged as one of the most promising lead-free ferroelectric systems for 
electromechanical applications at room temperature. 
The highest piezoresponse of equimolar ($x=0.5$)  BZCT at room temperature is around 
\SI{620}{\pico\coulomb\per\newton} which exceeds that of 
PZT~\cite{liu2009large,acosta2014relationship,brandt2014mechanical,bao2010modified}. 
Moreover, structural studies of BZCT and PZT 
revealed a unique similarity in thermodynamic characteristics -
both show morphotropic phase transition below the Curie temperature 
at or around the equimolar 
composition~\cite{noheda1999monoclinic,noheda2000stability,noheda2000tetragonal}. 
Morphotropic phase region (MPR) bounded by morphotropic phase boundaries (MPB)  
is a region of high electromechanical activity in a ferroelectric solid solution 
and consists of a linking phase between the terminal solid solutions. Presence of an intermediate phase having lower crystallographic symmetry than the terminal ones introduces tricritical points marking the coexistence of the ferroelectric phases and 
increases polarization rotation by reducing the energy barrier for transition between the phases.  Moreover, since low crystallographic symmetry of the intermediate phase 
increases the number of polar variants which are also ferroelastic, 
there is enhancement of strain accommodation within the MPR that softens  
effective elastic moduli of the ferroelectric thereby increasing the effective electromechanical moduli~\cite{cox2001universal,zhang2015advantages}.

Although initial structural studies of MPR in PZT and BZCT described these systems as a mixture of terminal phases with rhombohedral ($R$) and tetragonal ($T$) crystal structures below the Curie temperature
~\cite{liu2009large,singh1995coexistence,boutarfaia1995study}, 
later investigations using high-energy diffraction techniques revealed 
the coexistence of a bridging phase in both systems 
at the morphotropic 
composition~\cite{noheda1999monoclinic,noheda2000stability,noheda2000tetragonal}. 
The MPR in PZT is characterized by a narrow monoclinic region around the equimolar composition between terminal $T$ and $R$ 
phases~\cite{pan2020observation,chen2020designing,gupta2020phase}, 
whereas that in BZCT 
shows an wider region containing an intermediate phase with orthorhombic symmetry\cite{keeble2013revised}. However, there exists 
uncertainty in determining phase coexistence at the morphotropic region
of BZCT at the room temperature~\cite{keeble2013revised,brajesh2016structural,liu2009large,brandt2014mechanical,bao2010modified,brajesh2015relaxor}.
Keeble et al.~used high-resolution synchrotron X-ray diffraction to characterize
the structure of BZCT, prepared using conventional solid state reaction technique, 
for the entire range of alloy compositions ($0 \leq x \leq 1$) and temperatures varying between 
100 \si{\kelvin} and 500 \si{\kelvin}~\cite{keeble2013revised}. 
They established a diffusionless phase stability map from their diffraction studies 
and showed the existence of an intermediate orthorhombic phase for the composition range  $(0.45<x<0.55)$ at room temperature.  
Brajesh et al.~implemented a novel ``powder poling'' 
technique to study electric field induced structural transformations in BZCT at the MPB.
Structural analysis of poled BZCT powder using X-ray diffraction and Rietveld refinement 
showed the coexistence
of $T$, $O$, and $R$ phases at the morphotropic compositions ($0.45<x<0.55$) 
at room temperature~\cite{brajesh2015relaxor}. 
In a later study, they showed that a stress-induced ferroelastic transformation above the Curie temperature
 precedes the ferroelectric to paraelectric transformation in BZCT. 
Therefore, they annealed BZCT at $400 \si{\degreeCelsius}$ (well above $T_c=120\si{\degreeCelsius}$) 
to relieve stresses associated with high-temperature ferroelasticity. They observed a change in phase coexistence where the amount of $R$ reduces significantly with a subsequent 
increase in the fractions of $T$ and $O$~\cite{brajesh2016structural}. 
These studies of BZCT reveal the role of anisotropy of electrostriction induced by processing 
on the coexistence of phases within the MPR.

Recently Jeon et al.~reported how the variability in processing conditions (such as poling, annealing, quenching, and milling) can introduce changes in phase transition in a relaxor ferroelectric PMN-PT~\cite{jeon2020effect}.
Since coefficients of electrostriction are inherently related to 
oxygen octahedral structure in perovskite oxides,  
any structural change in oxygen octahedra due to electromechanical processing will affect the electrostrictive 
coefficients~\cite{yamada1972electromechanical}. 
Moreover, spontaneous strain being a function of electrostriction and spontaneous polarization, 
change in electrostrictive coefficients will change the spontaneous strain that can consequently 
alter the thermodynamic stability of polar phases because the free energy of ferroelectric materials 
is a function of both spontaneous polarization and spontaneous strain~\cite{lines2001principles}.     
Moreover, physical properties of the parent paraelectric phase with a perovskite crystal structure, represented by fourth rank or higher even-rank tensors, such as electrostriction or elasticity, can show cubic anisotropy at the most (Neumann's principle)~\cite{newnham2005properties}. For example, bulk barium titanate and lead-based perovskite solid solutions show large cubic anisotropy at room temperature when the anisotropy parameter is defined as: $Q_a=\frac{Q_{11} - Q_{12}}{Q_{44}} > 1$, 
where $Q_{11}$, $Q_{12}$, $Q_{44}$ are the independent electrostrictive coefficients~\cite{li2014electrostrictive}.    
Change in anisotropy in spontaneous strain during the paraelectric-to-ferroelectric transition can 
affect the switching behaviour manifested in large differences between the measured transverse $(d_{31})$ and 
longitudinal $(d_{33})$ piezo-coefficients~\cite{budimir2003piezoelectric,damjanovic2009comments,iwata2002anisotropy,li2014electrostrictive}.

To relate changes in domain evolution and switching properties in BZCT 
due to variations in external thermal, electrical, and mechanical fields, 
we require a thermodynamic potential integrated with elastic and electrostatic interactions 
that can not only predict phase stability in the stress-free, electrically neutral state as a function of temperature and composition but also changes in stability with application of electromechanical loading~\cite{chen1998applications, chen2008phase, heitmann2010thermodynamics, heitmann2014thermodynamics}. 
Cao and Cross made the first attempt to develop a thermodynamic description of a ferroelectric solid solution where they combined the classical Ginzburg-Landau-Devonshire formalism for ferroelectrics 
with a regular solution model describing interactions between the components of the ferroelectric system. They proposed a two-parameter free energy model where the total free energy of the solid solution is expressed as weighted sum of the Landau free energies of the terminal components, where the mole fraction of each component is the weight. They added an excess energy associated with mixing of the components using a regular solution formalism. Landau free energy for each of the terminal phases (components) was described using a unique order parameter~\cite{cao1993theoretical}. 
Bell and Furman modified the regular-solution based coupling term and included additional coupling between the polarization order parameters and Landau free energy coefficients~\cite{bell2003two}. The modified model could successfully describe phase coexistence at the MPB of PZT. Li et al.~made further modifications to 
the thermodynamic potential using a single order parameter free energy for the entire composition range and introduced composition- and temperature-dependent Landau coefficients~\cite{li2005ferroelectric}. 
The model was used to study ferroelectric/ferroelastic domain evolution in epitaxially grown PZT films. 
Later, Heitmann and Rossetti developed a generalized thermodynamic model for ferroelectric 
solid solutions with MPB/MPR, wherein they incorporated anisotropy in polarization associated with low-symmetry ferroelectric phases and redefined the Landau polynomial in terms of isotropic and anisotropic contributions~\cite{heitmann2010thermodynamics}. 
Since MPB defines the coexistence of low-symmetry phases marked by vanishing polarization anisotropy at the triple point, 
their model 
could accurately predict both location and shape of MPBs in several lead-based and lead-free solid solutions~\cite{heitmann2014thermodynamics}. Following~\cite{heitmann2010thermodynamics, heitmann2014thermodynamics},  
Yang et al.~developed a thermodynamic potential for BZCT based on the polar anisotropy theory of Heitmann and Rossetti~\cite{yang2016mechanisms}. Although the model could accurately predict the stability of 
orthorhombic phase within MPR of BZCT around the equimolar composition, it does 
not show correspondence with the phenomenological 
Landau-Ginzburg-Devonshire (LGD) theory
and does not correlate phase stability with electromechanical response as a function of underlying domain configuration.
Since ferroelctric phases are also ferroelastic in nature, 
accurate prediction of phase stability requires coupling of electrostatic and elastic interactions. Recently Huang et al.~\cite{huang2020thermodynamic} developed a thermodynamic potential for barium zirconate titanate (BZT) including coupling between electrostatic and elastic interactions that showed excellent agreement with experimental phase stability data. Here, we build upon the thermodynamic potential proposed by Yang et al.~\cite{yang2016mechanisms} to develop a free energy based on LGD formalism~\cite{heitmann2014thermodynamics} to study phase stability, domain evolution and polarization switching behaviour in BZCT system.  
Since electromechanical anisotropy can affect spontaneous strain field defined by $\varepsilon_{ij}^0(\mathbf{r})=Q_{ijkl}P_k(\mathbf{r})P_l(\mathbf{r})$, where $Q_{ijkl}$ is the electrostrictive coefficient tensor and $P_i\;(i=1,2,3)(\mathbf{r})$ are the components of spontaneous polarization order parameter field, we have systematically varied the anisotropy of electrostriction (defined with respect to the paraelectric cubic phase) and studied its role on 
phase stability within MPR, domain evolution and the resulting electromechanical response. In each case, we compare the effective piezoresponse coefficients, $d_{33}$ and $d_{31}$, computed from the simulated phase loops, with those measured experimentally~\cite{keeble2013revised}. 

The paper is organized as follows:
In the following section, we present our formulation where we derive a thermodynamic potential for BZCT solid solution incorporating electrostatic and elastic interactions and develop a phase-field model based on this potential to study domain evolution in BZCT under applied electromechanical fields. In the subsequent section, we present our results correlating the changes in the predicted diffusionless phase diagrams with the change in electromechanical anisotropy. We also present our results of three-dimensional phase field simulations of domain evolution and relate them to the switching characteristics in BZCT due to applied electromechanical fields. We have compared our simulated data on phase stability and effective piezoelectric coefficients with the best available experimental measurements. Finally, we summarize the key conclusions from this study.

\section{Model formulation}
We begin with the description of energetics of BCZT system using the Landau-Ginzburg-Devonshire (LGD) thermodynamic formalism. We draw equivalence between the LGD free energy and a thermodynamic potential based on the theory of polar anisotropy to derive thermodynamic criteria defining the MPBs in stress-free, electrically neutral BZCT system~\cite{heitmann2014thermodynamics, yang2016mechanisms}. Next, we describe electrostatic, elastic and external electromechanical field contributions to the total free energy of the system and develop a 
three-dimensional phase-field model to study domain evolution in BZCT. 
The phase-field model consists of a set of Allen-Cahn equations describing the spatiotemporal evolution of the polarization order parameter field $\mathbf{P}(\mathbf{r},t)$ coupled with an electrostatic equilibrium equation for the electric field and a 
mechanical equilibrium equation for the strain field. Since the model includes external field effects, it can be used to correlate domain configuration with the polarization switching characteristics. In what follows, we have used indicial notations (with Einstein summation convection) to describe vector and tensor quantities in terms of their components. Otherwise, we denote vector fields with bold letters and higher-order tensor fields using standard matrix notations.
\subsection*{Thermodynamic potential}
The total free energy $\mathcal{F}$ of a ferroelectric system is expressed as follows~\cite{li2001phase,hlinka2006phenomenological,chen2008phase,cao2010piezoelectric,zhang2005computational}:
\begin{equation}
  \mathcal{F}(P_i, \varepsilon_{ij},  E_i) = \int_V\left(f_{\textrm{bulk}} + f_{\textrm{electric}} + f_{\textrm{elastic}} + f_{\textrm{gradient}}\right)dV\;\;  (i,j=1,2,3),
\label{TotalF}
\end{equation} 
where $f_{\textrm{bulk}}$, $f_{\textrm{electric}}$, $f_{\textrm{elastic}}$, 
and $f_{\textrm{gradient}}$ denote the bulk, electric, elastic and gradient energy
contributions to the total free energy, $P_i$ are the components of spontaneous polarization order parameter 
field,  $\varepsilon_{ij}$ denotes the coefficients of spontaneous strain field related to $P_i$ through the third-order piezoelectric tensor $d_{ijk}$ and fourth order electrostrictive tensor $Q_{ijkl}$ with $i,j,k,l=1,2,3$.

Using the unpolarized, stress-free and centrosymmetric paraelectric state (cubic) as the reference state, $f_{\textrm{bulk}}$ for BZCT is expressed using a sixth-order Ginzburg-Landau polynomial~\cite{chen2007appendix}: 
\begin{equation}
\begin{split}
    f_{\textrm{bulk}} &=\frac{1}{2}\alpha_{1}(P_{1}^{2} + P_{2}^{2} + P_{3}^{2}) 
    +\frac{1}{4}\alpha_{11}(P_{1}^{4} + P_{2}^{4} + P_{3}^{4}) + \frac{1}{6}\alpha_{111}(P_{1}^{6} + P_{2}^{6} + P_{3}^{6})\\
    &+ \frac{1}{2}\alpha_{12}(P_{1}^{2}P_{2}^{2} + P_{2}^{2}P_{3}^{2} + P_{3}^{2}P_{1}^{2})
    +\frac{1}{2} \alpha_{112}(P_{1}^{4}(P_{2}^{2} + P_{3}^{2}) + P_{2}^{4}(P_{3}^{2} + P_{1}^{2})\\
    &+ P_{3}^{4}(P_{1}^{2} + P_{2}^{2})) 
    +\frac{1}{6} \alpha_{123}P_{1}^{2}P_{2}^{2}P_{3}^{2},
\end{split}
\label{fbulk}
\end{equation}
where $P_1, P_2, P_3$ are the components of the polarization field $\mathbf P(\mathbf{r})$. The phenomenological Landau expansion coefficients ($\alpha_1, \alpha_{11}, \alpha_{12}, \alpha_{111}, \alpha_{112}, \alpha_{123}$) are functions of composition ($x$) and temperature ($\theta$), and determine the energy of the stress-free, electroneutral state of the system. The coefficients are  chosen appropriately to ensure a first-order transition ($\alpha_1<0$, $\alpha_{11}<0$, $\alpha_{111}>0$) between the paraelectric and ferroelectric states. 

To derive thermodynamic stability conditions during paraelectric to ferroelectric phase transition 
and compute diffusionless phase diagrams of BZCT system, 
 Eq.~\eqref{fbulk} is expressed in an alternate form based on polar anisotropy theory 
where we separate isotropic part of the  free energy from the direction-dependent  anisotropic part~\cite{heitmann2010thermodynamics,heitmann2014thermodynamics}.
Therefore, we define the spontaneous polarization field $\mathbf{P}$ as a product of the magnitude of spontaneous polarization $P=|\mathbf{P}|$ and a unit vector $\mathbf{n}=(n_1,n_2,n_3)$ along the direction of spontaneous polarization: $\mathbf{P} = \mathbf{n}P$. 
Thus, Eq.~\eqref{fbulk} becomes:
\begin{equation}
\begin{split}
    f_{\textrm{bulk}}^{\textrm{modified}} &=\frac{1}{2}\alpha_{1}(n_{1}^{2} + n_{2}^{2} + n_{3}^{2})P^2 
    +\frac{1}{4}\alpha_{11}(n_{1}^{4} + n_{2}^{4} + n_{3}^{4})P^4 \\
    &+ \frac{1}{6}\alpha_{111}(n_{1}^{6} + n_{2}^{6} + n_{3}^{6})P^6
    + \frac{1}{2}\alpha_{12}(n_{1}^{2}n_{2}^{2} + n_{2}^{2}n_{3}^{2} + n_{3}^{2}n_{1}^{2})P^4\\
    &+\frac{1}{2} \alpha_{112}(n_{1}^{4}(n_{2}^{2} + n_{3}^{2}) + n_{2}^{4}(n_{3}^{2} + n_{1}^{3})
    + n_{3}^{4}(n_{1}^{2} + n_{2}^{2}))P^6\\
    &+\frac{1}{6} \alpha_{123}n_{1}^{2}n_{2}^{2}n_{3}^{2}P^6,
\end{split}
\label{fbulknew}
\end{equation}
where $f_{\textrm{bulk}}^{\textrm{modified}}$ is the alternate form of the bulk free energy which separates isotropic and anisotropic contributions. 
Note that the isotropic part of the energy 
describes transition from a non-polar phase to a polar glassy state 
with no preferential direction, while the anisotropic part defines the directional dependence of free energy surface due to the spontaneous polarization vector~\cite{heitmann2010thermodynamics,heitmann2014thermodynamics}.
The polar anisotropic contribution to the free energy is given by the cross terms of Eq.~\eqref{fbulknew}.
Since $(n_1^{2} + n_2^{2} + n_3^{2})^{m} = 1$ for any exponent $m$,
the powers of the expansion terms in Eq.~\eqref{fbulknew} for $m=2,3$ can be written as:
\begin{subequations}
\begin{align}
     (n_1^{2} + n_2^{2} + n_3^{2})^{2} &= (n_{1}^{4} + n_{2}^{4} + n_{3}^{4}) 
             + 2(n_{1}^{2}n_{2}^{2} + n_{2}^{2}n_{3}^{2} + n_{3}^{2}n_{1}^{2}),\label{bulkterm1}\\
    (n_1^{2} + n_2^{2} + n_3^{2})^{3} &= (n_{1}^{6} + n_{2}^{6} + n_{3}^{6})\nonumber
             + 3(n_{1}^{4}(n_{2}^{2} + n_{3}^{2}) + n_{2}^{4}(n_{3}^{2} + n_{1}^{2}) \\
             &+n_{3}^{4}(n_{1}^{2} + n_{2}^{2})) 
             + 6n_{1}^{2}n_{2}^{2}n_{3}^{2}.
             \label{bulkterm2}         
\end{align}
\label{bulkterm} 
\end{subequations}
Substituting the relations in Eq.~\eqref{bulkterm} in Eq~\eqref{fbulknew}, the modified free energy becomes:
\begin{equation}
\begin{split}
     f_{\textrm{bulk}}^{\textrm{modified}} &=\frac{1}{2}\alpha_{1}P^2 
     +\frac{1}{4}(\alpha_{12} + (\alpha_{11} - \alpha_{12})(n_{1}^{4} + n_{2}^{4} + n_{3}^{4}))P^4 \\
    &+ \frac{1}{6}(\alpha_{112} + (\alpha_{111} - \alpha_{112})(n_{1}^{6} + n_{2}^{6} + n_{3}^{6}))P^6
    +\frac{1}{6} (\alpha_{123}-6\alpha_{112})n_{1}^{2}n_{2}^{2}n_{3}^{2}P^6.
\end{split}
\label{fbulknewaniso2}
\end{equation}
Separating the isotropic and anisotropic parts, we rewrite the modified bulk free energy as
\begin{equation}
\begin{split}
     f_{\textrm{bulk}}^{\textrm{modified}} &=f_{\textrm{bulk}}^{\textrm{iso}}+f_{\textrm{bulk}}^{\textrm{aniso}}, \textrm{where}\\
 f_{\textrm{bulk}}^{\textrm{iso}} &=\frac{1}{2}\alpha_{1}P^2 + \frac{1}{4}\beta_1 P^4 + \frac{1}{6}\gamma_1 P^6 \\
 f_{\textrm{bulk}}^{\textrm{aniso}} &=\frac{1}{4}\beta_{2}(n_{1}^{4} + n_{2}^{4} + n_{3}^{4})P^4+
       \frac{1}{6}\left[\gamma_{2}(n_{1}^{6} + n_{2}^{6} + n_{3}^{6}) + \gamma_{3}n_{1}^{2}n_{2}^{2}n_{3}^{2}\right]P^6,
\end{split}
\label{fbulknewaniso3}
\end{equation}
where $\beta_{1}= \alpha_{12}$, $\beta_{2}= \alpha_{11} -\alpha_{12}$, $\gamma_{1}= \alpha_{112}$,
$\gamma_{2}= \alpha_{111} -\alpha_{112}$ and $\gamma_{3}= \alpha_{123} -6\alpha_{112}$
are the modified Landau coefficients. 
Table~\ref{table:comparison_landau} lists the coefficients of unmodified LGD energy (Eq.~\eqref{fbulk}) and the modified version of the energy (Eq.~\eqref{fbulknewaniso3}).
\begin{table}[!ht]
\caption{Relation between the original Landau coefficients and the modified ones} 
\centering 
\begin{tabular}{| m{4cm}| | m{5cm}|}
\hline\hline 
{\small Unmodified coefficients} & {\small Modified coefficients} \\ [2ex] 
\hline 
 $\alpha_{1}$& $\alpha_{1}$\\ %
 \hline
 $\alpha_{12}$& $\beta_{1}$\\
 \hline
 $\alpha_{11}$& $\beta_{1} + \beta_{2}$\\
 \hline
 $\alpha_{112}$& $\gamma_{1}$\\
 \hline
 $\alpha_{111}$& $\gamma_{1} + \gamma_{2}$\\
\hline
 $\alpha_{123}$& $\gamma_{3} + 6\gamma_{1}$\\
 \hline 
\end{tabular}
\label{table:comparison_landau} %
\end{table}
Assuming the paraelectric cubic state ($\mathbf{n}=\mathbf{0}$) as the reference state,
we use Eq.~\eqref{fbulknewaniso3} to define free energies 
of the paraelectric cubic phase $C (P=0)$ and the ferroelectric $T (n_1, n_2, n_3 = \pm 1, 0, 0)$, 
$O (n_1, n_2, n_3 = \pm1/\sqrt{2},\pm 1/\sqrt{2}, 0)$ and $R (n_1, n_2, n_3 = \pm 1/\sqrt{3}, \pm 1/\sqrt{3}, \pm 1/\sqrt{3})$ phases of stress-free BZCT 
as follows:
\begin{subequations}
\begin{align}
     f_{\textrm{C}}^{\textrm{modified}} &= 0\label{fcub}\\
     f_{\textrm{T}}^{\textrm{modified}} &=\frac{1}{2}\alpha_{1}P^2 
    +\frac{1}{4}(\beta_{1} + \beta_{2})P^4 
    + \frac{1}{6}(\gamma_{1} + \gamma_{2})P^6,\label{ftetra}\\
    f_{\textrm{O}}^{\textrm{modified}} &=\frac{1}{2}\alpha_{1}P^2 
     +\frac{1}{4}(2\beta_{1} + \beta_{2})P^4 
     + \frac{1}{24}(4\gamma_{1} + \gamma_{2})P^6,\label{fortho}\\
   f_{\textrm{R}}^{\textrm{modified}} &=\frac{1}{2}\alpha_{1}P^2 
     +\frac{1}{12}(3\beta_{1} + \beta_{2})P^4 
    + \frac{1}{162}(27\gamma_{1} + 3\gamma_{2} + \gamma_{3})P^6.
    \label{frhom}
\end{align}
\label{free_ph}
\end{subequations}


Minimization of Eqns.~\eqref{ftetra}-~\eqref{frhom} with respect to $P$ yields
equilibrium spontaneous polarization $P_{s,\phi}\;{\phi=T,O,R}$ of the ferroelectric phases, $T$, $O$ and $R$ 
as a function of temperature and composition:
\begin{subequations}
\begin{align}
    P_{s,T}^{2} &= \frac{1}{2}\frac{-(\beta_1 + \beta_2)\pm\sqrt{(\beta_1 + \beta_2)^2 
    - 4\alpha_1(\gamma_1 + \gamma_2)}}{\gamma_1 + \gamma_2},\label{polaT}\\
    P_{s,O}^{2} &= \frac{-(2\beta_1 + \beta_2)\pm\sqrt{(2\beta_1 + \beta_2)^2 
    - 4\alpha_1(4\gamma_1 + \gamma_2)}}{4\gamma_1 + \gamma_2},\label{polaO}\\
     P_{s,R}^{2} &= \frac{3}{2}\frac{-(9\beta_1 + 3\beta_2)\pm\sqrt{(9\beta_1 + 3\beta_2)^2 
    - 12\alpha_1(27\gamma_1 + 3\gamma_2 + \gamma_3)}}{27\gamma_1 + 3\gamma_2 + \gamma_3}.
    \label{polaR}
\end{align}
\label{sp_ph}
\end{subequations}
The equilibrium free energies of $T$, $O$ and $R$ in terms of temperature and composition are obtained by substituting the
expressions of spontaneous polarization of these phases (Eq.~\eqref{sp_ph}) in Eq~\eqref{free_ph}.

Following Yang et al.~\cite{yang2016mechanisms}, we express the 
composition and temperature dependent Landau free energy coefficients as follows: 
\begin{equation}
\begin{split}
    \alpha_{1}(\theta,x) &= \alpha_{0}(\theta-\theta_{c}(x)),\\
    \beta_{1}(\theta,x) &= \beta_{11}(x-x_{quad}) + \beta_{12}(\theta-\theta_{quad}),\\
    \gamma_{1}(\theta,x) &= \gamma_{11} + \gamma_{12}(x-x_{quad}),\\
     \beta_{2}(\theta,x) &= \beta_{21}(x-x_{quad}) + \beta_{22}(\theta-\theta_{quad}),\\
     \gamma_{2}(\theta,x) &= \gamma_{21}(x-x_{quad}),\\
    \gamma_{3}(\theta,x) &= \gamma_{31}(x-x_{quad}) + \gamma_{32}(\theta-\theta_{quad}),
\end{split}
\label{landau_coeff}
\end{equation}
where $\alpha_{0} = 4.142\times 10^{5}$, $\beta_{11}=-1.2\times 10^{8}$,
$\beta_{12}=7.56\times 10^{5}$, $\gamma_{11}=7.764\times10^{8}$, $\gamma_{12}=4\times 10^{7}$,
$\beta_{21}=-1.2\times 10^{8}$, $\beta_{22}=-7.56\times10^{5}$,
$\gamma_{21}=-2.2\times10^8$, $\gamma_{31}=1.0\times 10^{11}$, $\gamma_{32}=2.1\times10^8$
are the values in SI units, $\alpha_0=1/(\epsilon_0C_0)$,
$\epsilon_0=8.854\times10^{-12}\si{\coulomb\tothe{2}\newton\tothe{-1}\meter\tothe{-2}}$ is the permittivity of free space, 
$C_0$ is the average Curie constant, $\theta$ denotes the temperature in Kelvin, $x$ denotes the composition of BZT (in mole fraction),
$\theta_{c}(x) = \theta_{C}^{\textrm{BZT}} + (\theta_{C}^{\textrm{BCT}}-\theta_{C}^{\textrm{BZT}})x$ 
is the composition-dependent Curie temperature of the BZCT system, where 
$\theta_{C}^{\textrm{BZT}}$ is the Curie temperature of pure BZT ($x=0$), 
$\theta_{C}^{\textrm{BCT}}$ is the Curie temperature of BCT ($x=1$),
$\theta_{\textrm{quad}}$ and $x_{\textrm{quad}}$ denote the temperature and the composition
at the quadruple point defined by the coexistence of the paraelectric cubic phase and ferroelectric $T$, $O$ and $R$ phases in BZCT. The Landau coefficients are obtained by fitting the experimental values of $\theta_{\textrm{quad}}=335 \si{\kelvin}$, $x_{\textrm{quad}}=0.35$, $\theta_c^{BZT}$ = $293$ $\si{\kelvin}$, $\theta_c^{BCT}$ = $393$ $\si{\kelvin}$ 
and equilibrium spontaneous polarization $P_{s,\phi}\;\phi=T,O,R$ of 
the polar phases measured at room temperature. 


The electric energy density $f_{\textrm{electric}}$ in Eq.~\eqref{TotalF} is given as 
\begin{equation}
    f_{\textrm{electric}} = -\frac{1}{2}\epsilon_0\epsilon_{b} (E_{1}^{2} + E_{2}^{2} + E_{3}^{2})
     - (P_{1}E_{1} + P_{2}E_{2} + P_{3}E_{3}), 
    \label{felec}
\end{equation}
where $E_{1}$, $E_{2}$, $E_{3}$ are the components of the total electric field 
$\mathbf{E}(\mathbf{r})$ that comprises an internal depolarization field $\mathbf{E}^\textrm{d}(\mathbf{r})$ 
resulting from dipole-dipole interactions ($f_{\textrm{dipole}}$), an
externally applied field $\mathbf{E}^{\textrm{ext}}$, and 
a random field $\mathbf{E}^{\textrm{random}}(\mathbf{r})$ 
associated with compositional heterogeneity of the polar solid~\cite{wang2016phase}.
The depolarization field $\mathbf{E}^{\textrm{d}}(\mathbf{r})$ is defined as~\cite{hlinka2006phenomenological}
\begin{equation}
    E^{\textrm{d}}_{i}(r_{k}) = \frac{1}{4\pi\epsilon_{0}\epsilon_{b}}
     \left(\frac{P_{i}(r_{k})}{|\mathbf{r}|^{3}} - \frac{[3P_{j}(r_{k}) {r_{j}}]r_{i}}{|\mathbf{r}|^{5}}\right),
    \label{dipole}
\end{equation}
where $\epsilon_{0}$ is the permittivity of free space and $\epsilon_{b}$ denotes the 
background dielectric permittivity due to dielectric screening~\cite{tagantsev1986role,tagantsev2008landau,levanyuk2016background}.
In the absence of external and random fields, $E^{\textrm{d}}_{i}(\mathbf{r})$
is the solution to electrostatic equilibrium equation defined as follows:
\begin{equation}
\mathbf{\nabla}_{i}{D}_{i}=0,
\label{poisson}
\end{equation}
where $D_{i}=\epsilon_{0}\epsilon_{b}E^{\textrm{d}}_{i} + P_{i}$ is the 
electric displacement vector. 
We define electric energy (Eq.~\eqref{felec}) 
in accordance with ``spontaneous
polarization order parameter (SPOP)'' approach, 
wherein we separate the spontaneous polarization field $\mathbf{P}(\mathbf{r})$
from the polarization $\mathbf{P}^{ext}(\mathbf{r})$ induced by the 
externally applied field (including dielectric screening effect)~\cite{woo2008depolarization,zheng2009thermodynamic}. On the other hand, ``total
polarization order parameter (TPOP)'' approach uses total polarization 
field $\mathbf{P}^{T}(\mathbf{r})$ as the order parameter where
the dielectric displacement vector is defined as $D_{i}=\epsilon_{0}{E}_{i}^d + P_{i}^{T}$~\cite{kretschmer1979surface}. 
In the latter, we cannot define a background 
dielectric constant ($\epsilon_{b}$) and the contributions to the electric energy from the external and internal fields are given separately ~\cite{woo2008depolarization,zheng2009thermodynamic,kretschmer1979surface}:
\begin{equation}
      f_{\textrm{electric}} = -\frac{1}{2}P^T_{i}E_{i}^d -P^T_{i}E_{i}^{ext}.
    \label{tpopelec}
\end{equation}
Note that both approaches are equivalent and should yield the same electric energy density. 

Since spontaneous polarization in a ferroelectric crystal is a result of
displacement of ions in the lattice from the reference paraelectric state, it engenders spontaneous strain~\cite{lines2001principles}. 
The spontaneous strain order parameter field $\varepsilon^0(\mathbf{r})$ for a stress-free crystal is defined as follows~\cite{chen2008phase}:
\begin{equation}
    \varepsilon_{ij}^{0}= d_{ijk}P_{k} + Q_{ijkl}P_{k}P_{l},
    \label{sponstrain}
\end{equation}
where $d_{ijk}$ denotes the rank-3 piezoelectric tensor and
$Q_{ijkl}$ denotes the rank-4 electrostrictive coefficient tensor. 
Since the centrosymmetric paraelectric state is the reference state in 
the Landau expansion of free energy, the rank-3 piezoelectric coefficients, that linearly couple polarization and strain in our free energy, become zero. Thus, in the Cartesian frame of reference, the spontaneous strain components are given as 
\begin{equation}
\begin{split}
    \varepsilon_{11}^{0} &= Q_{11}P_{1}^{2} + Q_{12}(P_{2}^{2} + P_{3}^{2}),\\
    \varepsilon_{22}^{0} &= Q_{11}P_{2}^{2} + Q_{12}(P_{1}^{2} + P_{3}^{2}),\\
    \varepsilon_{33}^{0} &= Q_{11}P_{3}^{2} + Q_{12}(P_{1}^{2} + P_{2}^{2}),\\
    \varepsilon_{12}^{0} &= Q_{44}P_{1}P_{2},\; \varepsilon_{13}^{0} = Q_{44}P_{1}P_{3},\; \varepsilon_{23}^{0} = Q_{44}P_{2}P_{3}.
\end{split}
\label{eigenstrain}
\end{equation}

Using the definition of stress-free strain or eigenstrain (Eq.~\eqref{sponstrain}), 
elastic energy density $f_{\textrm{elastic}}$ is defined as:
\begin{equation}
    f_{\textrm{elastic}} =\frac{1}{2}C_{ijkl}
    (\varepsilon_{ij}^T-\varepsilon_{ij}^{0})(\varepsilon_{kl}^T-\varepsilon_{kl}^{0}),
    \label{felas}
\end{equation}
where $C_{ijkl}$ is the elastic stiffness tensor and  $\varepsilon_{ij}^T$ 
denotes the total strain at a point. 
Based on homogenization theory for structurally inhomogeneous solids, 
we express $\varepsilon_{ij}^T(\mathbf{r})$ as the sum of spatially-invariant
homogeneous strain $\bar{\varepsilon}_{ij}$  
and position-dependent heterogeneous strain field $\delta\varepsilon_{ij}(\mathbf{r})$ that vanishes when integrated over the total volume~\cite{li2001phase}: 
\begin{equation}
\varepsilon_{ij}^T(\mathbf{r})= \bar{\varepsilon}_{ij} + \delta\varepsilon_{ij}(\mathbf{r});\, 
\int_V\delta\varepsilon_{ij}(\mathbf{r})d^3\mathbf{r}=0.
\label{strdef}
\end{equation}
We use Khachaturyan's microelasticity theory with a homogeneous modulus approximation
to solve the mechanical equilibrium equation~\cite{khachaturyan1969theory}
\begin{equation}
\begin{split}
\frac{\partial\sigma_{ij}}{\partial r_{j}} =
C_{ijkl}\frac{\partial\left[\varepsilon_{kl}^T(\mathbf{r})-
Q_{klmn}P_m(\mathbf{r})P_n(\mathbf{r})\right]}{\partial r_{j}}=0,
\end{split}
\label{meche}
\end{equation}
in Fourier space (assuming periodicity in local displacement and strain fields).
In the absence of applied stress, homogeneous strain
$\bar{\varepsilon}_{ij}$ is simply given by the weighted mean of total eigenstrain field:
\begin{equation}
   \bar{\varepsilon}_{ij} = \sum\limits_{p=0}^{5}\varepsilon_{ij}^{p}\Psi_{p}, 
   \label{homostrain}
\end{equation}
where 
$\Psi_{p} = (1/V)\int_V\left[\psi_{p}(\mathbf{r})-\bar{\psi}_{p}\right]dV$ 
and $\psi_{p}(\mathbf{r})=P_{i}(\mathbf{r})P_{j}(\mathbf{r})\; (i,j=1,2,3)$
and $p=0\ldots5$. 
Local displacement field in Fourier space, $\tilde{\mathbf{u}}(\mathbf{k})$, 
is obtained by solving Eq.~\eqref{meche}
\begin{equation}
    \tilde{u}_{k}(\mathbf{k}) =
    -\textrm{I} |\mathbf{k}|^{-1}n_j\Omega_{ik}(\mathbf{n})\sum\limits_{p=0}^{5}{\sigma_{ij}^{p}
     \Delta\tilde{\psi}_{p}(\mathbf{k})}.
    \label{displacement}
\end{equation}
Here, $\mathbf{k}$ denotes the wave vector in the reciprocal space, $\mathbf{n} = \mathbf{k}/|\mathbf{k}|$, and $\textrm{I}=\sqrt{-1}$. 
$\Omega_{ik}^{-1}(\mathbf{n})= C_{ijkl}n_{j}n_{l}$ is the inverse of Green's function,
$\varepsilon_{ij}^{p}$ is the position-independent part of the eigenstrain
tensor (Eq.~\eqref{eigenstrain}) associated with the field $\psi_{p}(\mathbf{r})$, and
$\Delta \psi_{p}(\mathbf{r}) = \psi_{p}(\mathbf{r}) - \bar{\psi}_{p}$, 
and $\sigma_{ij}^{p} = C_{ijkl}\varepsilon_{kl}^{p}$.
Using the displacement field from Eq.~\eqref{displacement}, we express the   
elastic energy in reciprocal space as follows
\begin{equation}
    \mathcal{F}_{\textrm{elastic}} = \frac{1}{2}\sum_{p,q=0}^{5}\int\frac{d^{3}\mathbf{k}}{(2\pi)^3}
                B_{pq}(\mathbf{n})\tilde{\psi}_{p}(\mathbf{k})\tilde{\psi}_{q}^{\ast}(\mathbf{k}),
\label{elasticenergy}
\end{equation}
where $B_{pq}(\mathbf{n}) = C_{ijkl}\varepsilon_{ij}^{p}\varepsilon_{kl}^{q} 
- n_{i}\sigma_{ij}^{p}\Omega_{jk}(\mathbf{n})\sigma_{kl}^{q}n_{l}$, 
and $\tilde{\psi}^{\ast}$ refers to the complex conjugate of $\tilde{\psi}$.
A volume of $(2\pi)^3/V$ about $\mathbf{k} = 0$ is excluded from the integration in Eq.~\eqref{elasticenergy}.

Assuming the domain wall energies to be isotropic, the gradient energy density $f_{\textrm{gradient}}$ in Eq.~\eqref{TotalF} is written as~\cite{zhang2005computational}
\begin{equation}
f_{\textrm{gradient}} = \frac{1}{2}G_{11}(P_{1,1}^{2} +P_{1,2}^{2}+P_{1,3}^{2} +P_{2,1}^{2}+P_{2,2}^{2}+P_{2,3}^{2}+P_{3,1}^{2}+P_{3,2}^{2}+P_{3,3}^{2}),
\label{fgrad}
\end{equation}
where $P_{i,j}$ denotes $\partial P_i/ \partial x_j$. $G_{11}$ is a positive gradient energy coefficient associated with the 
gradients in polarization field. Although gradient energy coefficients generally form a fourth-rank tensor whose components 
can be determined using first-principles calculations~\cite{neaton2005first,lubk2009first}, 
thus far no such calculation is reported for BZCT. Therefore, we assume 
a scalar gradient energy coefficient in our model.  

To include the effects of external fields in our thermodynamic stability analysis, we introduce external electromechanical fields in our model. For example, for a mechanically constrained (clamped) system that is not allowed to deform along any direction ($\bar{\varepsilon}_{ij}=0$), we introduce a uniform stress field $\bar{\sigma}_{ij}$ whose magnitude increases quadratically with polarization~\cite{chen2008phase}: 
\begin{align}
\bar{\sigma}_{ij}&=\frac{1}{V}\int_V{C_{ijkl}\varepsilon_{kl}^{0}(\mathbf{r})}dV \\ 
                 &=\frac{1}{V}\int_V{C_{ijkl}Q_{klmn}P_m(\mathbf{r})P_n(\mathbf{r})}dV \\
                 &=q_{ijmn}\langle {P_m P_n} \rangle  \; (\because q_{ijmn}=C_{ijkl}Q_{klmn}).
\end{align}
Here, $\langle \cdot \rangle$ denotes the volume average of the quantity inside the angular brackets. 
When the macroscopic average stress $\bar{\sigma}_{ij}$ is zero everywhere in the system, we call it 
stress-free or unconstrained. To define a mechanically constrained state where the system is 
clamped in all directions, we set  $\bar{\varepsilon}_{ij}=0$. 
Considering bulk and elastic contributions to the total free energy density given in  Eqns.~\eqref{fbulk} and~\eqref{felas})~\cite{hlinka2006phenomenological}, 
we construct the thermodynamic potential $f^{\textrm{constrained}}$ 
for mechanically constrained BZCT, given as:
\begin{equation}
\begin{split}
     f^{\textrm{constrained}} &=\frac{1}{2}\alpha_{1}(P_{1}^{2} + P_{2}^{2} + P_{3}^{2}) 
    +\frac{1}{4}\alpha_{11}^{e}(P_{1}^{4} + P_{2}^{4} + P_{3}^{4}) + \frac{1}{6}\alpha_{111}(P_{1}^{6} + P_{2}^{6} + P_{3}^{6})\\
    &+ \frac{1}{2}\alpha_{12}^{e}(P_{1}^{2}P_{2}^{2} + P_{2}^{2}P_{3}^{2} + P_{3}^{2}P_{1}^{2})
    +\frac{1}{2} \alpha_{112}(P_{1}^{4}(P_{2}^{2} + P_{3}^{2}) + P_{2}^{4}(P_{3}^{2} + P_{1}^{2})\\
    &+ P_{3}^{4}(P_{1}^{2} + P_{2}^{2})) 
    +\frac{1}{6} \alpha_{123}P_{1}^{2}P_{2}^{2}P_{3}^{2},
\end{split}
\label{ftotalconstrained}
\end{equation}
where
\begin{subequations}
\begin{align}
    \alpha_{11}^{e} &= \alpha_{11} + 4\left\{\frac{1}{6}\bigg[\frac{\hat{q}_{11}^{2}}{\hat{C}_{11}}
       + 2\frac{\hat{q}_{22}^{2}}{\hat{C}_{22}}\bigg]\right\}\label{effec_a11},\\
    \alpha_{12}^{e} &= \alpha_{12} + 2\left\{\frac{1}{6}\bigg[\frac{2\hat{q}_{11}^{2}}{\hat{C}_{11}}
        - 2\frac{\hat{q}_{22}^{2}}{\hat{C}_{22}} + 3\frac{q_{44}^{2}}{C_{44}}\bigg]\label{effec_a12}\right\},
\end{align}
        \label{consunconsrelations}
\end{subequations}
with
\begin{equation}
\begin{split}
    &\hat{C}_{11} = C_{11} + 2C_{12},\\
    &\hat{C}_{22} = C_{11} - C_{12},\\
    &\hat{q}_{11} = q_{11} + 2q_{12},\\
    &\hat{q}_{22} = q_{11} - q_{12}.
    \end{split}
    \label{modi_cq}
\end{equation}
Here, $\alpha_{11}^e$ and $\alpha_{12}^e$ are the modified Landau coefficient for a clamped system. 
The effective electrostrictive coefficients $q_{ij}$ in Eq.~\eqref{modi_cq} are 
defined as  $q_{11}=C_{11}Q_{11}+2C_{12}Q_{12}$, $q_{12}=C_{11}Q_{12}+C_{12}(Q_{11}+Q_{12})$, 
and $q_{44}=2C_{44} Q_{44}$. 
Thus, free energies of $C$, $T$, $O$, $R$ phases in mechanically constrained BZCT can be obtained as 
\begin{subequations}
\begin{align}
f_{\textrm{C}}^{\textrm{constrained}} &= 0, \label{first:a}\\
f_{\textrm{T}}^{\textrm{constrained}} &= \frac{1}{2}\alpha_{1}P^{2} + \frac{1}{4}\alpha_{11}^{e}P^{4} 
+ \frac{1}{6}\alpha_{111} P^{6}, \label{first:b}\\
f_{\textrm{O}}^{\textrm{constrained}} &= \frac{1}{2}\alpha_{1}P^{2} + \bigg(\frac{\alpha_{11}^{e}}{8} + \frac{\alpha_{12}^{e}}{8}\bigg)P^{4} + \bigg(\frac{\alpha_{111}}{24} + \frac{\alpha_{112}}{8}\bigg)P^{6},\label{first:c}\\
f_{\textrm{R}}^{\textrm{constrained}} &= \alpha_{1}P^{2} + \bigg(\frac{\alpha_{11}^{e}}{12} + \frac{\alpha_{12}^{e}}{3}\bigg)P^{6} + \bigg(\frac{\alpha_{111}}{54} + \frac{2\alpha_{112}}{9} + \frac{\alpha_{123}}{162}\bigg)P^{6}. \label{first:d}
\end{align}
\label{f_phcons}
\end{subequations}
Minimization of the expressions in Eq.~\eqref{f_phcons} with respect to $P$ yields the equilibrium spontaneous polarization of each phase in constrained BZCT as a function of temperature and composition. 
In Section~\ref{res}, we will use the comparison between coefficients $\alpha_{11}$, $\alpha_{12}$ for the stress-free system and $\alpha_{11}^e$, $\alpha_{12}^e$ for the constrained system to determine electrostrictive anisotropy effects on diffusionless phase diagrams and to examine changes in phase stability with the application of constraint.   

To compare polarization switching characteristics between stress-free and constrained BZCT,  
we include an additional term $\mathbf{P}\cdot\mathbf{E_\textrm{ext}}$ in Eqns.~\eqref{fbulknewaniso3} and~\eqref{ftotalconstrained}~\cite{chandra2007landau} to derive analytical expressions 
relating externally applied electric field $\mathbf{E_\textrm{ext}}$ to polarization 
and measure ``theoretical'' $P-E$ loops analytically. Minimization of total energy with respect to polarization provides relations 
between external electric field and polarization for stress-free and constrained systems:
\begin{align}
E_{ext}^{\textrm{stress-free}}(\phi) &= \frac{\partial f_{\phi}^{\textrm{stress-free}}}{\partial P}\nonumber\\
E_{ext}^{\textrm{constrained}}(\phi) &= \frac{\partial f_{\phi}^{\textrm{constrained}}}{\partial P},
\end{align}
where $\phi=T,O,R$.  However, one should note that the analytically measured 
characteristics ignore spatial variations in the polarization field and the local interactions. 

\subsection*{Phase-field model}
To incorporate spatial interactions between the fields, we derive the following Euler-Lagrange equation 
for the polarization field $\mathbf{P}(\mathbf{r},t)$ that minimizes the total free energy of the system (Eq.~\eqref{TotalF}) at a given temperature and composition:
\begin{align}
    &\frac{\delta \mathcal{F}}{\delta{P_i}} = 0 \nonumber\\
   \implies&\frac{\partial f_{\textrm{bulk}}}{\partial P_i}\Big|_{P_i^{eq}} - E_i^{eq} 
   + \sigma^{eq}_{ij}\frac{\partial \varepsilon_{ij}^{el}}{\partial P_{i}}\Big|_{P_i^{eq}}
    - G_{11}\frac{\partial^2 P_i}{\partial r_j^2}\Big|_{P_i^{eq}}
    =0\; (i=1,2,3).
    \label{variational}
\end{align}
Here, $E_i^{eq}$ and $\sigma^{eq}_{ij}$ are obtained by solving electrostatic 
and mechanical equilibrium equations (Eqns.~\eqref{poisson} and~\eqref{meche}).
The variational derivative $\delta\mathcal{F}/\delta{P_i}$ in Eq.~\eqref{variational} defines the 
driving force for domain evolution in the ferroelectric system.  

Thus, the Allen-Cahn equation governing spatiotemporal evolution of $\mathbf{P}(\mathbf{r},t)$ is given as
\begin{equation}
    \frac{\partial P_{i}(\mathbf{r},t)}{\partial t} 
     = -L\frac{\delta \mathcal{F}}{\delta P_{i}}\; (i=1,2,3),
\label{ginzburg}
\end{equation}
where $L$ is a relaxation coefficient related to domain wall mobility in an overdamped system.

We solve Eq.~\eqref{ginzburg} coupled with electrostatic and mechanical equilibrium equations (Eqns.~\eqref{poisson},~\eqref{meche}) in three-dimensions using a semi-implicit Fourier spectral method~\cite{chen1998applications}.
Eq.~\eqref{ginzburg} in Fourier space is given as
 \begin{equation}
     \frac{\partial \Tilde{P}_{i}(\mathbf{k},t)}{\partial t} 
     = -L\left[\frac{\delta \mathcal{F}}{\delta P_{i}}\right]_{k}, 
     \label{fourier_tdgl}
 \end{equation}
where $\Tilde{P}_{i}(\mathbf{k},t)$ is the Fourier transform of $P_{i}(\mathbf{r},t)$,
and $\left[\frac{\delta \mathcal{F}}{\delta P_{i}}\right]_{k}$ represents the Fourier transform of the
driving force given in Eq.~\eqref{variational}.

We numerically approximate Eq.~\eqref{fourier_tdgl} using a semi-implicit Fourier spectral scheme for spatial discretization and a forward Euler scheme for temporal discretization~\cite{chen1998applications}: 
\begin{equation}
\begin{split}
    \Tilde{P}_{i}^{n+1}(\mathbf{k},t) &= \frac{1}{1 + L\Delta t G_{11}}\Bigg[\Tilde{P}_{i}^{n}(\mathbf{k},t)
     - L\Delta t \Bigg\{\left(\frac{\partial f_{bulk}}{\partial P_{i}}\right)_{k} \\&+ \left(\frac{\partial f_{electric}}{\partial P_{i}}\right)_{k} + \left(\frac{\partial f_{elastic}}{\partial P_{i}}\right)_{k}\Bigg\}\Bigg].
     \end{split}
     \label{discritization_tdgl}
\end{equation}
Eqns.~\eqref{poisson} and~\eqref{meche} are also solved in the Fourier space to obtain
$(\partial f_{electric}/\partial P_{i})_k$ and $(\partial f_{elastic}/\partial P_{i})_k$ at each time step.
We use FFTW library along with OpenMP parallelization to numerically implement our phase-field model.  

\section{Results and discussion}
\label{res}
In this section, we present the results of thermodynamic stability analysis 
of ferroelectric domains in BZCT system using the potential given in Eq.~\eqref{fbulknewaniso3}.
We also present the results of three-dimensional phase-field simulations of 
domain evolution in equimolar BZCT and the corresponding switching behaviour as a function of applied electromechanical fields. 

We scale and nondimensionalize all parameters used in our study using 
characteristic values of length ($L_c$), energy ($E_c$), charge ($q_c$) and time ($\tau_c$). 
Using the experimental values of spontaneous polarization $P_{s}$=\SI{0.2}{\coulomb\per\meter\tothe{2}} and the reciprocal of dielectric
susceptibility $|\alpha_{1}|_{\theta=298 \si{\kelvin}} = 2.2781\times 10^{7}\si{\joule\meter\coulomb\tothe{-2}}$ of equimolar BZCT at room temperature (298 \si{\kelvin}), 
we obtain $E_c=4.1\times10^{-21}\si{\joule}$ and $L_c=1.65 \si{\nano\meter}$~\cite{liu2009large,turygin2015domain}. 
We use the factor $|\alpha_{1}|_{\theta=298 \si{\kelvin}}P_{s}^{2} = 9.11\times 10^{5} \si{\joule\meter\tothe{3}}$ to normalize all parameters appearing in the governing equations 
(Eqns.~\eqref{poisson},~\eqref{meche},~\eqref{ginzburg}). The characteristic time $\tau$ is determined using the relation $|\alpha_1|_{\theta=298 \si{\kelvin}} L \tau = 1$ where $L$ denotes the dimensional value of relaxation coefficient.  
The dimensional  gradient energy coefficient $G_{110}$ is given as
$G_{110}= L_c^2|\alpha_{1}|_{\theta=298 \si{\kelvin}} = 6.2 \times 10^{-11}\si{\joule\metre\tothe{3}\per\coulomb\tothe{2}}$ corresponding to 
a nondimensional value $G_{110}^{\prime}=1$.



Phase-field simulations are carried out in a $200\Delta \times 200 \Delta \times 200\Delta$ simulation box where 
$\Delta =$\SI{0.32}{\nano\meter} is the grid spacing (corresponding to a non-dimensional spacing $\Delta^{\prime} = 1$).
We choose a 
nondimensional time step $\Delta t^{\prime} = 0.01$ to ensure high spatiotemporal accuracy.

Since our model uses spontaneous polarization as the order parameter field, 
we specify a nondimensional background dielectric constant $\epsilon_{b}= 8$ 
to describe dielectric screening effects of high
frequency polar phonon modes (e.g., electronic polarization)~\cite{marton2006simulation}. 

\begin{figure}[htbp]
    \centering
    \subfloat[]{\label{electrostrictive}\includegraphics[width=.5\linewidth]{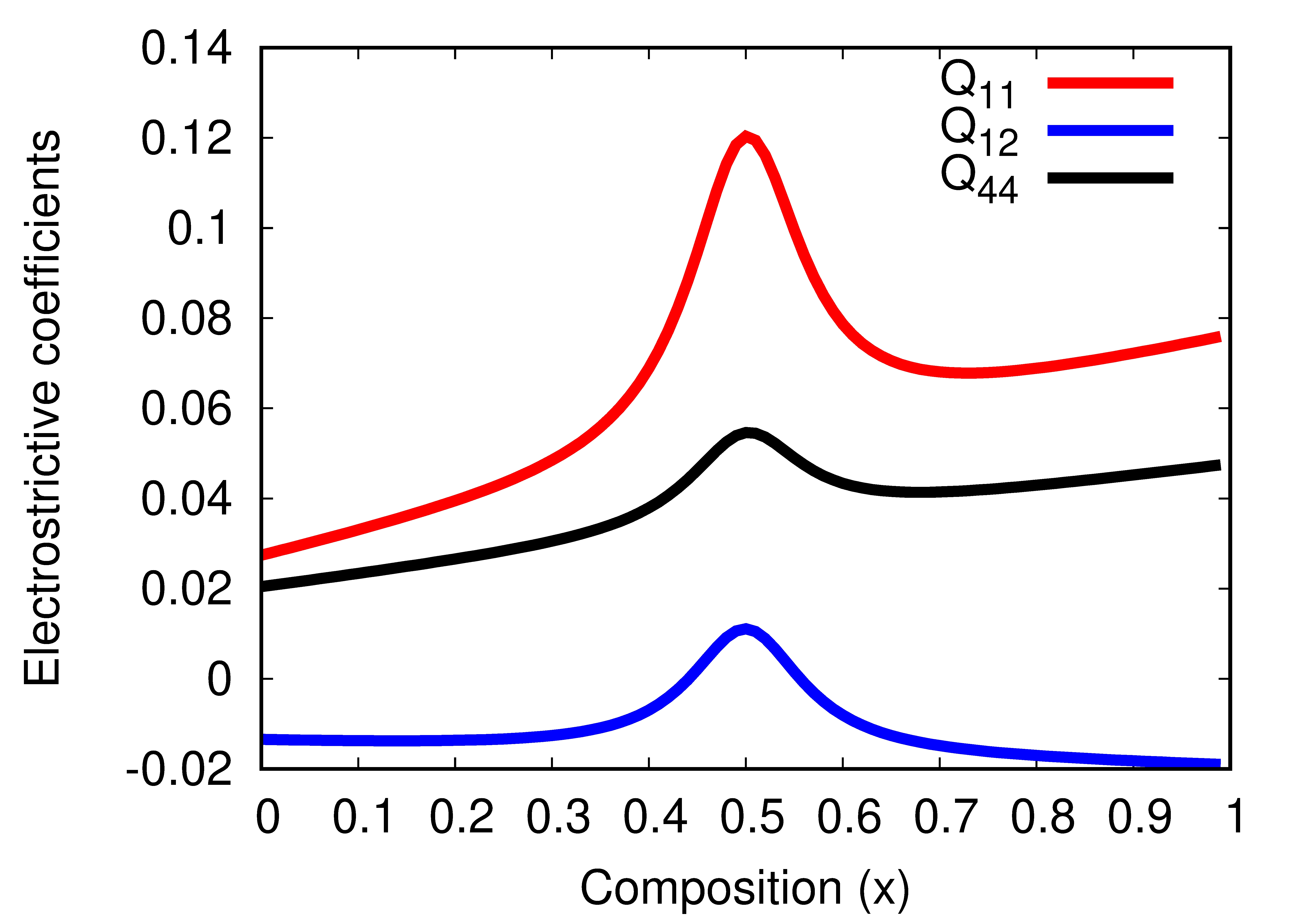}}\hfill
     \subfloat[]{\label{electrostrictive1}\includegraphics[width=.5\linewidth]{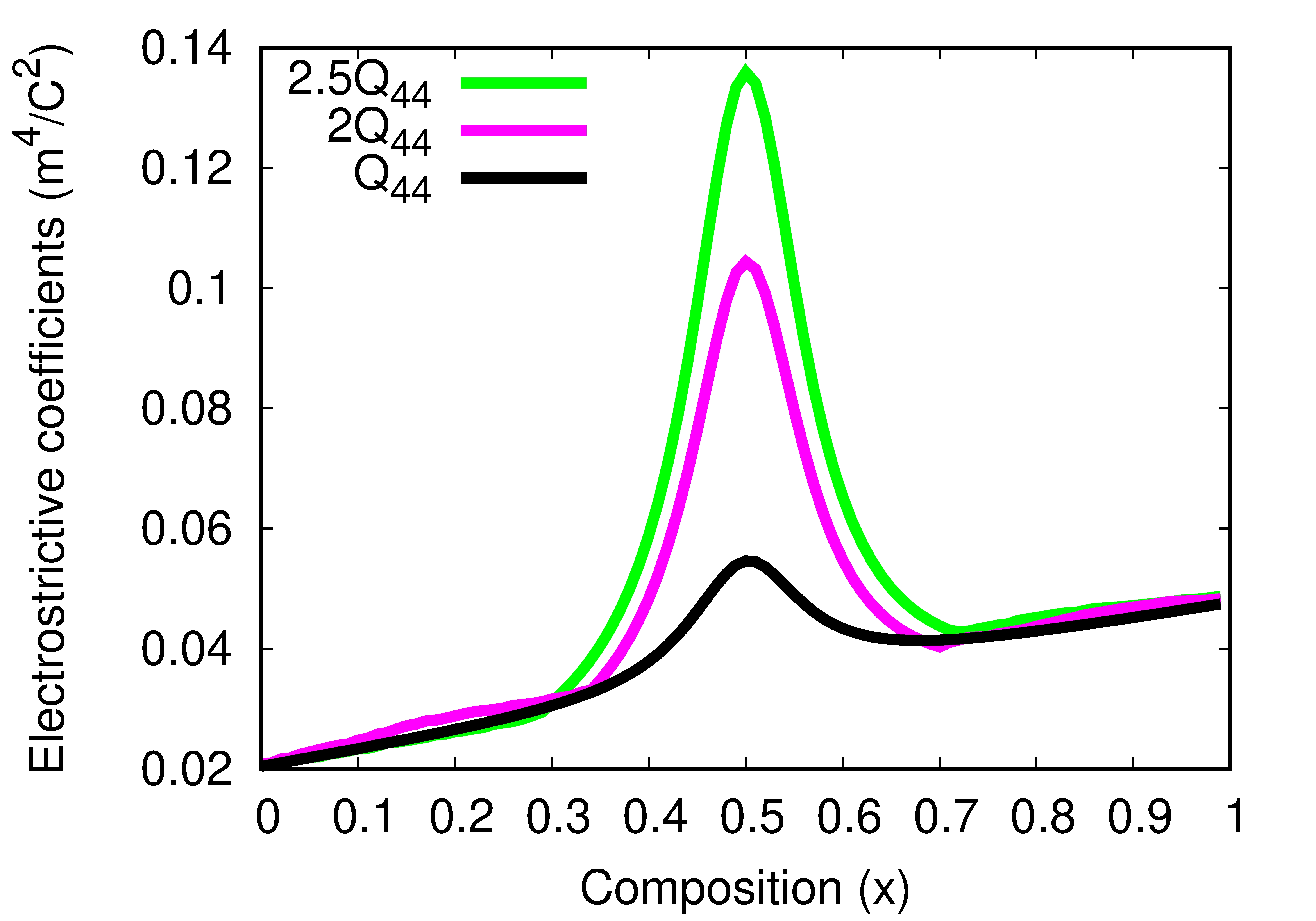}}\hfill
    \subfloat[]{\label{elastic}\includegraphics[width=.5\linewidth]{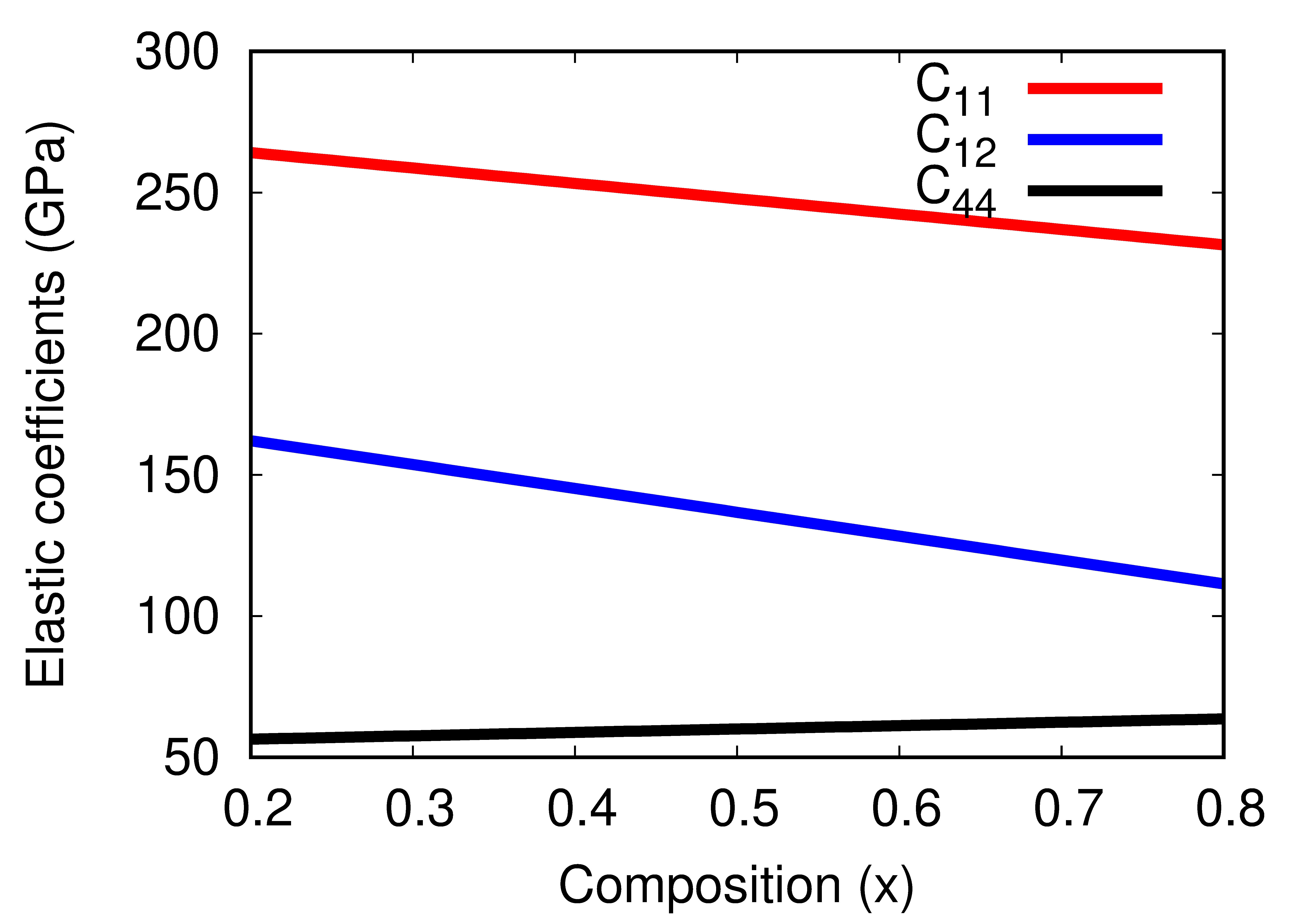}}
    \caption{(a) Variation of $Q_{11}$, $Q_{12}$, $Q_{44}$ with composition $x$ when
     $Q_{z}=1$. (b) Change in $Q_z=\frac{2Q_{44}}{Q_{11}-Q_{12}}$ is realized through the
     change in $Q_{44}$ keeping $Q_{11}$ and $Q_{12}$ unchanged: Case 1: $Q_z = 1, 
     Q_{44}=Q_{44}$, Case 2: $Q_z=2, Q_{44}=2Q_{44}$, Case 3: $Q_z=2.5, Q_{44}=2.5Q_{44}$. 
     (c) Variation of elastic moduli $C_{11}$ and $C_{12}$ as a function of composition $x$.}
    \label{fig:coefficients}
\end{figure}

Moreover, we assume the elastic and electrostrictive coefficients 
($C_{ij}$ and $Q_{ij}$ $i,j=1,\ldots,6$) to 
be invariant with temperature $\theta$ as long as $\theta$ does not 
exceed the Curie temperature $\theta_c(x)$ 
for a given composition $x$.  Due to limited experimental data across entire composition range of 
BZCT system, we have assumed the functional forms of composition dependency of these coefficients 
to be similar to those used in PZT. Using experimental values of piezoelectric voltage constants 
and elastic moduli for the terminal and equimolar compositions ($x=0,1,0.5$) of BZCT as fitting 
parameters~\cite{xue2011elastic,newnham2005properties}, we  arrive at the following relations for 
$C_{ij}(x)$ and $Q_{ij}(x)$:
\begin{equation}
\begin{split}
     &C_{ij}(x)=x C_{ij}^{\textrm{BZT}} + (1-x) C_{ijkl}^{\textrm{BCT}},\\
     &Q_{11}(x) = 0.04895x + 0.02605 + 0.069778/(1 + 200(x-0.5)^{2}),\\
     &Q_{12}(x) = -0.0056x - 0.01400 + 0.0279/(1 + 200(x-0.5)^{2}),\\
     &Q_{44}(x) = 0.02728x + 0.02002 + 0.02095/(1 + 200(x-0.5)^{2}).
\end{split}
    \label{elastic_coeff}
\end{equation}
Fig.~\ref{fig:coefficients} shows the variation of $Q_{ij}$ and $C_{ij}$ with composition. Although the elastic moduli $C_{ij}$ vary linearly with composition, the electrostrictive coefficients $Q_{11}$, $Q_{12}$ and $Q_{44}$ show a pronounced maximum at $x=0.5$.    

Both dimensional and nondimensional forms of all coefficients used in our study are listed in Table~\ref{table:normalized}. Here, all parameters are normalized using $|\alpha_1|_{\theta=298 K}P_0^2\si{\joule\meter\tothe{-3}}$ where $P_0=0.2\si{\coulomb\meter\tothe{-2}}$ is the experimentally determined spontaneous polarization of BCZT at $x=0.5,\theta=298K$, and the 
non-dimensional temperature $\theta^{\prime} = \theta/298$. 
\begin{table}[htpb]
\caption{Temperature and composition-dependent parameters used in the study} 
\centering 
\begin{tabular}{| m{3cm}| | m{5.5cm}|| m{5.5cm}|}
\hline\hline 
 Coefficients & Dimensional form & Non-dimensional form\\ [2ex] 
\hline 
 $\alpha_{1}(\theta,x)$~\cite{yang2016mechanisms} & 
 \small{$4.142\times10^{5}(\theta-\theta_{c}(x))$} $\si{\joule\metre\per\coulomb\tothe{2}}$ & \small{$0.0182(\theta^{\prime}-\theta^{\prime}_{c}(x))$} \\ %
 \hline
$\alpha_{11}(\theta,x)$~\cite{yang2016mechanisms} & \small{$1.512\times10^{6}(\theta - 
\theta_{\textrm{quad}}) - 2.4\times10^{8}(x - x_{\textrm{quad}})$} 
$\si{\joule\metre\tothe{5}\per\coulomb\tothe{4}}$ & \small{$0.00266 
(\theta^{\prime} - \theta^{\prime}_\textrm{quad})$}\small{$- 0.4214(x - x_{\textrm{quad}})$}
 \\
 \hline
 $ \alpha_{12}(\theta,x)$~\cite{yang2016mechanisms}& \small{$-2.4\times 10^{8}(x - x_{\textrm{quad}})$} $\si{\joule\metre\tothe{5}\per\coulomb\tothe{4}}$   
 & \small{$-0.4214(x - x_{\textrm{quad}})$}
 \\
 \hline
 $\alpha_{111}(\theta,x) $~\cite{yang2016mechanisms}& $2.329\times 10^{9} + 1.2\times 10^{8}(x - x_{\textrm{quad}})$ $\si{\joule\metre\tothe{9}\per\coulomb\tothe{6}}$& \small{$0.164 + 0.00843(x - x_{\textrm{quad}})$}\\
 \hline
 $\alpha_{112}(\theta,x)$~\cite{yang2016mechanisms}& $7.764\times 10^{8} - 1.8\times 10^{8}(x - x_{\textrm{quad}})$ $\si{\joule\metre\tothe{9}\per\coulomb\tothe{6}}$& \small{$0.0545 - 0.0126(x - x_{\textrm{quad}})$}\\
\hline
 $\alpha_{123}(\theta,x)$~\cite{yang2016mechanisms}& $-4.658\times 10^{9} + 2.1\times10^{8}(\theta - \theta_{\textrm{quad}}) 
                      + 9.976\times 10^{10}(x - x_{\textrm{quad}})$ $\si{\joule\metre\tothe{9}\per\coulomb\tothe{6}}$
                      & \small{$-0.327 + 0.01474(\theta^{\prime} - \theta^{\prime}_\textrm{quad})$}
                      \small{$+ 7.007(x - x_{\textrm{quad}})$}\\
 \hline
 $Q_{11}(T,x)$~\cite{li2005ferroelectric,xue2011elastic,haun1989electrostrictive}&$0.04895x + 0.02605 + 0.069778/(1 + 200(x-0.5)^{2})\si{\meter\tothe{4}\coulomb\tothe{-2}}$&$0.0000196x + 0.001042 + 0.00279/(1 + 200(x-0.5)^{2})$\\
 \hline
 $Q_{12}(T,x)$~\cite{li2005ferroelectric,xue2011elastic,haun1989electrostrictive}&$-0.0056x - 0.01400 + 0.0279/(1 + 200(x-0.5)^{2})\si{\meter\tothe{4}\coulomb\tothe{-2}}$&$-0.000224x - 0.00056 + 0.00117/(1 + 200(x-0.5)^{2})$\\
\hline 
$Q_{44}(T,x)$~\cite{li2005ferroelectric,xue2011elastic,haun1989electrostrictive}&$0.02728x + 0.02002 + 0.02095/(1 + 200(x-0.5)^{2})\si{\meter\tothe{4}\coulomb\tothe{-2}}$&$0.00109x + 0.0008 + 0.000838/(1 + 200(x-0.5)^{2})$\\
\hline 
 $C_{11}(T,x)$~\cite{garbarz2011elastic,xue2011elastic}&$220x + 275(1-x)\si{\giga\pascal}$&$241429.2x + 301786.5(1-x)$\\
 \hline
 $C_{12}(T,x)$~\cite{garbarz2011elastic,xue2011elastic}&$94x + 179(1-x)\si{\giga\pascal}$&$103156.2x + 196435.6(1-x)$\\
\hline 
$C_{44}(T,x)$~\cite{garbarz2011elastic,xue2011elastic}&$66x + 54(1-x)\si{\giga\pascal}$&$72428.8x + 59259.9(1-x)$\\
\hline 
\end{tabular}
\label{table:normalized} %
\end{table}

Several experimental studies have found a strong correlation between dielectric and piezoelectric anisotropy and attributed this to the intrinsic anisotropy in electrostrictive 
coefficients stemming from the change in structure of BO$_6$ 
oxygen octahedra in perovskite~\cite{li2014electrostrictive, budimir2003piezoelectric,damjanovic2009comments}. 
Thus, when processing conditions (involving change in chemistry and application of external electromechanical fields) trigger a change in the geometry of oxygen octahedra of the ferroelectric perovskite (manifested by change in tilt angle of BO$_6$ octahedra), there is a subsequent change in the inherent anisotropy associated with electrostrictive coefficients~\cite{yamada1972electromechanical,damjanovic2005contributions}. 
To understand the role of electromechanical anisotropy 
on the stability of ferroelectric domains and consequent switching dynamics, we define an electrostrictive anisotropy parameter $Q_z=\frac{2Q_{44}}{Q_{11}-Q_{12}}$ and systematically investigate the role of $Q_z$ on switching behaviour of BZCT. Studies also report  
another measure of electrostrictive anisotropy $Q_a=\frac{Q_{11}-Q_{12}}{Q_{44}}$~\cite{li2014electrostrictive}. Our definition of $Q_z$ 
is in the same spirit as the Zener anisotropy parameter $A_Z$ associated with elastic stiffness tensor that distinguishes between elastically soft and hard directions in orthotropic materials~\cite{newnham2005properties}. 
Moreover, one should note that electrostrictive anisotropy and  
anisotropy in spontaneous strain are related because 
the spatially-invariant part of spontaneous strain (eigenstrain $\varepsilon^0$) 
is solely a function of electrostrictive coefficients $Q$ (see Eq.~\ref{eigenstrain}).   

\begin{figure}[htbp]
    \centering
    \subfloat[Two dimensional representation of the  orientation dependence of electrostrictive 
    coefficient $Q_{33}^{*}$ for BZCT system.]{\label{2daniso}\includegraphics[width=.5\linewidth]{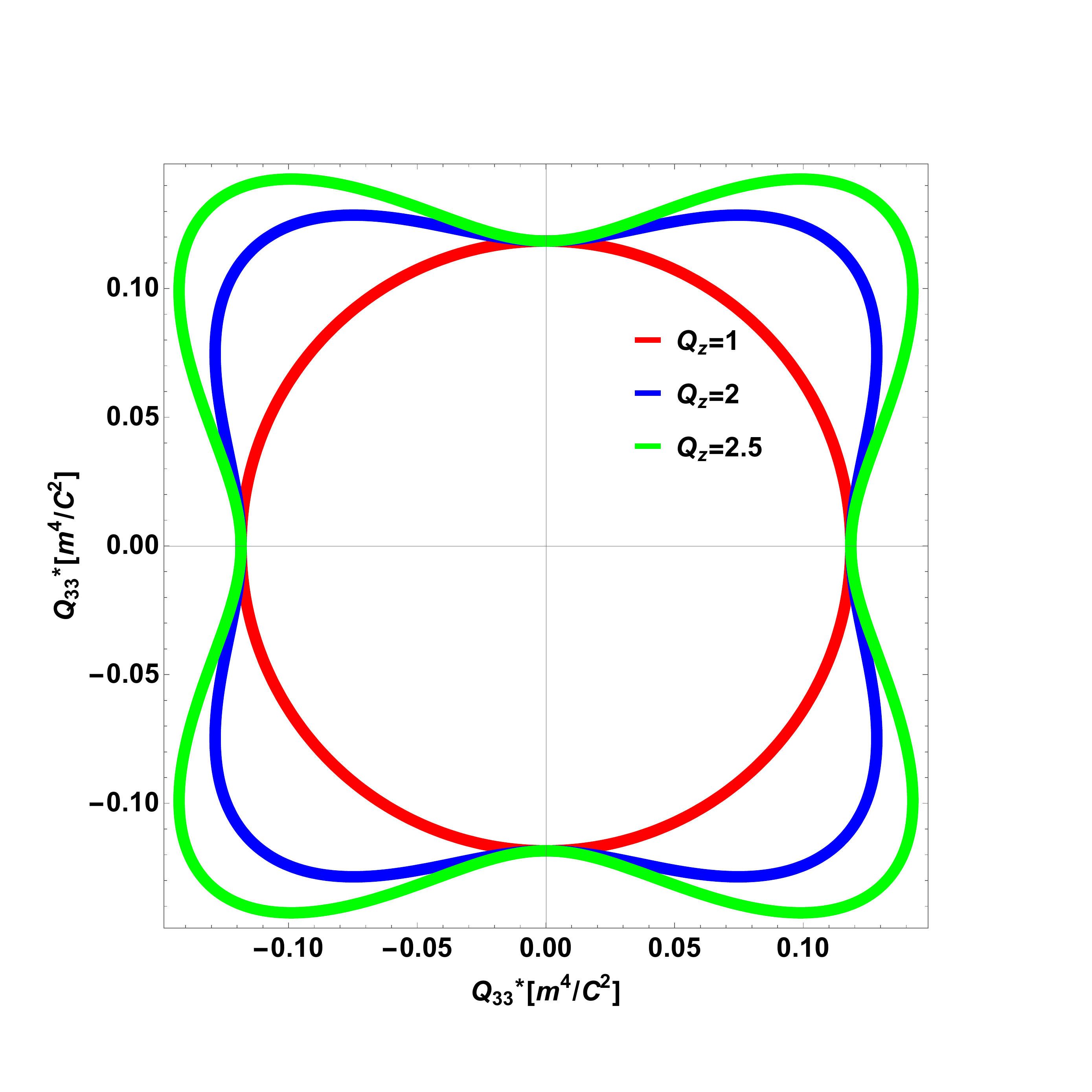}}\hfill
    \subfloat[Case 1]{\label{Q2}\includegraphics[width=.38\linewidth]{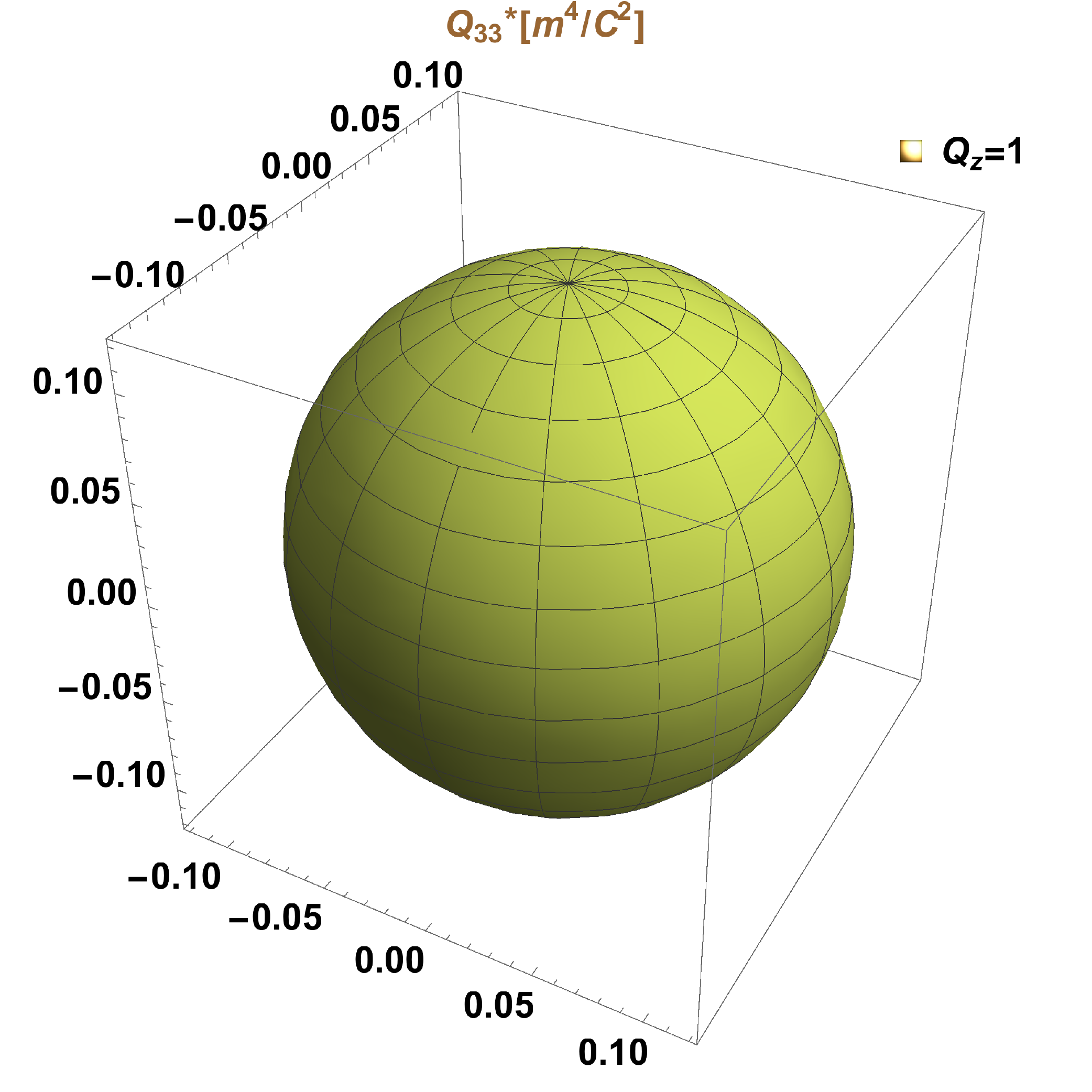}}
     \subfloat[Case 2]{\label{Q1}\includegraphics[width=.38\linewidth]{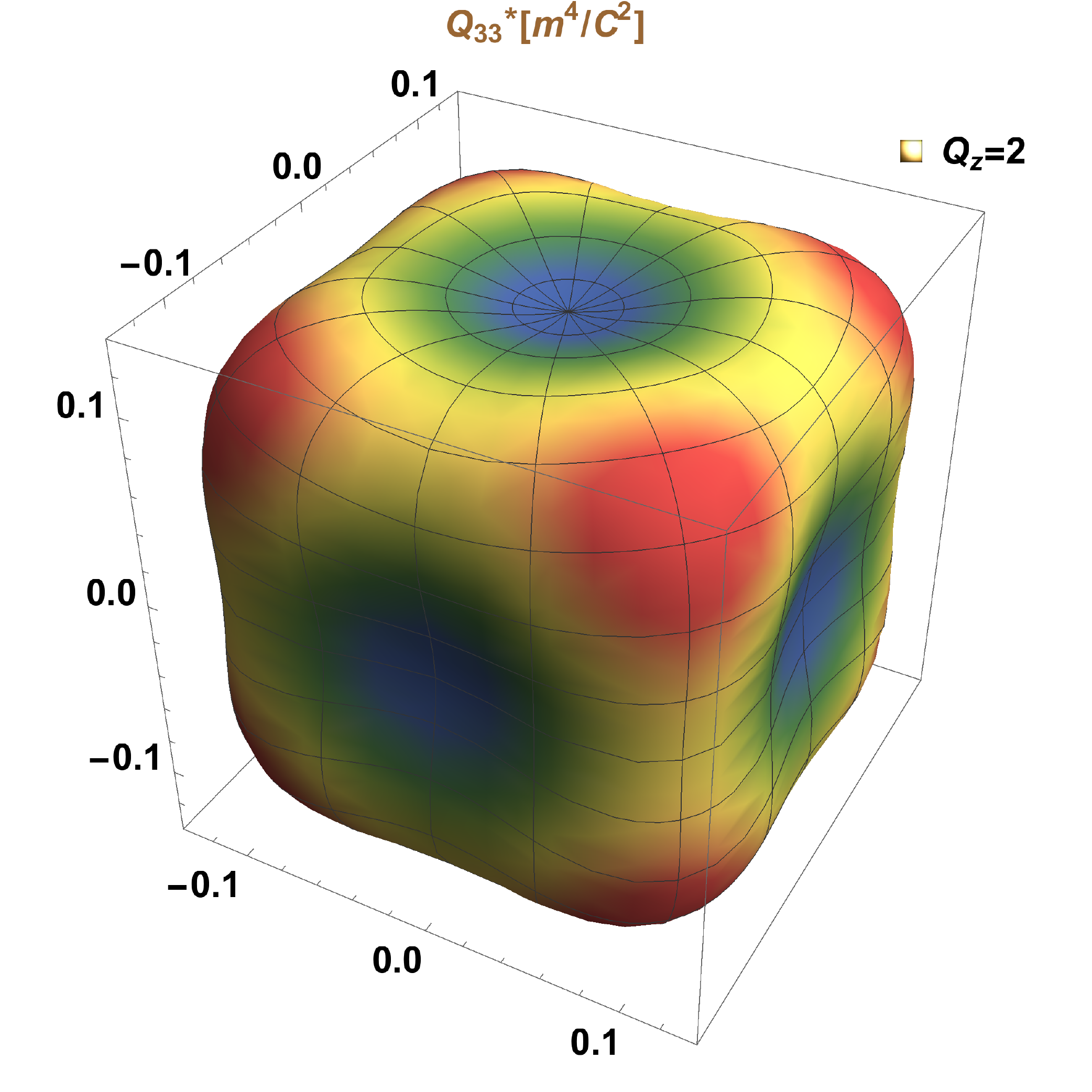}}
      \subfloat[Case 3]{\label{Q0.8}\includegraphics[width=.38\linewidth]{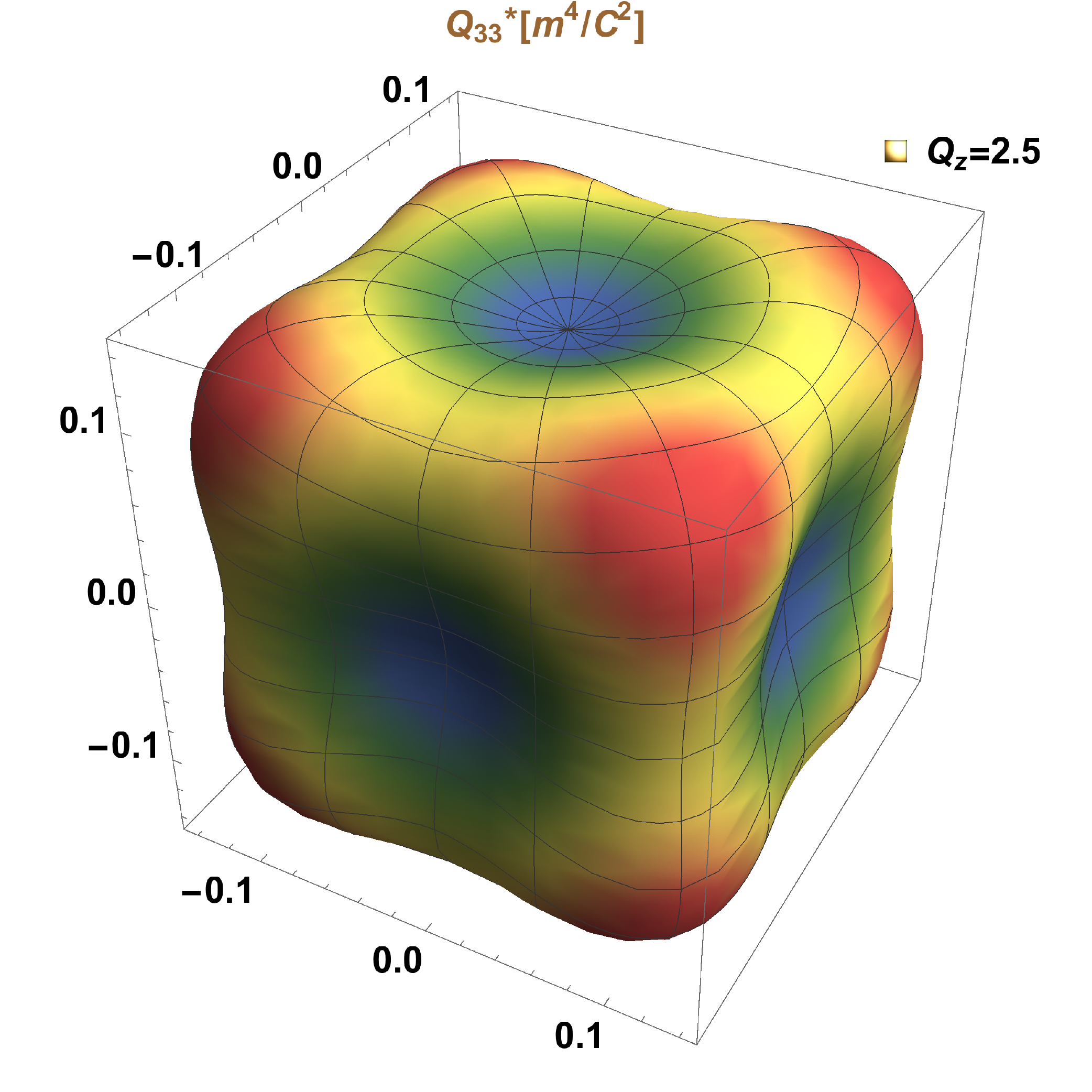}}
    \caption{The orientation dependence of electrostrictive coefficient $Q_{33}^{*}$ for BZCT system.
    for three different cases.}
    \label{fig:anisotropy}
\end{figure}

To demonstrate the correlations between electrostrictive anisotropy 
and domain stability/switching in BZCT, 
we choose three cases based on the value of $Q_z$:
\begin{itemize}
    \item Case 1: $Q_z=1$ ($Q_a=2$),
    \item Case 2: $Q_z=2$ ($Q_a=1$),
    \item Case 3: $Q_z=2.5$ ($Q_a=0.8$).
\end{itemize} 
We define an electrostrictive modulus $Q_{33}^{\ast}$  as follows
\begin{equation}
    \frac{1}{Q_{33}^{\ast}}=\frac{Q_{11}+Q_{12}}{(Q_{11}-Q_{12})(Q_{11}+2Q_{12})}
                          -\frac{2(Q_z - 1)}{Q_z(Q_{11}-Q_{12})}(l_1^2l_2^2 + l_2^2l_3^2+l_1^2l_3^2), 
    \label{elecmod}
\end{equation}
to show the orientation dependence of fourth-order electrostrictive tensor 
in two and three dimensions (Fig.~\ref{fig:anisotropy}).
When $Q_z=1$ (Case 1), the representation quadric is spherical indicating isotropic behaviour, whereas when $Q_z>1$ (Cases 2 and 3), the surfaces become anisotropic showing lower values of $Q_{33}^{\ast}$ along $\langle 100 \rangle$ directions. Note that $Q_z>1$ introduces 
anisotropy in electrostriction in the paraelectric state. Thus 
we need to use Eq.~\ref{eigenstrain} to determine 
the spontaneous strain components of the ferroelectric phases 
and define elastic energy according to Eq.~\eqref{elasticenergy} for a given $Q_z$. 
Since all ferroelectric variants are also ferroelastic, preferred orientations of 
variants are determined by the minimization of elastic interactions between the variants
(corresponding to cusps in $B_{pq}(\mathbf n)$).
However, one should note than the stable domain configuration in a stress-free, electrically neutral ferroelectric system requires minimization of total energy arising from coupled elastic and electric interactions.

\subsection*{Diffusionless phase diagram}
Minimization of $f_{\textrm{bulk}}^{\textrm{modified}}$ in Eq.~\eqref{fbulknewaniso3} determines 
the polarization states of the stable phases in electrically neutral, stress-free BZCT at any given temperature $\theta$ and composition $x$. 

Phase-coexistence conditions are derived as follows. When $O$ alone is the stable phase, phase stability condition using Eq~\eqref{free_ph} leads to an inequality
\begin{equation}
    \frac{\beta_{2}}{\gamma_{2} P_{eq}^{2}} > -1.
    \label{stableO}
\end{equation}
This implies that $|\beta_{2}| > \gamma_{2}P_{eq}^{2}$.
However, when two phases coexist (e.g., $T$ and $O$), 
the stability condition becomes an equality given as
\begin{equation}
    \frac{1}{2}\beta_{2} P_{eq}^{4} = -\frac{1}{12}\gamma_{2}P_{eq}^{6},
    \label{TOphaseequi}
\end{equation}
implying $|\beta_{2}| =\gamma_{2} P_{eq}^{2}$.
Similarly, phase stability conditions for a three-phase coexistence (i.e., a stable 
phase mixture of $T$, $R$ and $O$) is given as
\begin{equation}
    \frac{1}{2}\beta_{2}P_{eq}^{4} = -\frac{1}{12}\gamma_{2}P_{eq}^{6} = \frac{1}{36}\gamma_{3}P_{eq}^{6}.
    \label{TORphaseequi}
\end{equation}
For a given temperature and any composition lying within the MPR, 
we may assume either $\gamma_2$ or 
$\gamma_3$ to be a constant. Assuming a fixed value of $\gamma_2$ at a given 
temperature and composition and requiring $\gamma_2$ to be positive to ensure a first order transition, we find the absolute value of $\beta_2$ to decrease with a corresponding increase in the number of degenerate minima of Landau free energy corresponding to coexistence of polar phases. Since $|\beta_{2}|$ is a measure of the extent of polar anisotropy of the free energy, our analysis confirms reduction in polar anisotropy when more polar phases coexist (i.e., the free energies of the polar phases become degenerate). Moreover, we note that $\beta_1$=$\alpha_{12}$ remains unchanged and does not contribute to the polar anisotropy. 

We equate the anisotropic contributions 
to the free energy of each phase (Eq.~\eqref{free_ph}) 
to calculate the temperature-composition ($\theta-x$) relations for the ``T-O'' and ``O-R'' phase boundaries~\cite{heitmann2014thermodynamics}. These are given as follows:
\begin{equation}
    \theta_{OT} = \theta_{quad} - \frac{1}{\beta_{22}}[\beta_{21} + \gamma_{21}P_{eq}^{2}](x-x_{quad}),
    \label{TOboundary},
\end{equation}
\begin{equation}
    \theta_{OR} = \theta_{quad} - \frac{1}{\beta_{22}-\frac{4}{27}\gamma_{32}}\left[\beta_{21} + \left(\frac{5}{9}\gamma_{21}-\frac{4}{27}\right)P_{eq}^{2}\right](x-x_{quad}),
    \label{ORboundary}
\end{equation}
where the equilibrium polarization $P_{eq}$ values at the T-O and O-R phase boundaries are obtained
by assuming a weak first order transition along phase-coexistence lines. Thus, 
$P_{eq}^{T-O}=P_{eq,T}$ or $P_{eq,O}$, and $P_{eq}^{O-R}=P_{eq,O}$ or $P_{eq,R}$, where the equilibrium values of polarization are given in Eq.~\eqref{sp_ph}. 

To determine the interrelation between $Q_z$ and the Landau free energy coefficients $\alpha_{ij}$, we proceed as follows:  first,
we find $\beta_2=\alpha_{11}-\alpha_{12}$ from the 
phase-coexistence conditions (Eqns.~\eqref{stableO},~\eqref{TOphaseequi} and~\eqref{TORphaseequi}) keeping $\beta_1=\alpha_{12}$ fixed. Next, for $Q_z=1$, we find relations between stress-free $\alpha_{11}$, $\alpha_{12}$ and constrained $\alpha_{11}^{e}$, $\alpha_{12}^{e}$ from 
Eq.~\eqref{consunconsrelations}. Demanding the difference in the constrained coefficients, $\alpha_{11}^e-\alpha_{12}^e$, and the unconstrained coefficient $\alpha_{12}$ to be invariant for all $Q_z$, we find the change in $Q_{44}$ as a function of $\alpha_{11}$:
\begin{equation}
    Q_{44}=\left[\frac{1}{C_{44}} \left\{(\alpha_{11}-\alpha_{12}) - (\alpha_{11}^e-\alpha_{12}^e) + 2 \frac{\hat{q}_{22}}{\hat{C}_{22}}\right\} \right]^{\frac{1}{2}}.
\end{equation}
When $Q_z$ is greater than unity, $\alpha_{11}$ decreases with increasing $Q_z$.

\begin{figure}[htpb]
    \centering
    \subfloat[Case 1]{\label{phasediagnorm}\includegraphics[width=0.8\linewidth]{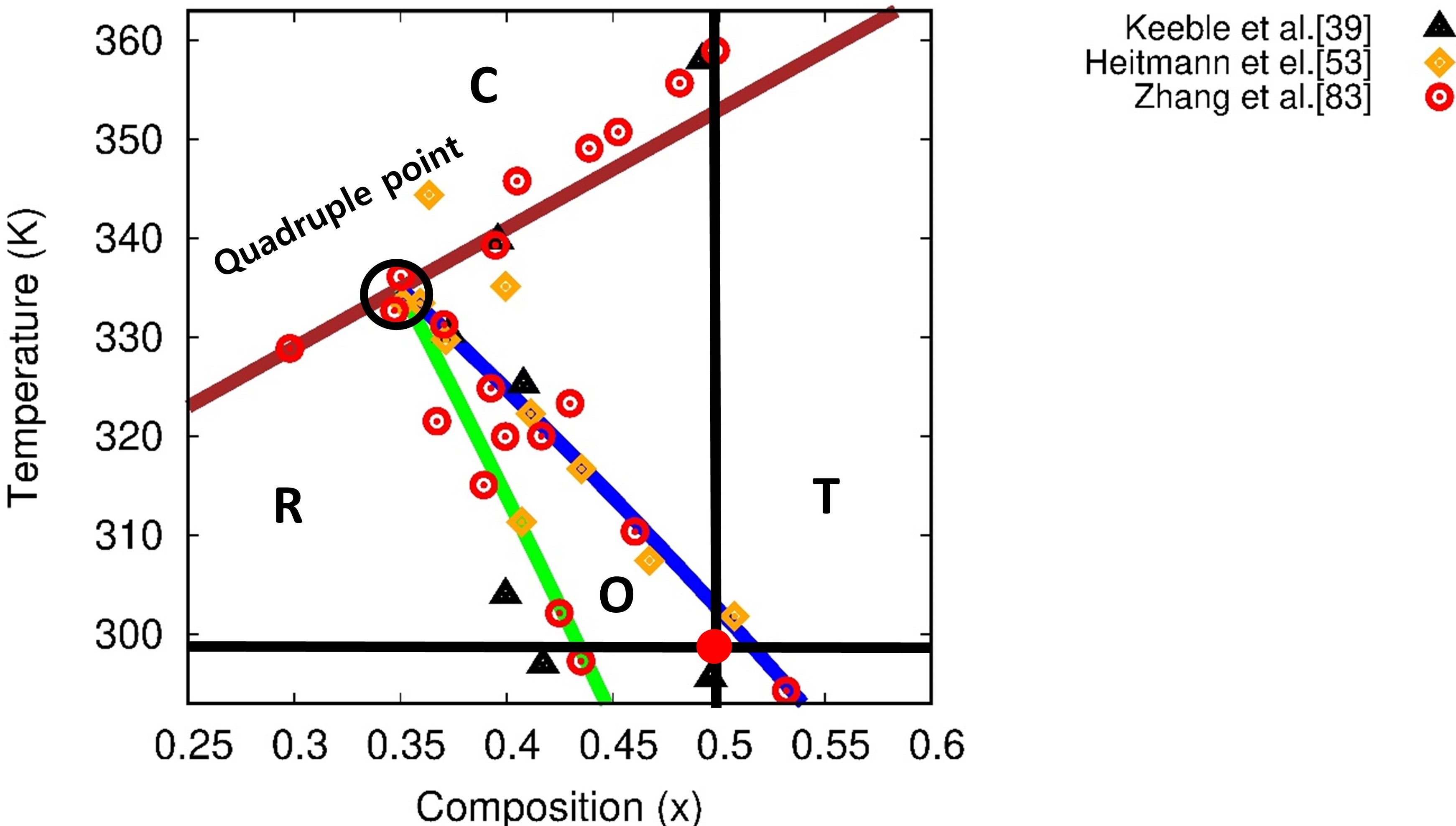}}\par
    \subfloat[Case 2]{\label{phasediag2phase}\includegraphics[width=.5\linewidth]{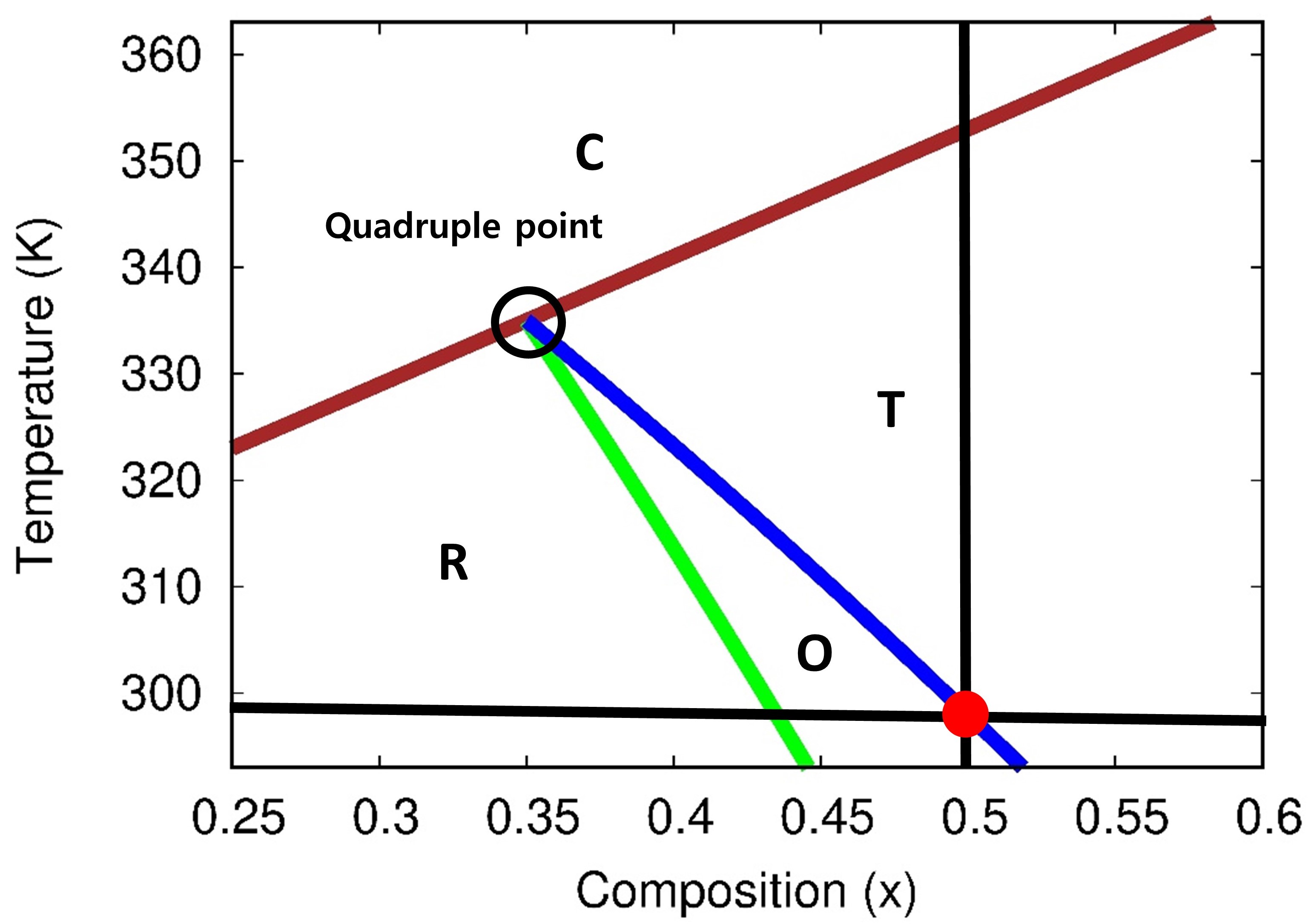}}\hfill
    \subfloat[Case 3]{\label{phasediag3phaseQ}\includegraphics[width=.5\linewidth]{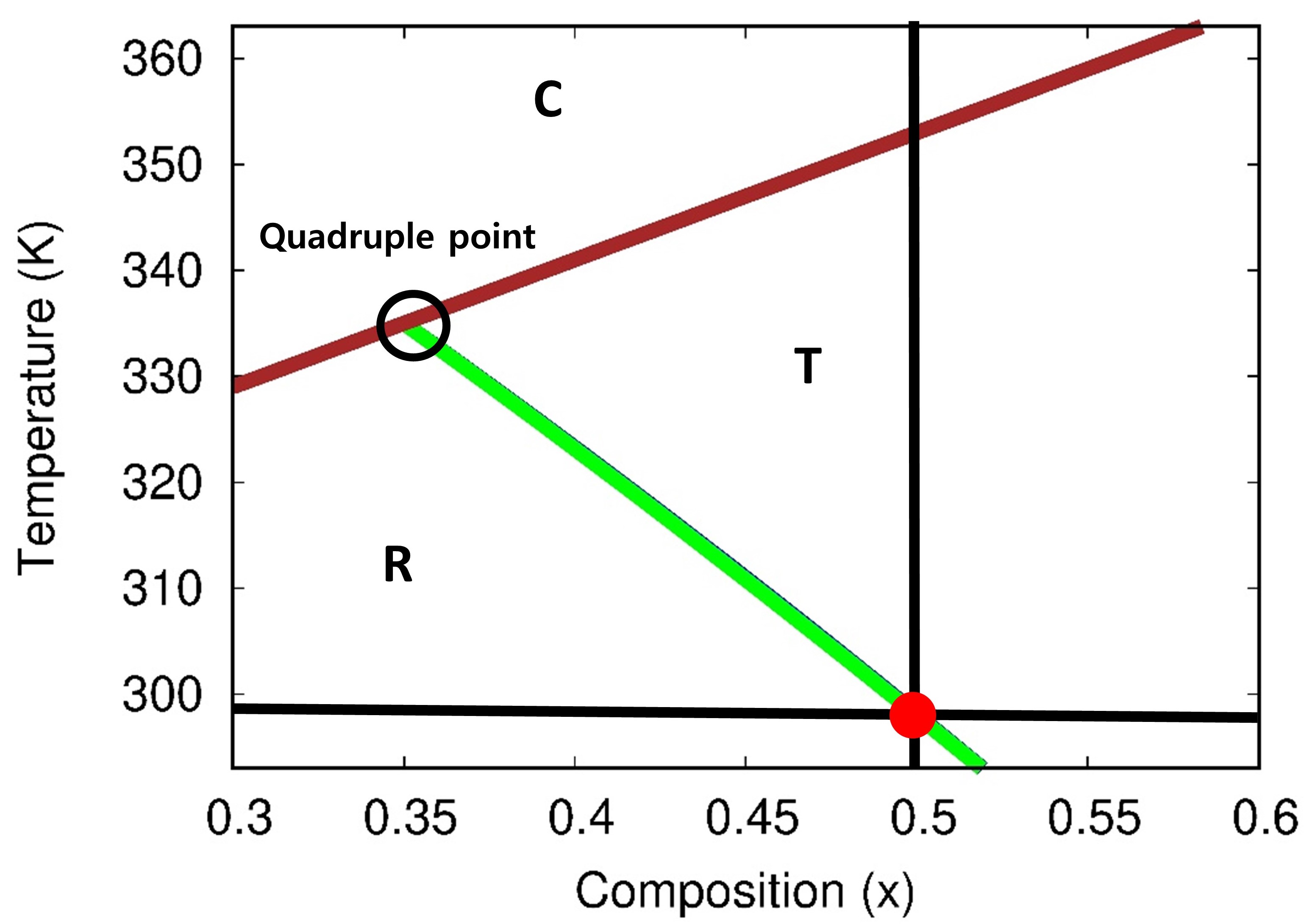}}
    \caption{Computed diffusionless phase diagrams of BZCT system as a function of electrostrictive anisotropy parameter $Q_z$ and their comparison with experimental data. Experimental data for comparison are obtained from~\cite{keeble2013revised,heitmann2014thermodynamics,zhang2014phase}. 
    Here, $C$, $T$, $O$ and $R$ denote cubic, tetragonal, orthorhombic and
    rhombohedral phases, respectively. 
    The black solid line and the red dot in each diagram correspond to the equimolar composition of BZCT at room temperature ($x=0.5, T=298\si{\kelvin}$).  Note that in Cases 2 and 3, the predictions of phase stability at the equimolar composition $x=0.5$ at ambient temperature and pressure ($T = 298\si{\kelvin}, P=1$ atm.) agree with the experimental observations~\cite{brajesh2016structural,brajesh2015relaxor}.}
    \label{fig:phase_diag}
\end{figure}

In Fig.~\ref{fig:phase_diag}, we present the computed temperature-composition phase diagrams for different values of $\beta_2$. 
For Case 1 ($Q_z=1, \beta^{\ast}=0.017$), the shapes of the computed $T-O$ and $O-R$ MPBs show good agreement with experimental data obtained from high resolution X-ray diffraction studies~\cite{keeble2013revised}, and the MPR contains only $O$
phase. 
When $Q_z$ becomes anisotropic, thermodynamic stability within the MPR changes from single phase $O$ to a mixture of $T$ and $O$ phases for Case 2 ($Q_z=2, \beta^{\ast}=0.012$,Fig.~\ref{phasediag2phase}), and a mixture of all three polar 
phases $T + R + O$ in Case 3 ( $Q_z=2.5, \beta^{\ast}=0.0022$, Fig.~\ref{phasediag3phaseQ}).
Thus, it is evident from the computed diagrams that the increase in electrostrictive 
anisotropy leads to a reduction in  
the polar anisotropic contribution to the free energy.   

\begin{figure}[htbp]
    \centering
    \subfloat[Case 1]{\label{normfree}\includegraphics[width=.5\linewidth]{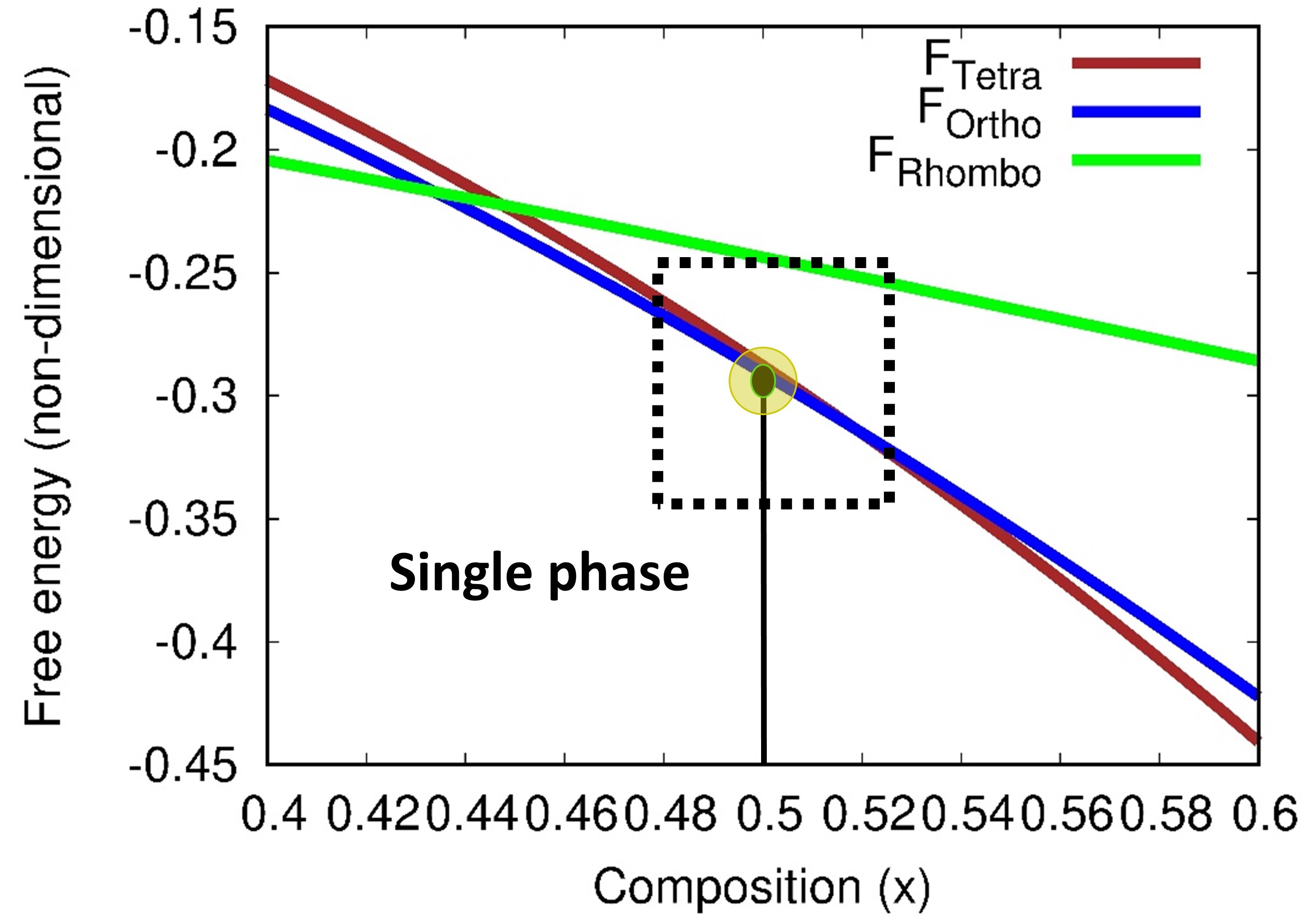}}\hfill
    \subfloat[Zoomed image of the marked portion for case 1]{\label{zoomed}\includegraphics[width=.5\linewidth]{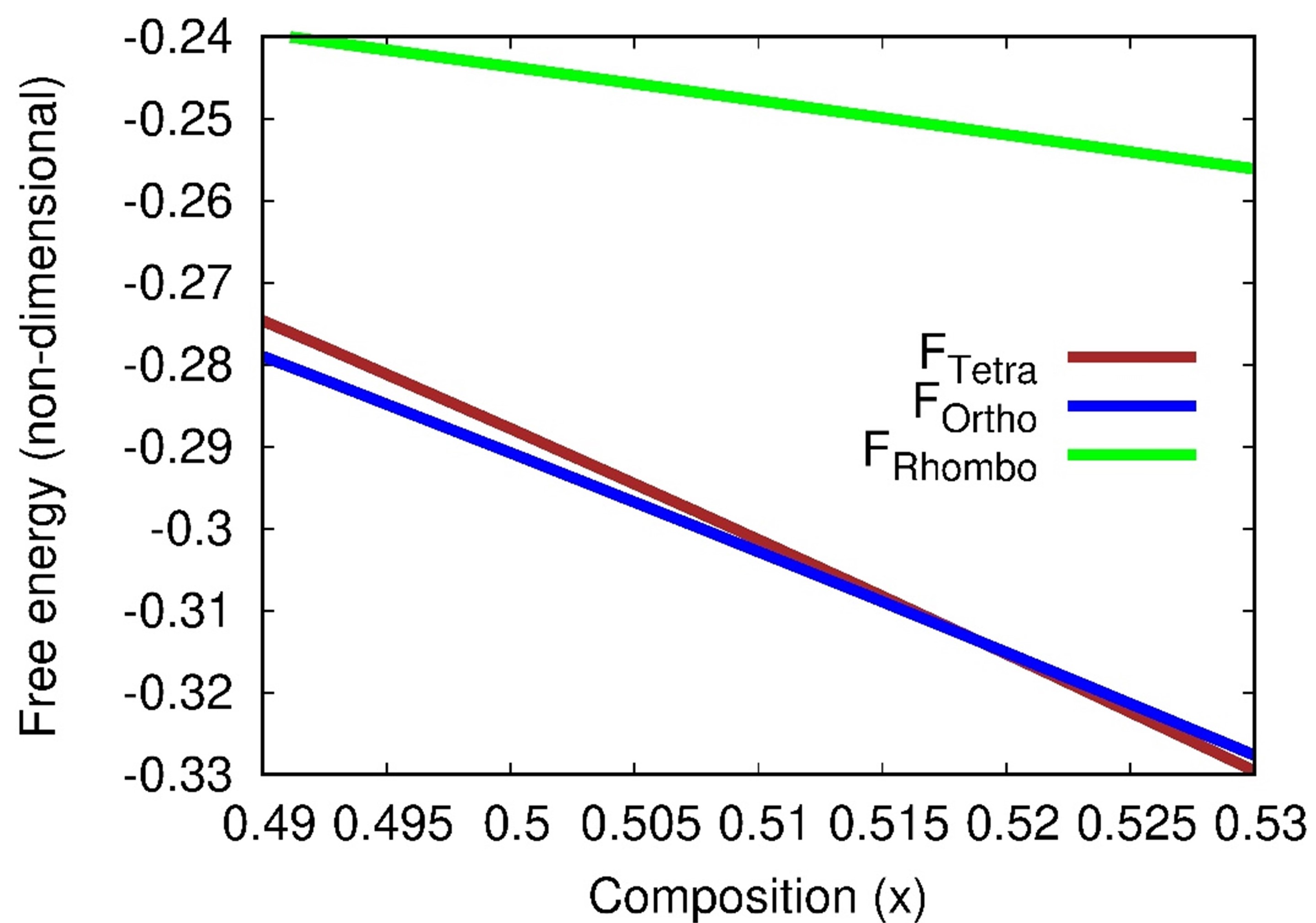}}\hfill
\subfloat[Case 2]{\label{free2}\includegraphics[width=.5\linewidth]{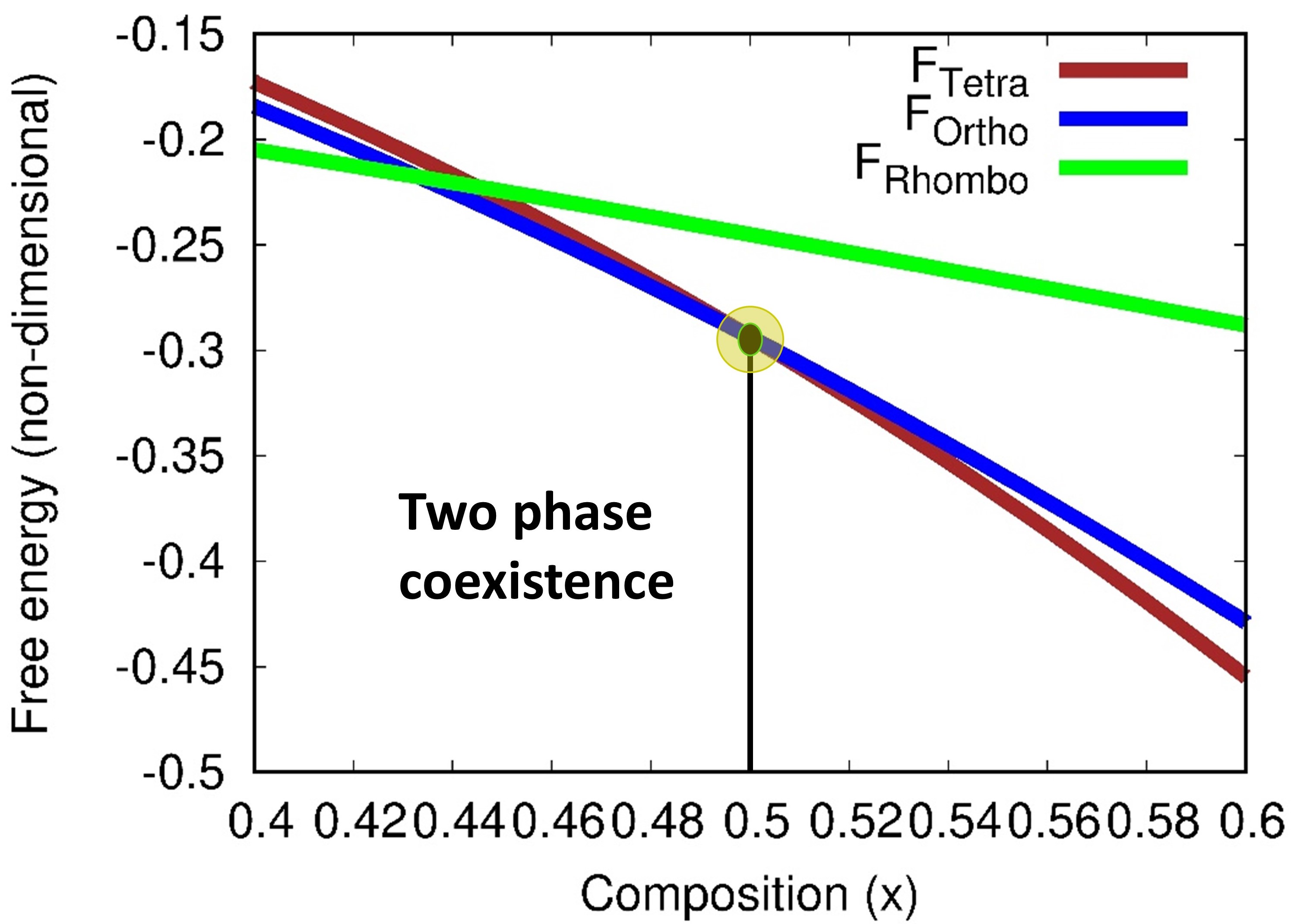}}\hfill 
\subfloat[Case 3]{\label{free3}\includegraphics[width=.5\linewidth]{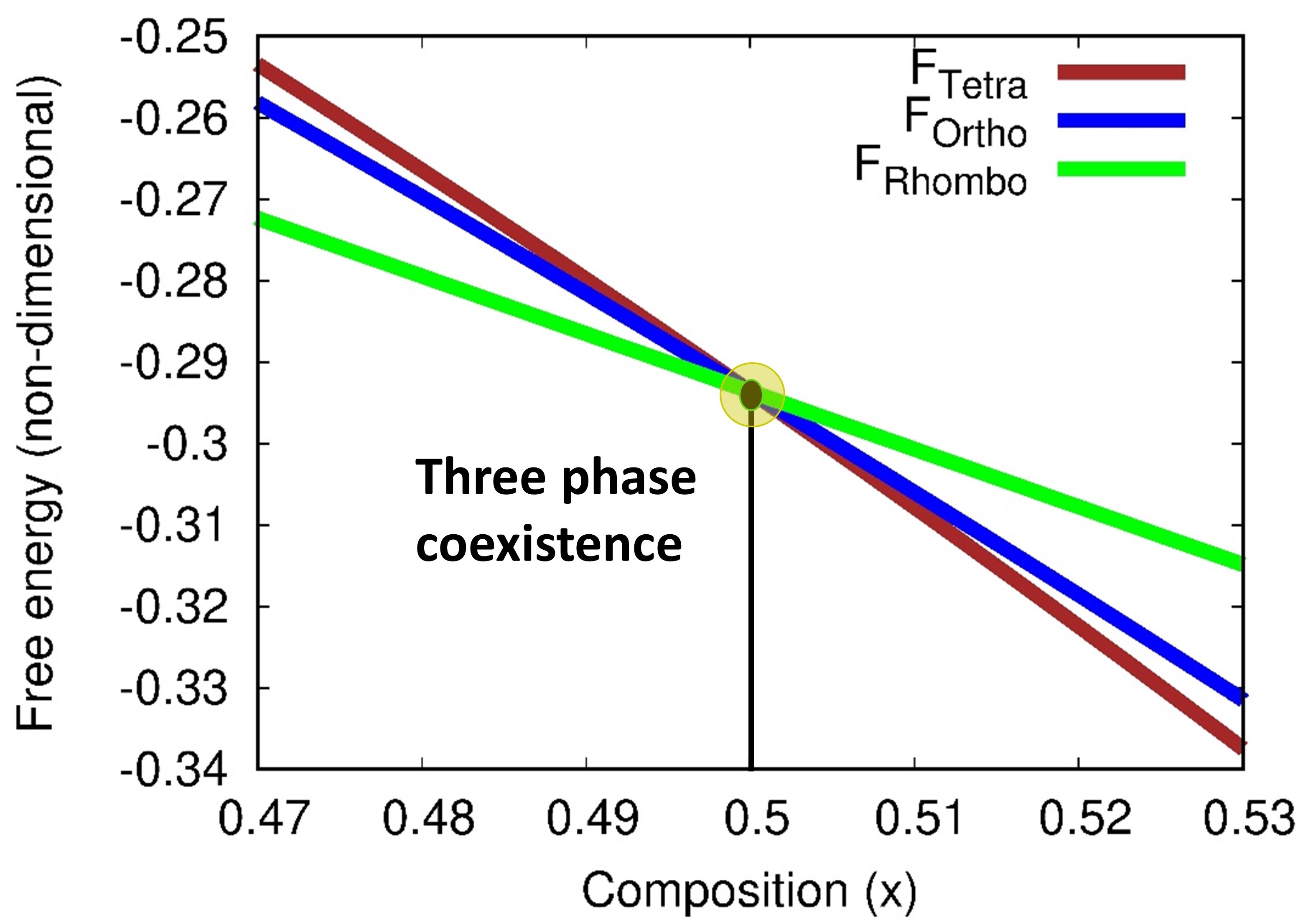}}
    \caption{Free energy – composition diagrams for different cases of electrostrictive anisotropy: (a,b) Case 1 with a zoomed region around the equimolar composition  
    indicating lowest energy for the orthorhombic phase at $x=0.5$
    (c) Case 2 showing $T+O$ coexistence (d) Case 3 showing $T+R+O$ coexistence.}
    \label{fig:free_comp}
\end{figure}

Having established the correspondence between $\beta_2$, $\alpha_{11}$ and $Q_z$, 
we plot the free energies of $T$, $O$ and $R$ phases as a function of $x$ at room 
temperature for all cases (Fig.~\ref{fig:free_comp}) .
We are particularly interested in examining 
phase stability at $x=0.5$ where most of the experimental reports are available~\cite{keeble2013revised, brajesh2015relaxor, brajesh2016structural, zhang2014phase, heitmann2014thermodynamics}. When $Q_z=1$, 
the $O$ phase has lowest free energy among all the polar phases of BZCT. 
As $Q_z$ increases, there is a reduction in the free energy of $T$ and $R$. Thus, when $Q_z=2$, we find the energies of $T$ and $O$ to be equal 
at $x=0.5 $. With a further increase in $Q_z$ to 2.5, we get equal free energies for $T$, $O$ and $R$ at $x=0.5$. 
Although the overall energy of the system at $x=0.5$ shows minimal variation with the change in $Q_z$, energies of $T$ and $R$ phases decrease with increasing $Q_z$ such that $T-O$ intersection ($f_T=f_O$) moves towards decreasing $x$ while the $O-R$  intersection ($f_O=f_R$) moves in the opposite direction. Brajesh et 
al.~\cite{brajesh2015relaxor,brajesh2016structural} showed a change 
in the phase stability of BZCT from three-phase ($T+O+R$) to two-phase ($T+O$) when it is  subjected to a stress-relief anneal at $400\si{\degreeCelsius}$ far above $\theta_c$. They attributed this change to a stress-induced phase transformation 
occurring at $400\si{\degreeCelsius}$. 
Change in the anisotropy of electrostrictive coefficients 
of the paraelectric phase correspond to such stress-induced transformations preceding paraelectric$\rightarrow$ferroelectric transition. Since the spatially invariant part of spontaneous strain tensor is a function of electrostrictive coefficients for any given $x$ (Eq.~\eqref{eigenstrain}), a 
change in $Q_z$ alters the spontaneous strain tensors associated with the 
ferroelectric phases thereby affecting phase stability. 

\begin{figure}
    \centering
    \subfloat[Computed spontaneous polarization for all the phases for case 1.]{\label{Case 1}\includegraphics[width=.5\linewidth]{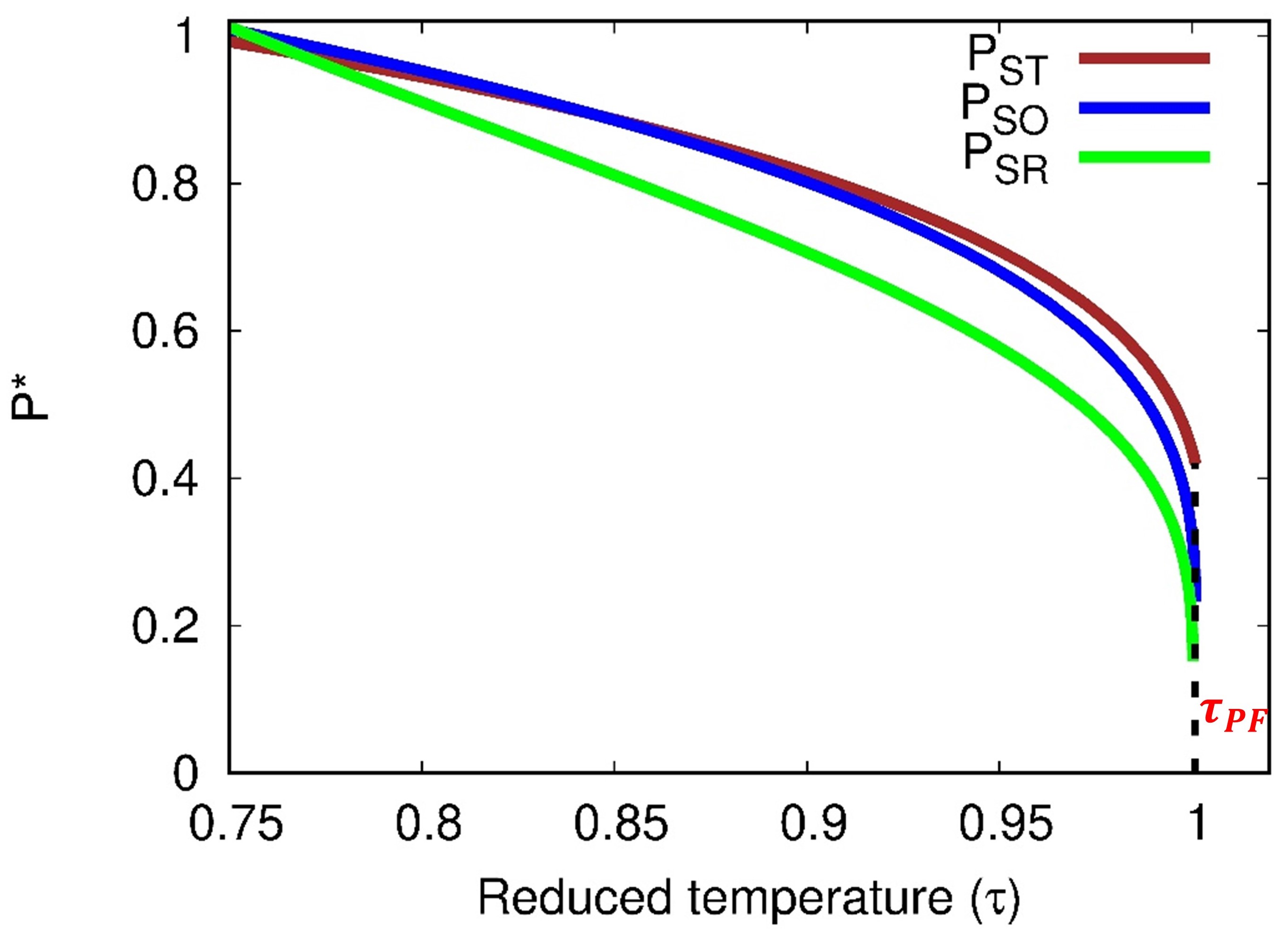}}\hfill
     \subfloat[Computed spontaneous polarization for all the phases for case 2.]{\label{Cae 2}\includegraphics[width=.5\linewidth]{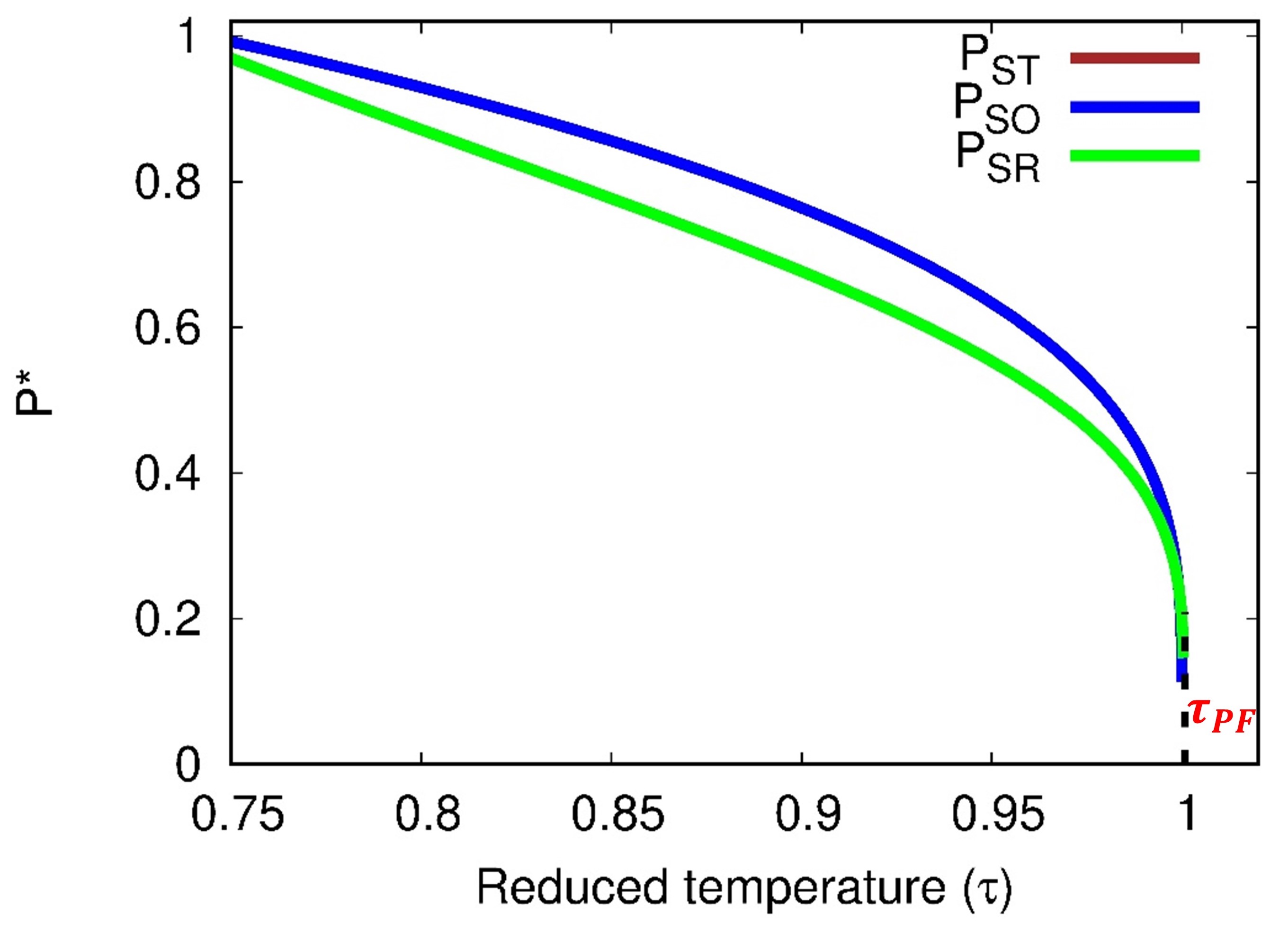}}\hfill
     \subfloat[Computed spontaneous polarization for all the phases case 3.]{\label{Case 3}\includegraphics[width=.5\linewidth]{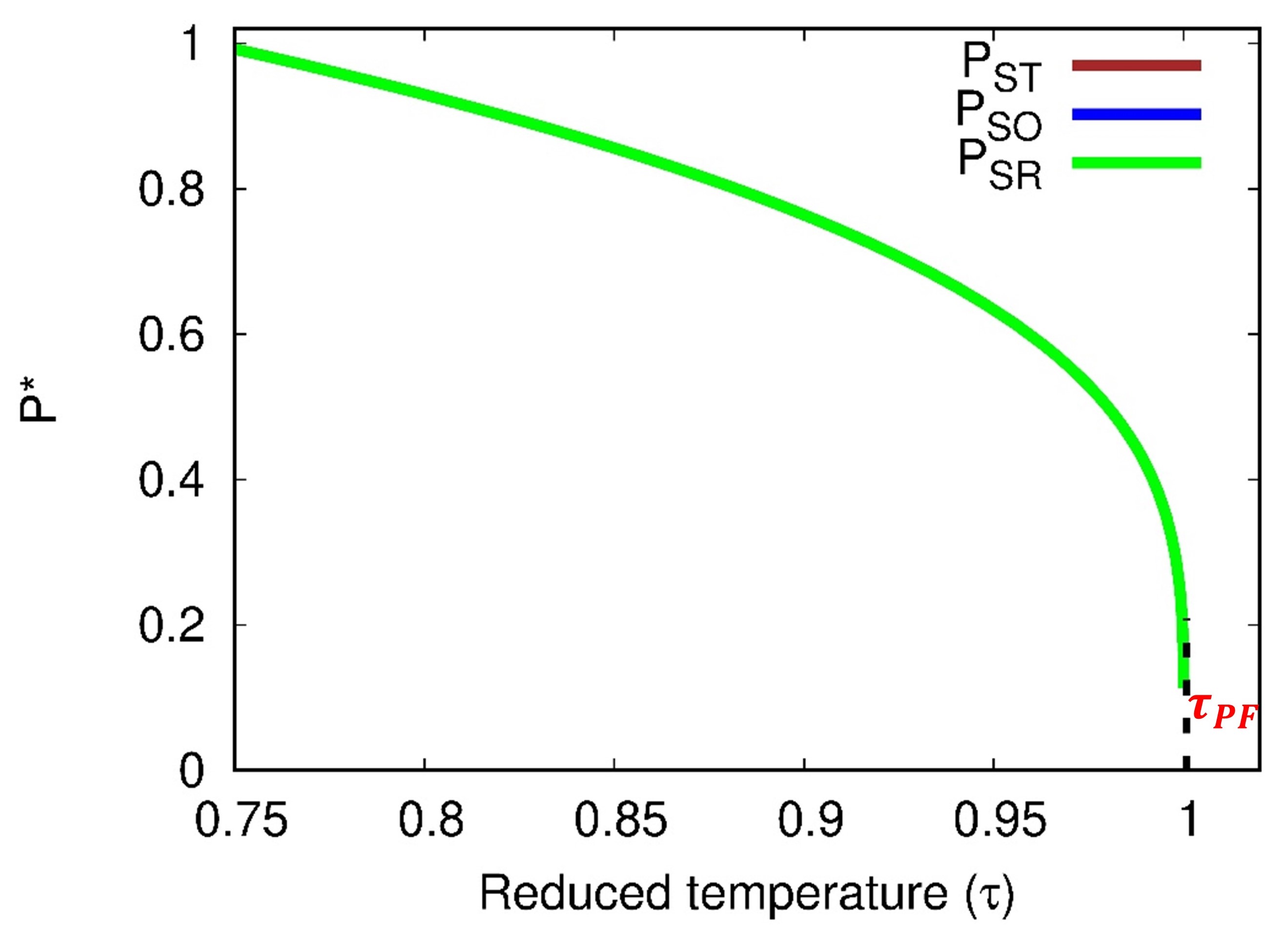}}\hfill
    \caption{Computed spontaneous polarization of ferroelectric phases $T$, $R$
    as a function of temperature at $x=0.5$ for BZCT solid solution for the three different cases.}
    \label{fig:Spon_pola_strain}
\end{figure}

Fig.~\ref{fig:Spon_pola_strain} shows the variation of scaled spontaneous polarization $P^{\ast}=P/P_s$  of $T$, $O$ and $R$ phases of equimolar 
as a function of reduced temperature $\tau=\theta/\theta_c$ (Eq.~\eqref{sp_ph}). When $Q_z=1$, the stable 
$O$ has the highest value of spontaneous polarization at room temperature 
($\theta = 298 \si{\kelvin}, \tau = 0.84$). On the other hand, for $Q_z=2$,  
$T$ and $O$ phases have the same $P^{\ast}$ at room temperature which 
is greater than the spontaneous polarization of $R$. While, for $Q_z=2.5$, $T$, $R$ and $O$ have the same $P^{\ast}$ at all temperatures up to $\theta=\theta_c$. This corroborates our observation of degeneracy of free energies of the ferroelectric phases with increasing $Q_z$ at the equimolar morphotropic composition $x=0.5$.

The corresponding components of spontaneous strain tensor for each polar phase can be calculated using Eq.~\eqref{sponstrain}.
Thus, two independent components of spontaneous strain associated with 
$T$ phase are $\varepsilon^{0}_{1} = Q_{11}P^{2}_{s,T},\; \varepsilon^{0}_{2} = \varepsilon^{0}_{3} = Q_{12}P^{2}_{s,T}$. The $O$ phase possesses three independent components:
$\varepsilon^{0}_{1} =  \varepsilon^{0}_{2} = \frac{1}{2}(Q_{11}+Q_{12})P^{2}_{s,O}, \;
      \varepsilon^{0}_{3} = Q_{12}P^{2}_{s,O},\;
       \varepsilon^{0}_{6} = \frac{1}{2}Q_{44}P^{2}_{s,O}$, 
and the $R$ phase possesses two independent components:
    $\varepsilon^{0}_{1} =  \varepsilon^{0}_{2}= \varepsilon^{0}_{3}=\frac{1}{3}(Q_{11}+Q_{12})P^{2}_{s,R}, \;
    \varepsilon^{0}_{4} = \varepsilon^{0}_{5} = \varepsilon^{0}_{6}= \frac{1}{3}Q_{44}P^{2}_{s,R}$.
Here, $P_{s,T}$, $P_{s,O}$ and $P_{s,R}$, given in Eqns.~\eqref{polaT},~\eqref{polaO} and~\eqref{polaR}, represent the spontaneous polarization of $T$, $O$ and $R$ phases,
respectively.
\subsection{Morphological evolution of domains}\label{evolution}
\begin{figure}[!htpb]
\begin{subfigure}{\textwidth}
  \centering
  \includegraphics[width=\linewidth]{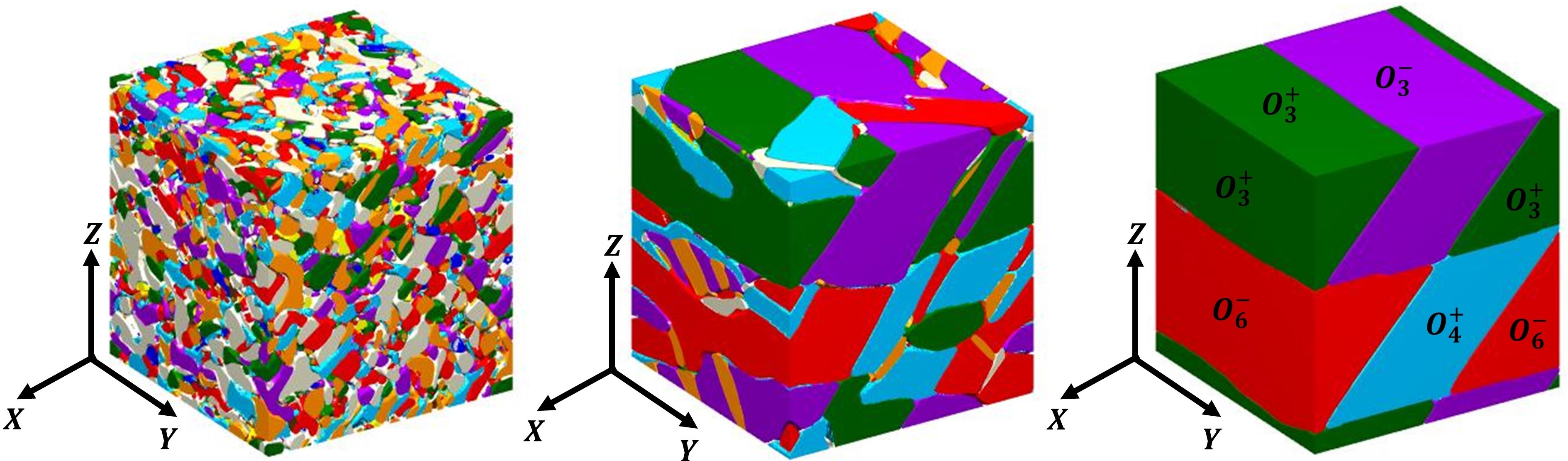}  
 \caption{\textbf{Case 1}}
  \label{fig:evol1}
\end{subfigure}
\begin{subfigure}{\textwidth}
  \centering
  \includegraphics[width=\linewidth]{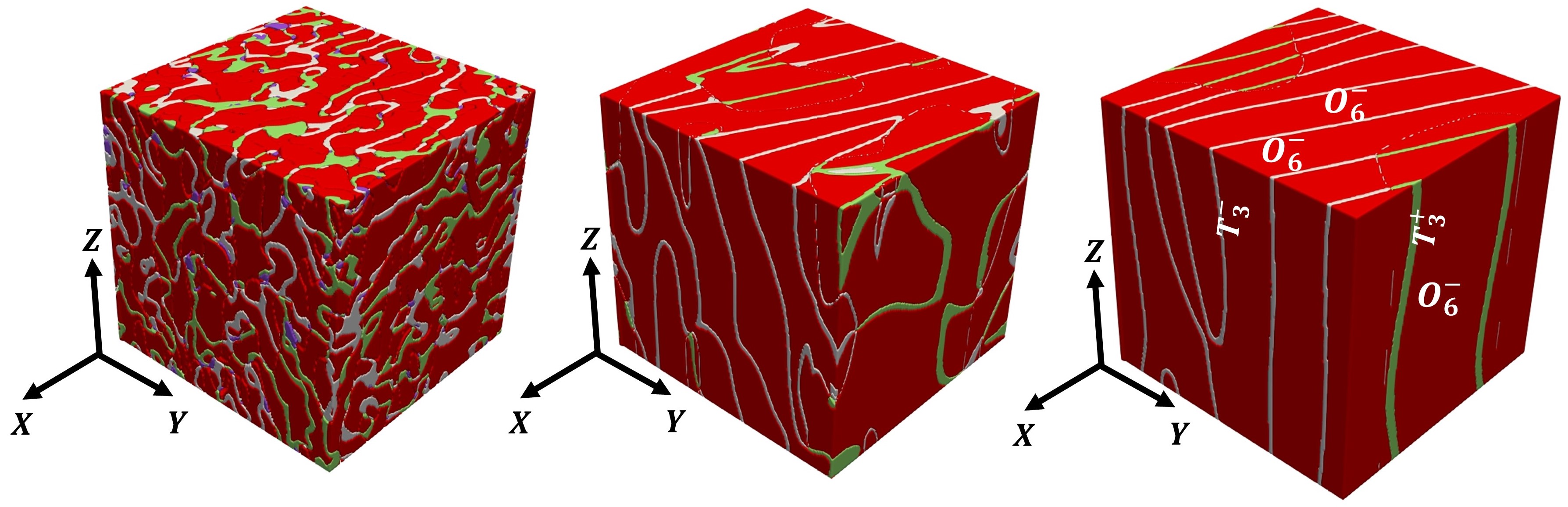}  
 \caption{\textbf{Case 2}}
  \label{fig:evol2}
\end{subfigure} 
\begin{subfigure}{\textwidth}
  \centering
  \includegraphics[width=\linewidth]{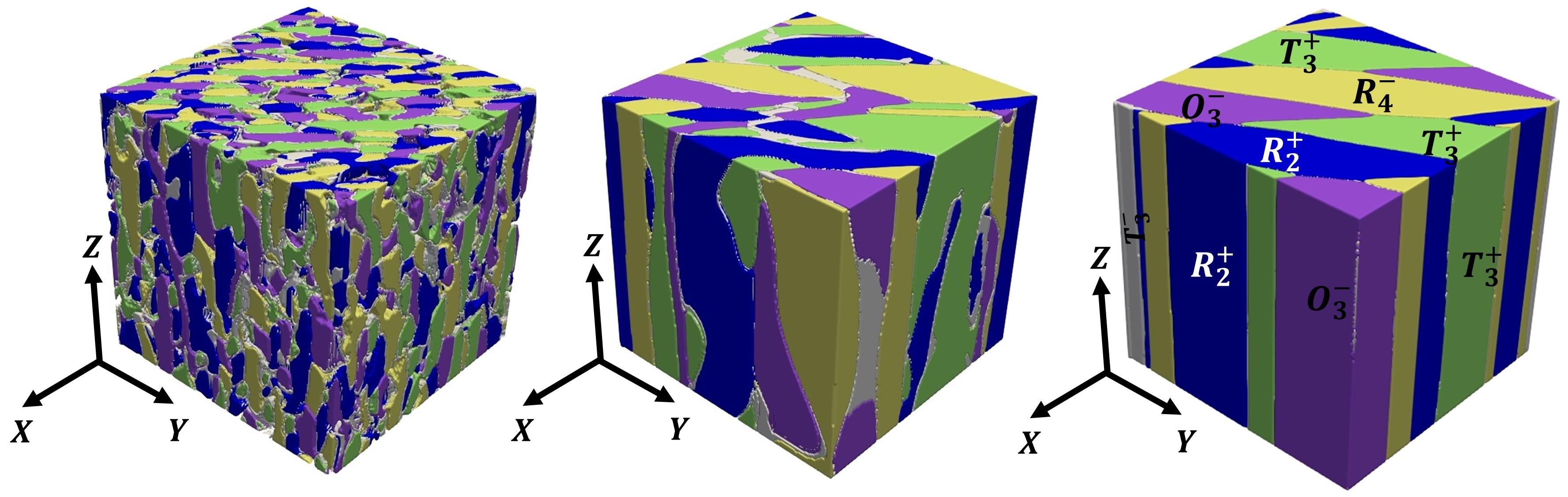}  
 \caption{\textbf{Case 3}}
  \label{fig:evol3}
\end{subfigure} 
\centering
\begin{subfigure}{0.3\textwidth}
  \centering
  \includegraphics[width=1.2\linewidth]{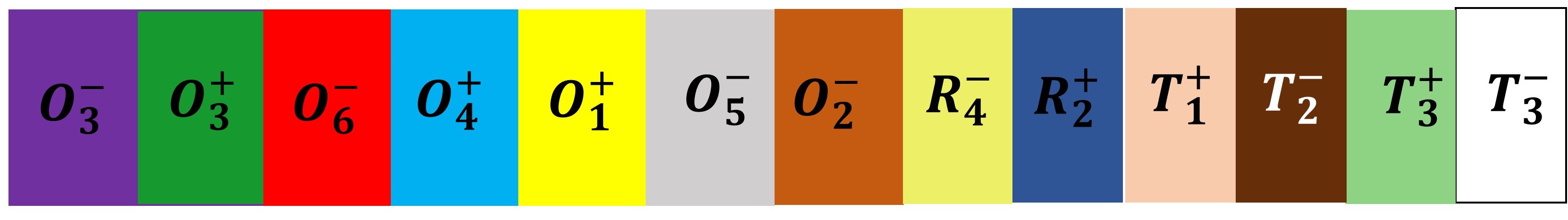}  
\caption{\textbf{}}
  \label{fig:color2}
\end{subfigure} 
 \caption{Evolution of domain structures of equimolar BZCT at room temperature 
 for three cases of electrostrictive anisotropy: (a) Case 1 ($Q_z=1$), (b) Case 2 ($Q_z=2$), (c) Case 3 ($Q_z=2.5$). In all cases, time snapshots of evolution are shown at nondimensional times $t = 10000, 50000$ and $150000$ (steady-state configuration). The colors distinguish between variants of $T$, $R$ and $O$. 
 (d) List of distinct colors corresponding to each variant. 
 Components of polarization vector corresponding to each variant are given below: 
 $O_{1}^{+}:[1 1 0]$, $O_{1}^{-}:[\bar{1} \bar{1} 0]$, $O_{2}^{+}:[0 1 1]$,
 $O_{2}^{-}:[0 \bar{1} \bar {1}]$, $O_{3}^{+}:[1 0 1]$, 
 $O_{3}^{-}:[\bar{1} 0 \bar{1}]$, $O_{4}^{+}:[\bar{1} 1 0]$, 
 $O_{4}^{-}:[1 \bar{1} 0]$, $O_{5}^{+}:[0 \bar{1} 1]$, 
 $O_{5}^{-}:[0 1 \bar{1}]$, $O_{6}^{+}:[\bar{1} 0 1]$, $O_{6}^{-}:[1 0 \bar{1}]$, 
 $T_{3}^{+}:[0 0 1]$, $T_{3}^{-}:[0 0 \bar{1}]$, $R_{2}^{+}: [1 \bar{1} \bar{1}]$, $R_{4}^{-}: [1 \bar{1} 1]$.
    }
    \label{fig:domains_evol}
\end{figure}

Fig.~\ref{fig:domains_evol} shows the evolution of ferroelectric domains to a 
steady state in equimolar BZCT at room temperature for all three cases of $Q_z$
in the absence of external electromechanical fields. Thus, in all cases, 
the system is assumed to be stress-free with periodic boundary conditions on polarization ($\mathbf{P}$), displacement ($\mathbf{u}$) and electric potential
($\phi$) fields. Note that for bulk ferroelectric systems the contribution to depolarization energy due to surface bound charge is zero (i.e., electrically unbounded domain: $\mathbf{E}\to\mathbf{0}$ as $\mathbf{r}\to\infty$). 
In all cases we start with the same random initial configuration which allows nucleation of any of the ferroelectric phases below $\theta_c$. 

\begin{figure}[!htpb]
\begin{subfigure}{0.3\textwidth}
  \centering
  \includegraphics[width=0.9\linewidth]{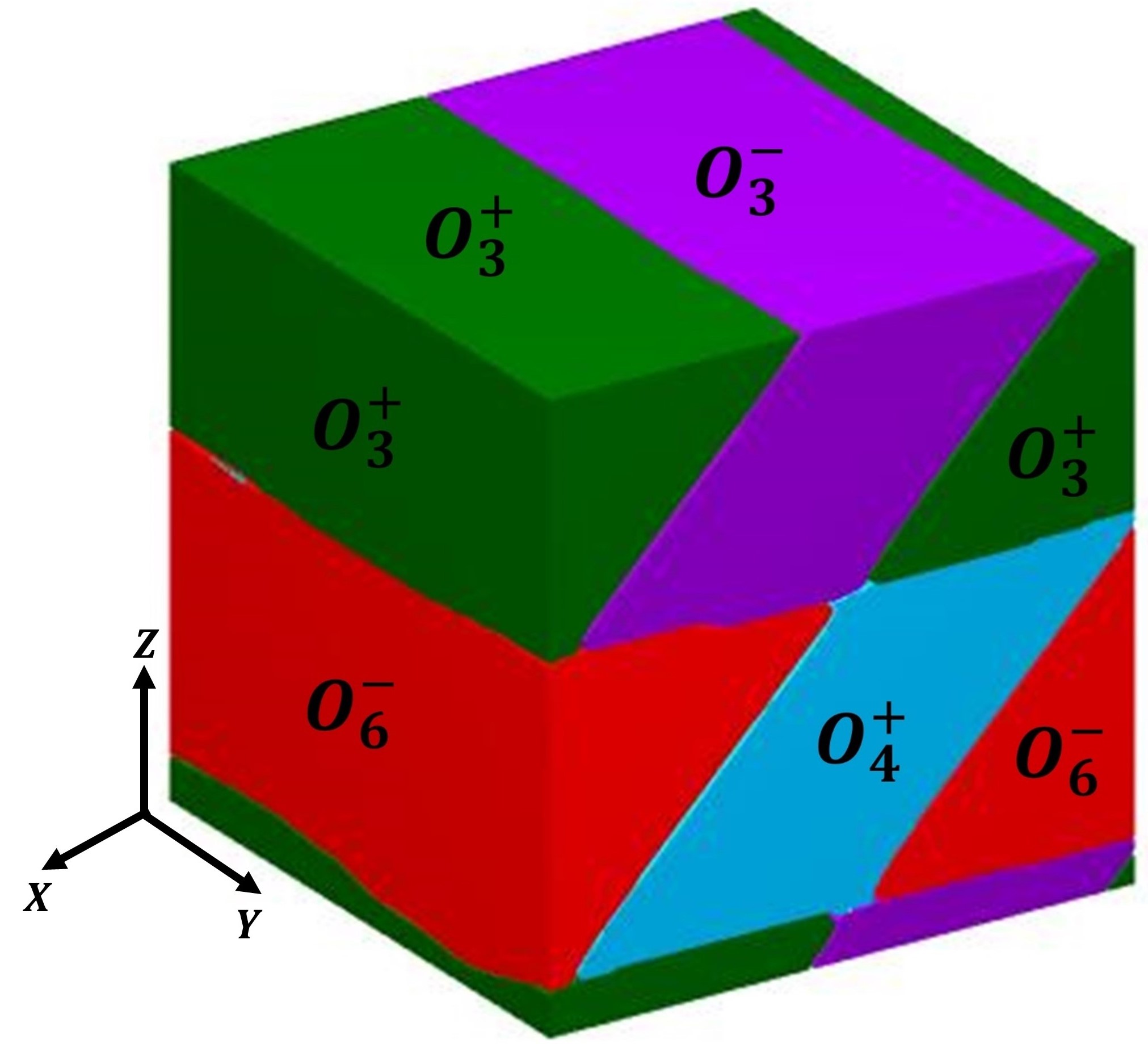}  
 \caption{\textbf{}}
  \label{fig:domain1}
\end{subfigure}
\begin{subfigure}{0.3\textwidth}
  \centering
  \includegraphics[width=0.9\linewidth]{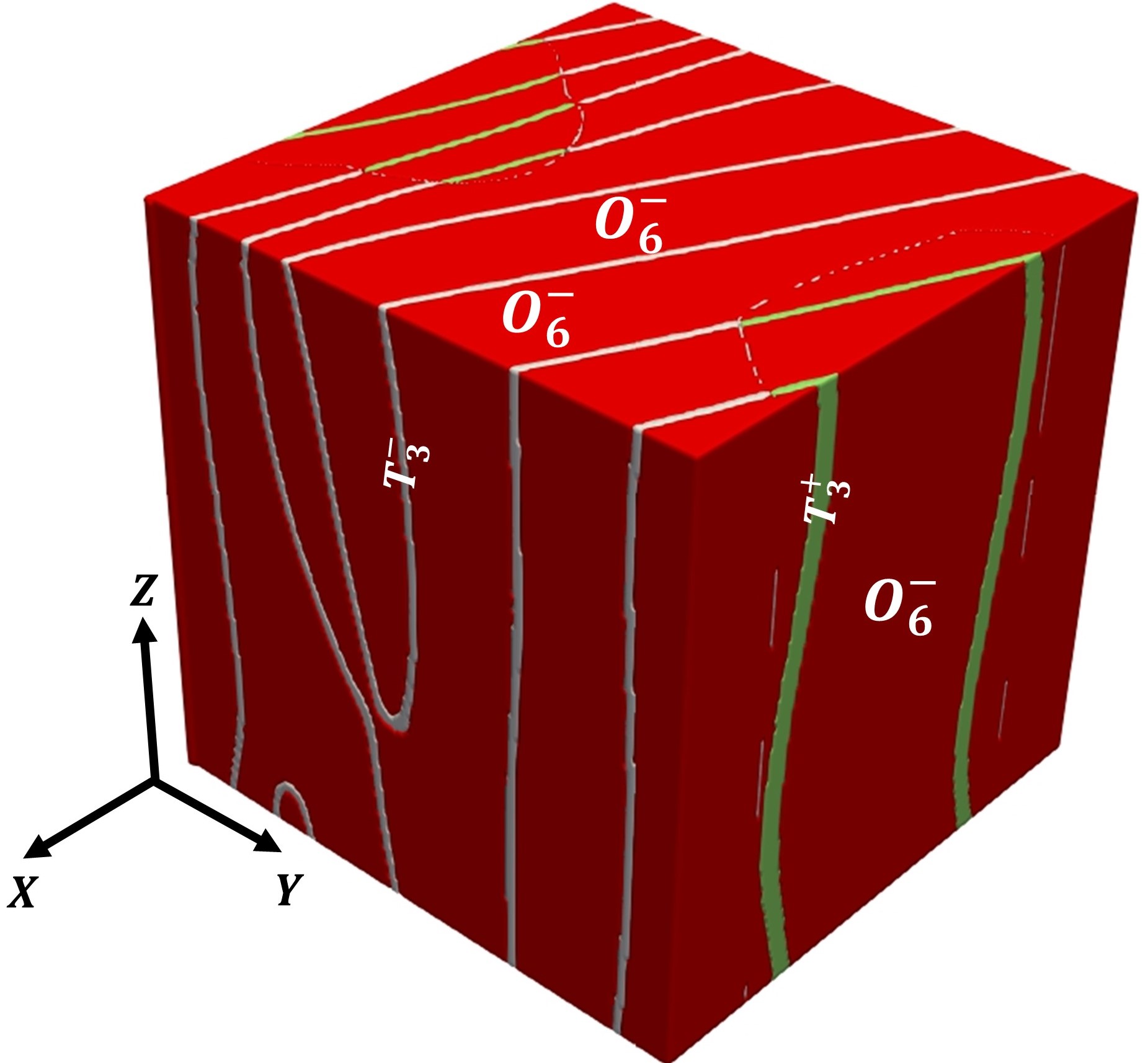}  
 \caption{\textbf{}}
  \label{fig:domain2}
\end{subfigure} 
\begin{subfigure}{0.3\textwidth}
  \centering
  \includegraphics[width=0.9\linewidth]{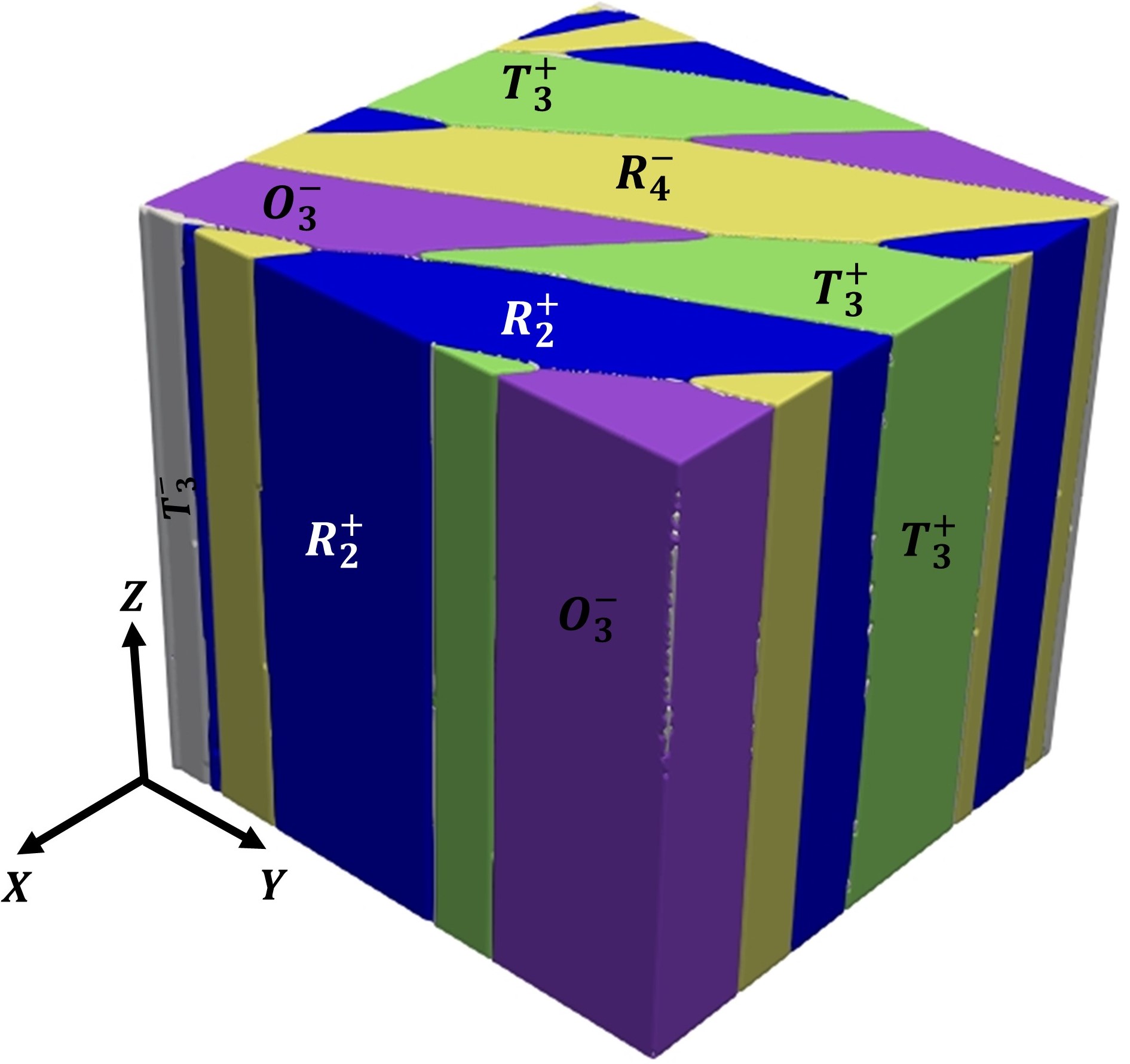}  
 \caption{\textbf{}}
  \label{fig:domain3}
\end{subfigure} 
\begin{subfigure}{0.3\textwidth}
  \centering
  \includegraphics[width=\linewidth]{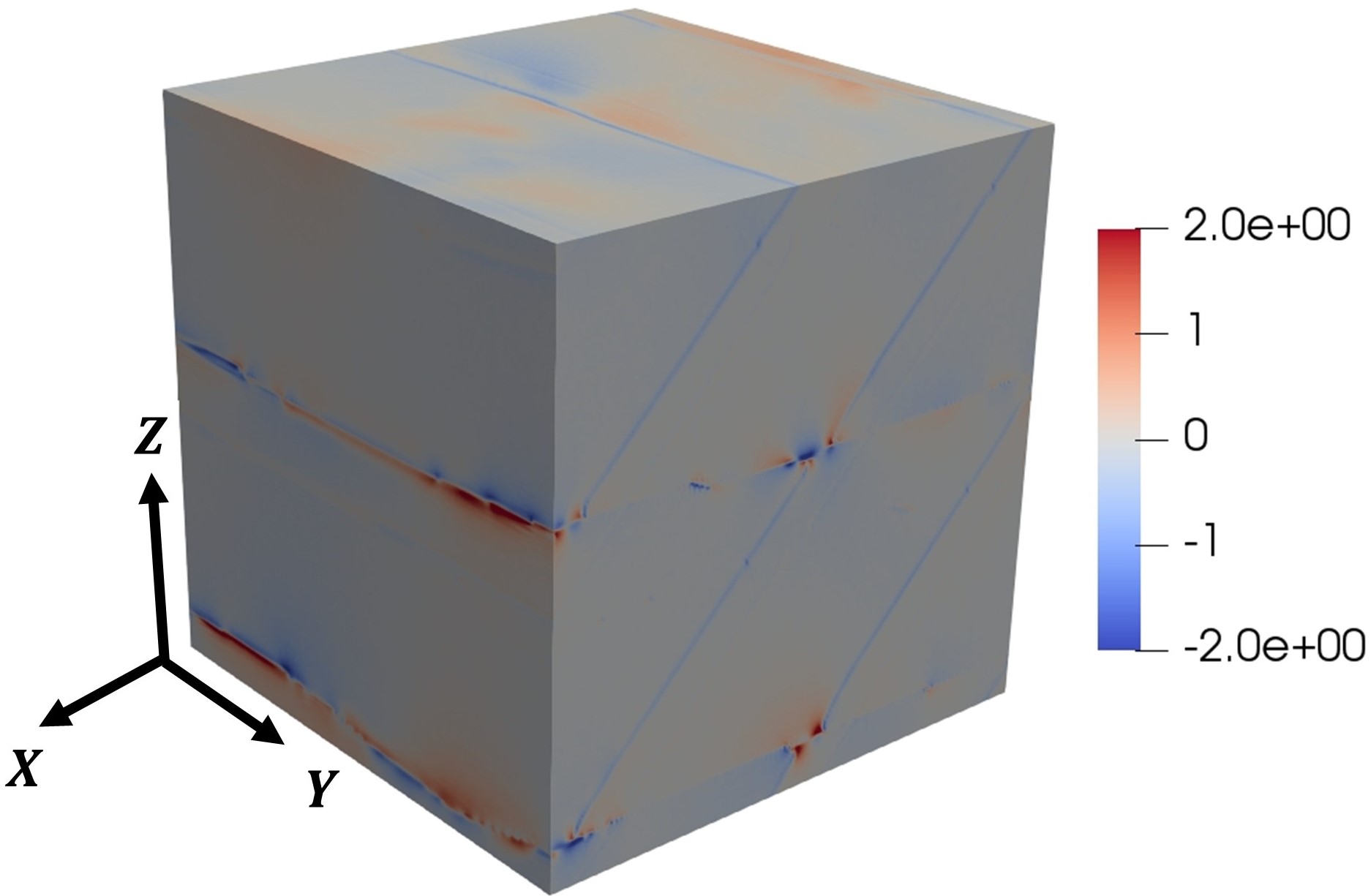}  
 \caption{\textbf{}}
  \label{fig:elec1}
\end{subfigure} 
\hspace{1cm}
\begin{subfigure}{0.3\textwidth}
  \centering
  \includegraphics[width=\linewidth]{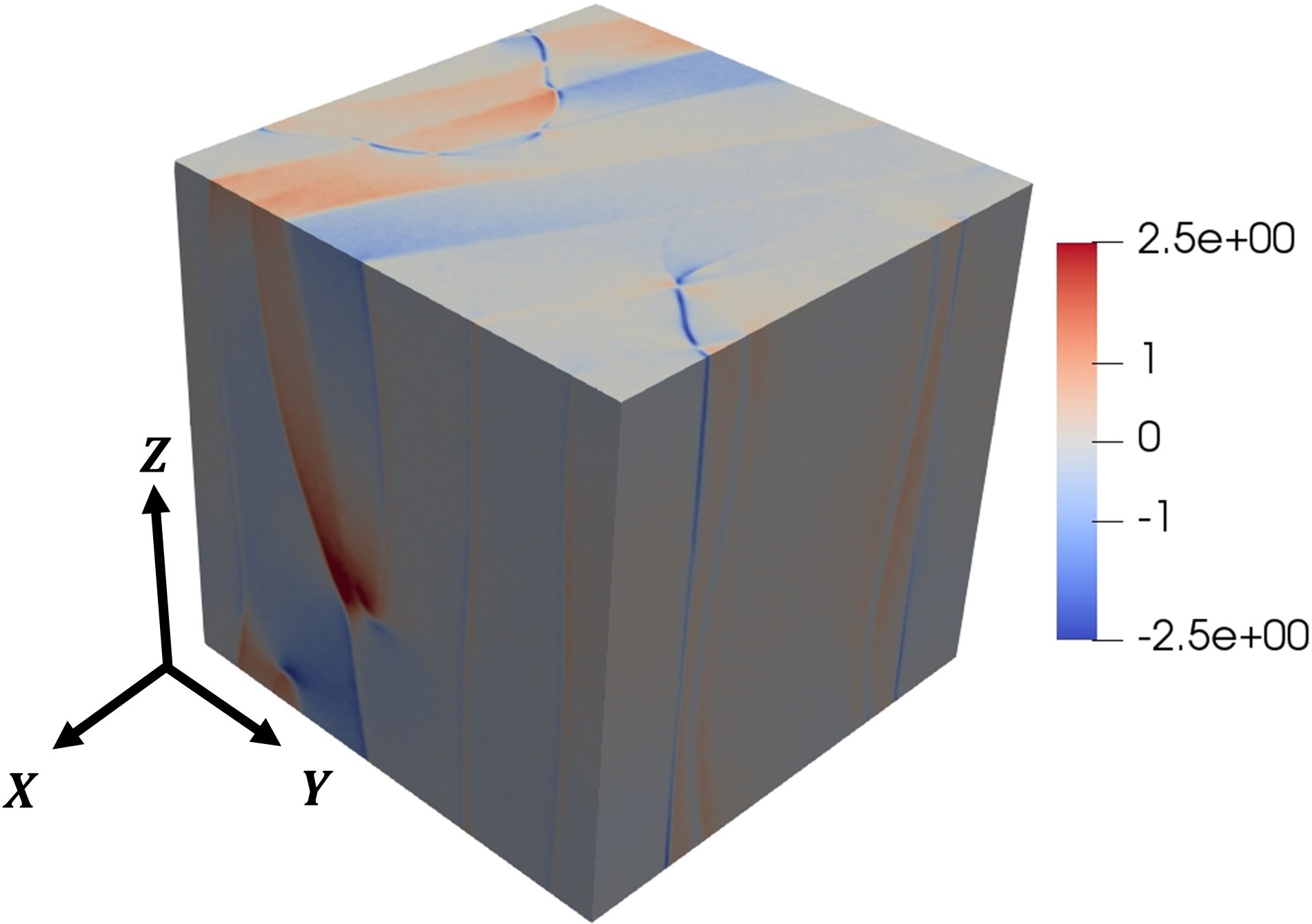}  
 \caption{\textbf{}}
  \label{fig:elec2}
\end{subfigure} 
\begin{subfigure}{0.3\textwidth}
  \centering
  \includegraphics[width=\linewidth]{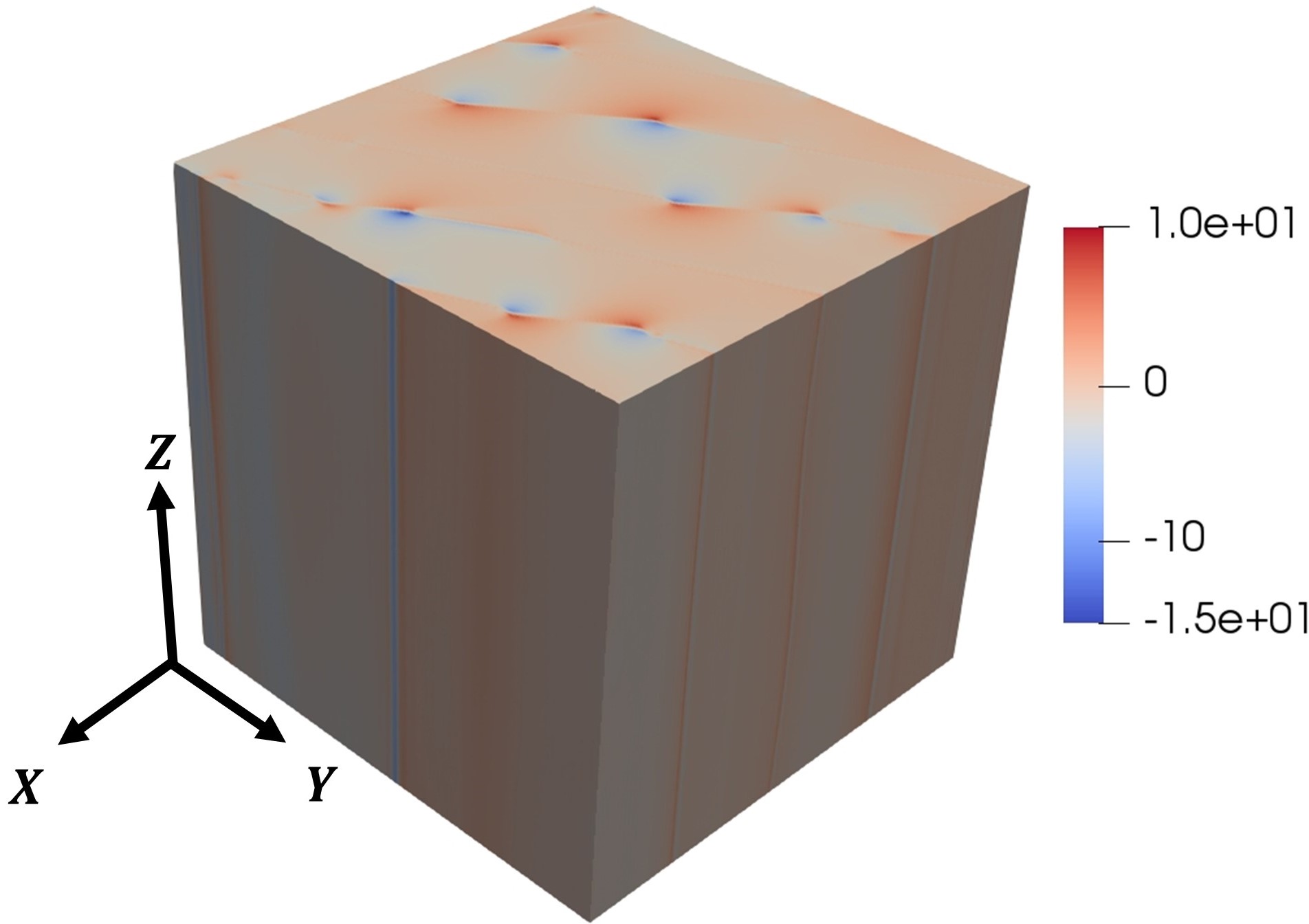}  
\caption{\textbf{}}
  \label{fig:elec3}
\end{subfigure} 
\begin{subfigure}{0.3\textwidth}
  \centering
  \includegraphics[width=\linewidth]{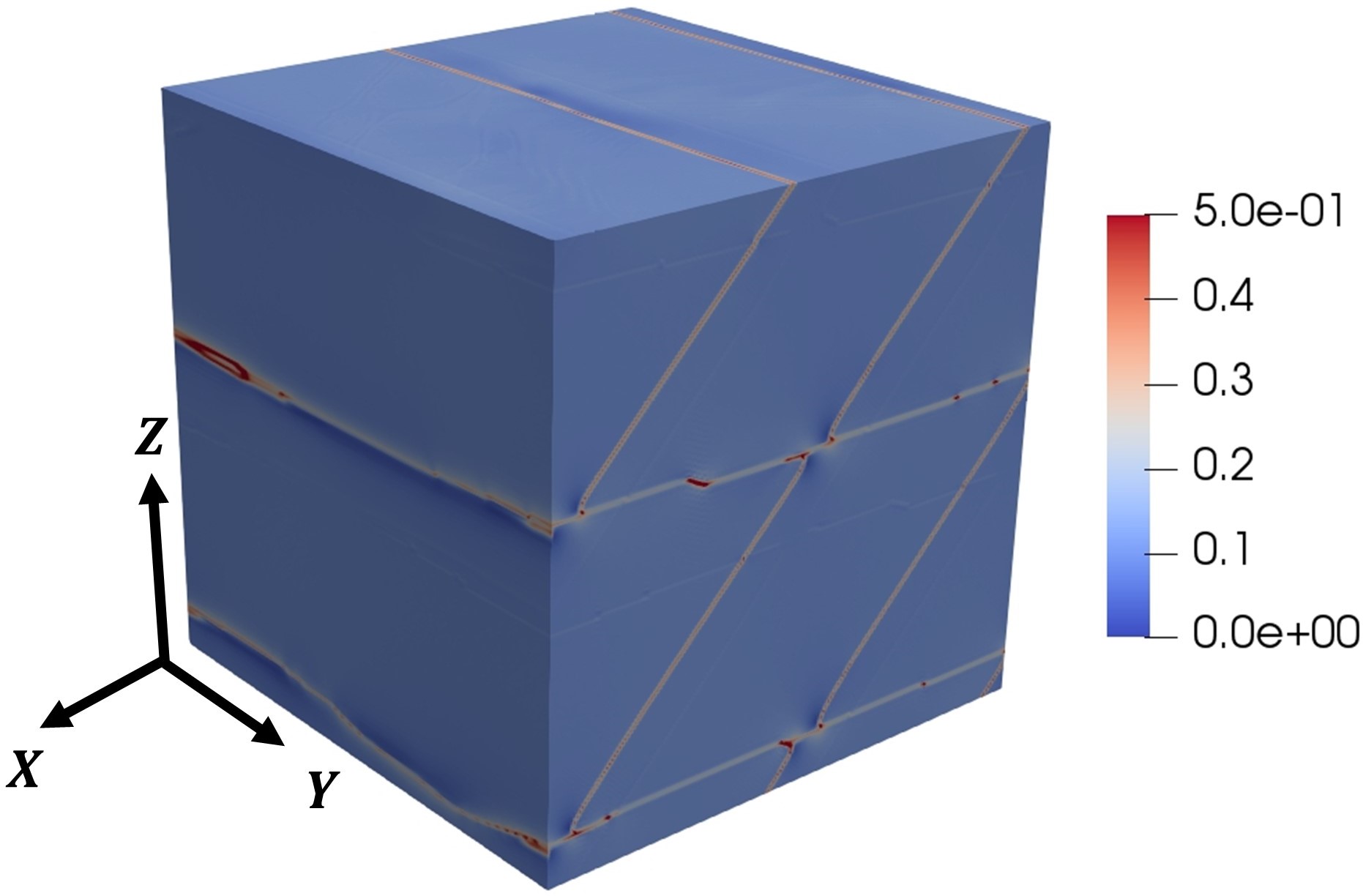}  
 \caption{\textbf{}}
  \label{fig:elas1}
\end{subfigure} 
\hspace{1cm}
\begin{subfigure}{0.3\textwidth}
  \centering
  \includegraphics[width=\linewidth]{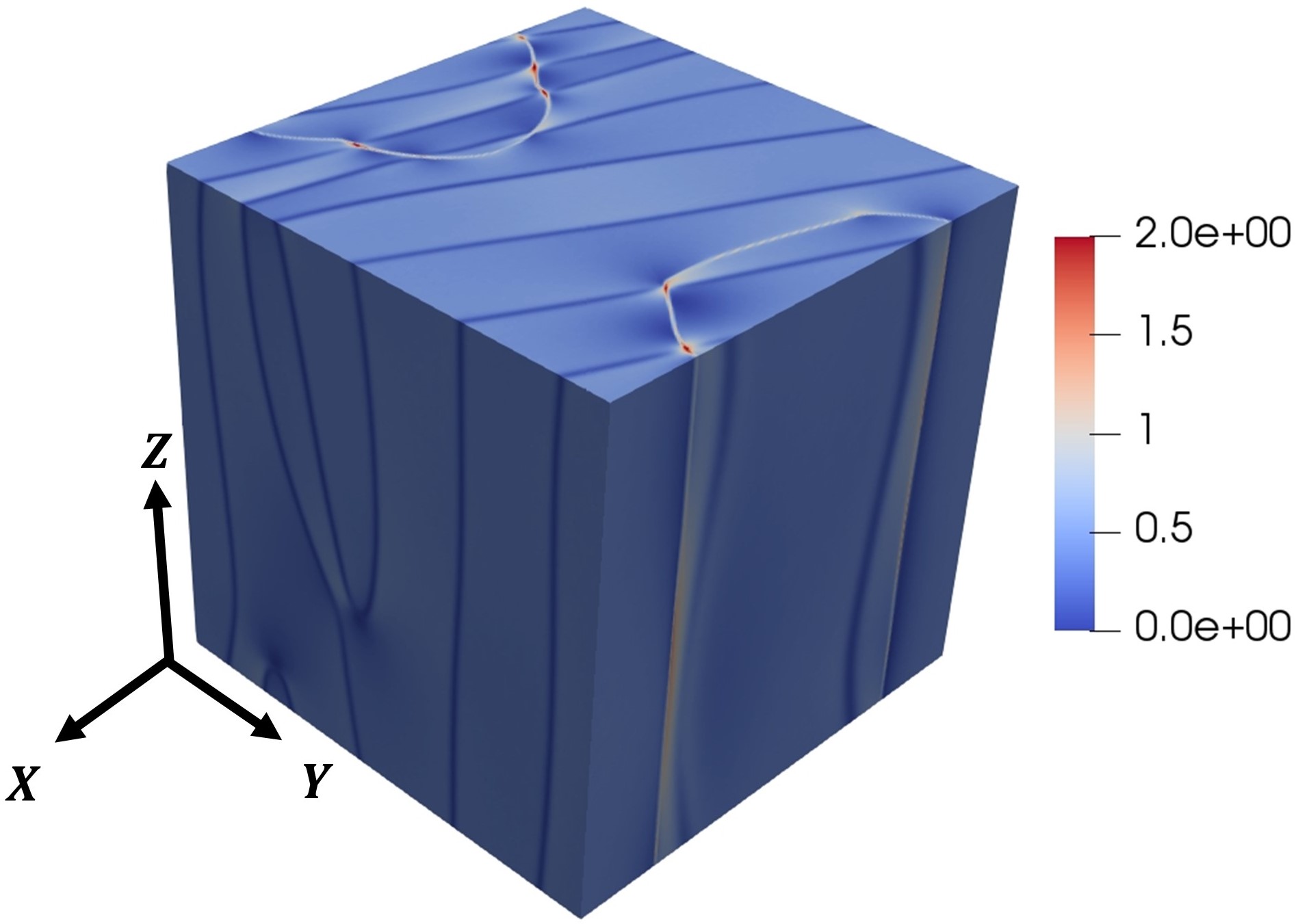}  
 \caption{\textbf{}}
  \label{fig:elas2}
\end{subfigure} 
\begin{subfigure}{0.3\textwidth}
  \centering
  \includegraphics[width=\linewidth]{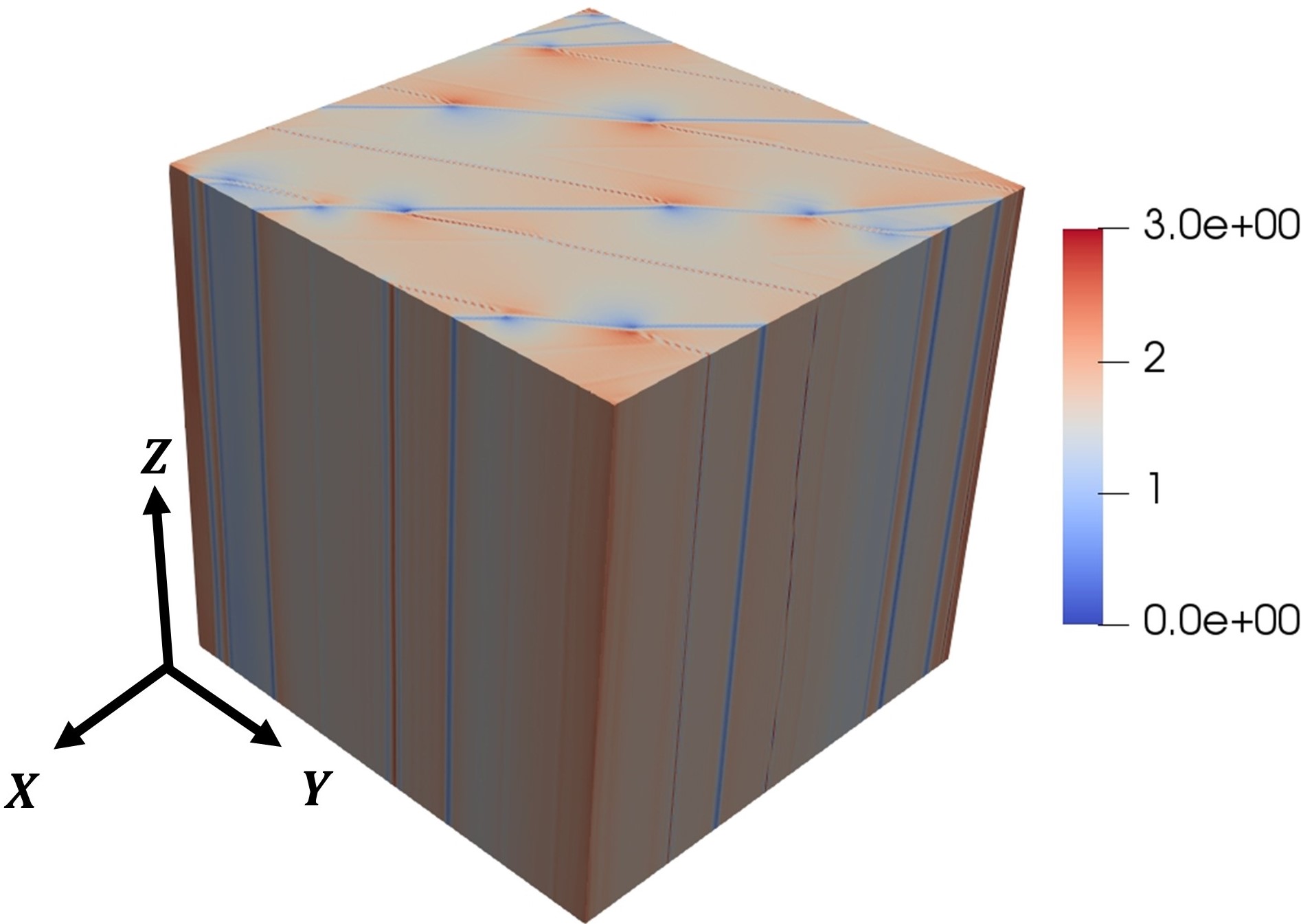}  
\caption{\textbf{}}
  \label{fig:elas3}
\end{subfigure} 
\begin{subfigure}{0.3\textwidth}
  \centering
  \includegraphics[width=\linewidth]{modi_fig/colormap2.jpg}  
\end{subfigure} 
 \caption{Simulated microstructures at room temperature showing steady-state spatial distribution of polar domains 
in stress-free equimolar BZCT as a function of electrostrictive anisotropy: (a) Case 1: single phase orthorhombic, (b) Case 2: coexistence of $T$ and $O$ domains, 
(c) Case 3: coexistence of $T$, $R$ and $O$ domains; (d, e, f) Electric energy distribution (nondimensional) corresponding to Cases 1, 2 and 3, respectively;
(g, h, i)  Corresponding nondimensional elastic energy distribution. 
    energy distribution (g, h, i)\\
    $O_{1}^{+}:[1 1 0]$, $O_{1}^{-}:[\bar{1} \bar{1} 0]$, $O_{2}^{+}:[0 1 1]$, $O_{2}^{-}:[0 \bar{1} \bar {1}]$,
    $O_{3}^{+}:[1 0 1]$, $O_{3}^{-}:[\bar{1} 0 \bar{1}]$, $O_{4}^{+}:[\bar{1} 1 0]$, $O_{4}^{-}:[1 \bar{1} 0]$, 
     $O_{5}^{+}:[0 \bar{1} 1]$, $O_{5}^{-}:[0 1 \bar{1}]$, $O_{6}^{+}:[\bar{1} 0 1]$, $O_{6}^{-}:[1 0 \bar{1}]$, 
     $T_{3}^{+}:[0 0 1]$, $T_{3}^{-}:[0 0 \bar{1}]$, $R_{2}^{+}: [\bar{1} 1 1]$, $R_{4}^{-}: [\bar{1} 1\bar{1}]$.
     }
     \label{fig:domains_stressfree}
\end{figure}

When $Q_z=1$, evolution leads to the formation of multiple variants of $O$ phase (Fig.~\ref{fig:domain1}). 
The domain walls of these variants show specific crystallographic orientation. Analysis
of mechanical compatibility using the difference in spontaneous strain between neighbouring
variants point to ferroelastic nature of the $60^{\circ}$ and $120^{\circ}$ domain walls.
Even the $180^{\circ}$ domain walls ($O_3^+/O_3^-$) show specific crystallographic orientation.
The steady state domain structure consists of a regular twin-related arrangement of plate shaped $O$ domains
separated by straight domain walls. Such a strain accommodating arrangement of plates leads to
reduction in elastic energy of the configuration. Absence of curvature of the domain walls indicates
stress-free nature of the domains and local electroneutrality at the domain walls. The trijunctions and quadrijunctions formed by the intersection of $60^{\circ}/120^{\circ}/180^{\circ}$ domain walls
are regions with increased electrostatic and elastic interactions (Figs.~\ref{fig:elec1},~\ref{fig:elas1}).

When $Q_z = 2$ (Case 2), evolution at early stages leads to formation of discrete islands of $T$ ($T_3^+/T_3^-$)
in $O$ ($O_6^-$) matrix (Fig.~\ref{fig:evol2}). These islands eventually get connected and arrange in
the form of thin striped network dividing the continuous $O$ domain into discrete parallel plates. Thus,
the steady state configuration shows a parallel plate geometry with the stripes 
of $T$ dividing the $O_6^-$
domain into discrete plates (Fig.~\ref{fig:domain2}). 
Although the straight walls separating the plates of $T+O$ are associated with low strain and electric
energy, there is a marked increase in electrostatic and elastic interactions when 
$T_3^+/T_3^-$ $180^{\circ}$ domain walls intersect with $O_6^-$ domains leading to an increase
in curvature of $T/O$ boundaries around the junction (Figs.~\ref{fig:elec2},~\ref{fig:elas2}).

In Case 3, evolution starts with clusters of all the three polar phases, $T$, $R$ and $O$, distributed
homogeneously throughout the volume (Fig.~\ref{fig:evol3}). 
Growth of these clusters leads to a steady state twinned pattern
wherein the plates of $T$, $R$, $O$ variants wedge into one another forming nearly equal number 
of $T-R$, $R-O$ and $T-O$ boundaries (Fig.~\ref{fig:domain3}). The domain pattern has the lowest 
average value of electric energy among these cases. However, at the 
triple junctions formed by the wedges of $T$, $R$ and $O$, 
we see an increase in electrostatic and elastic interactions (Figs.~\ref{fig:elec3},~\ref{fig:elas3}). Moreover, in all the cases, increase 
in elastic energy at the domain walls leads to a reduction in electric energy and 
vice versa indicating dual ferroelectric-ferroelastic character of these walls.

\begin{figure}[htbp]
\begin{subfigure}{0.3\textwidth}
  \centering
  \includegraphics[width=\linewidth]{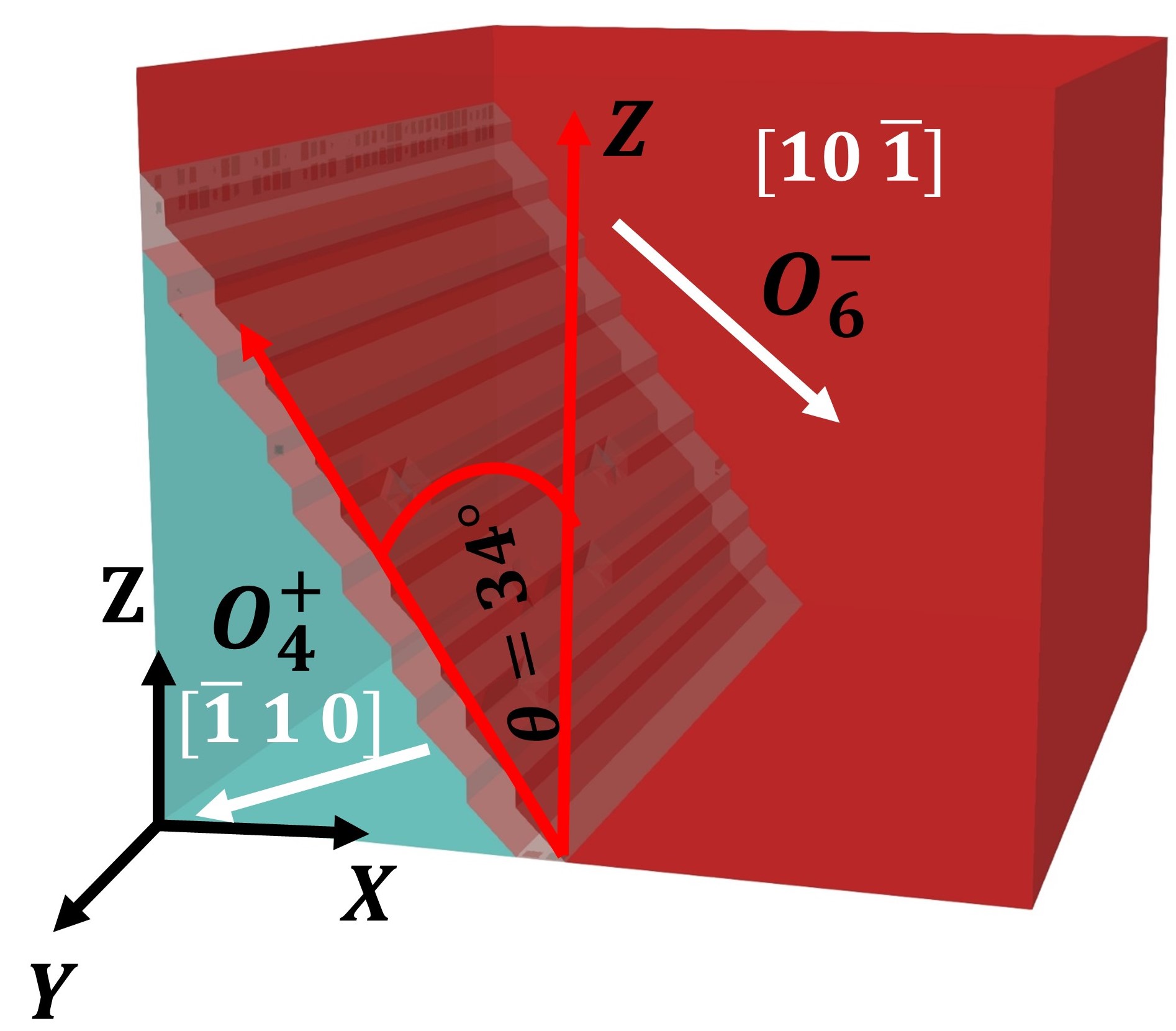}  
 \caption{\textbf{}}
  \label{fig:analysis5}
\end{subfigure}
\hspace{0.2 cm}
\begin{subfigure}{0.3\textwidth}
  \centering
  \includegraphics[width=1.13\linewidth]{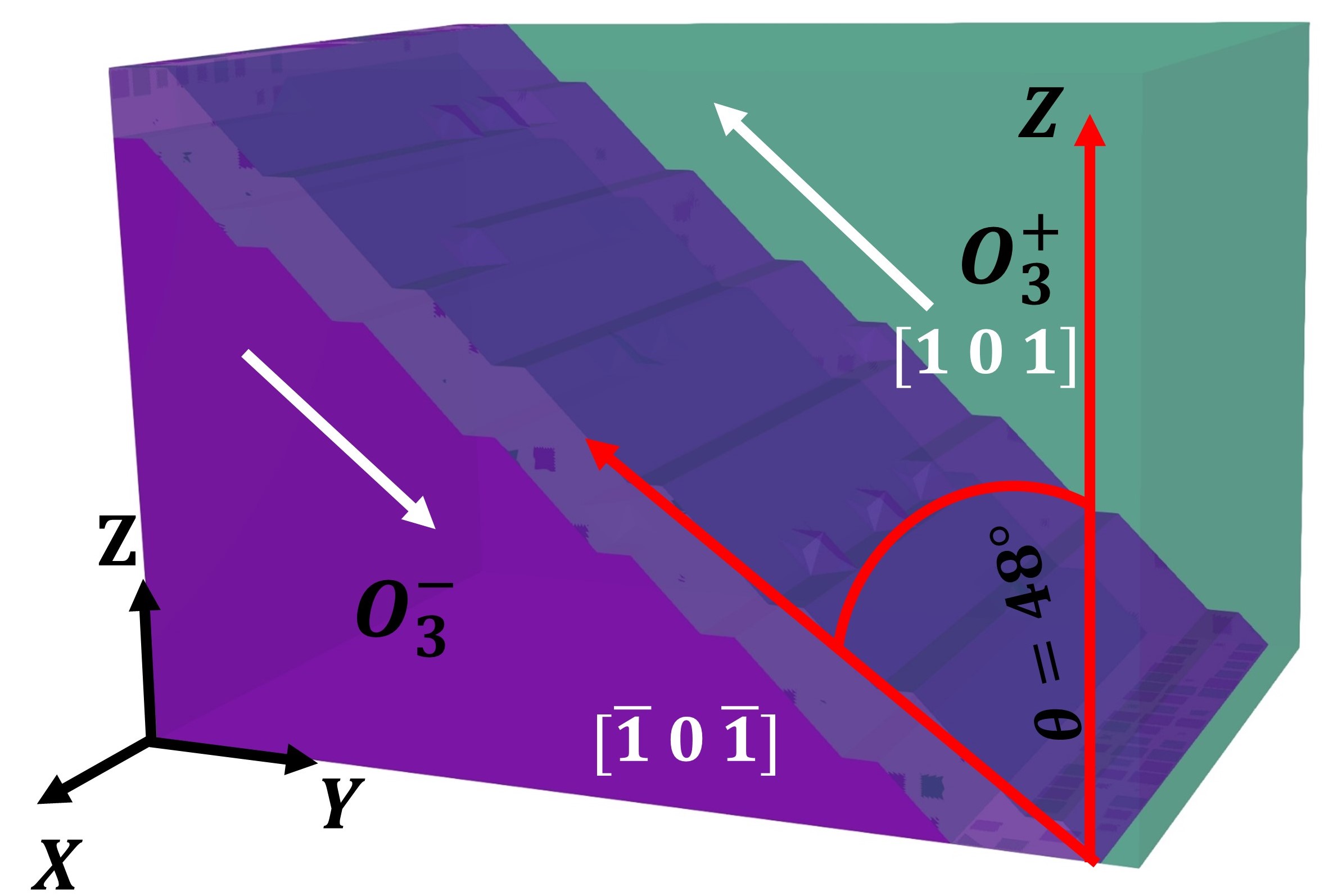}
  
 \caption{\textbf{}}
  \label{fig:analysis6}
\end{subfigure}
\hspace{0.6 cm}
\begin{subfigure}{0.3\textwidth}
  \centering
  \includegraphics[width=\linewidth]{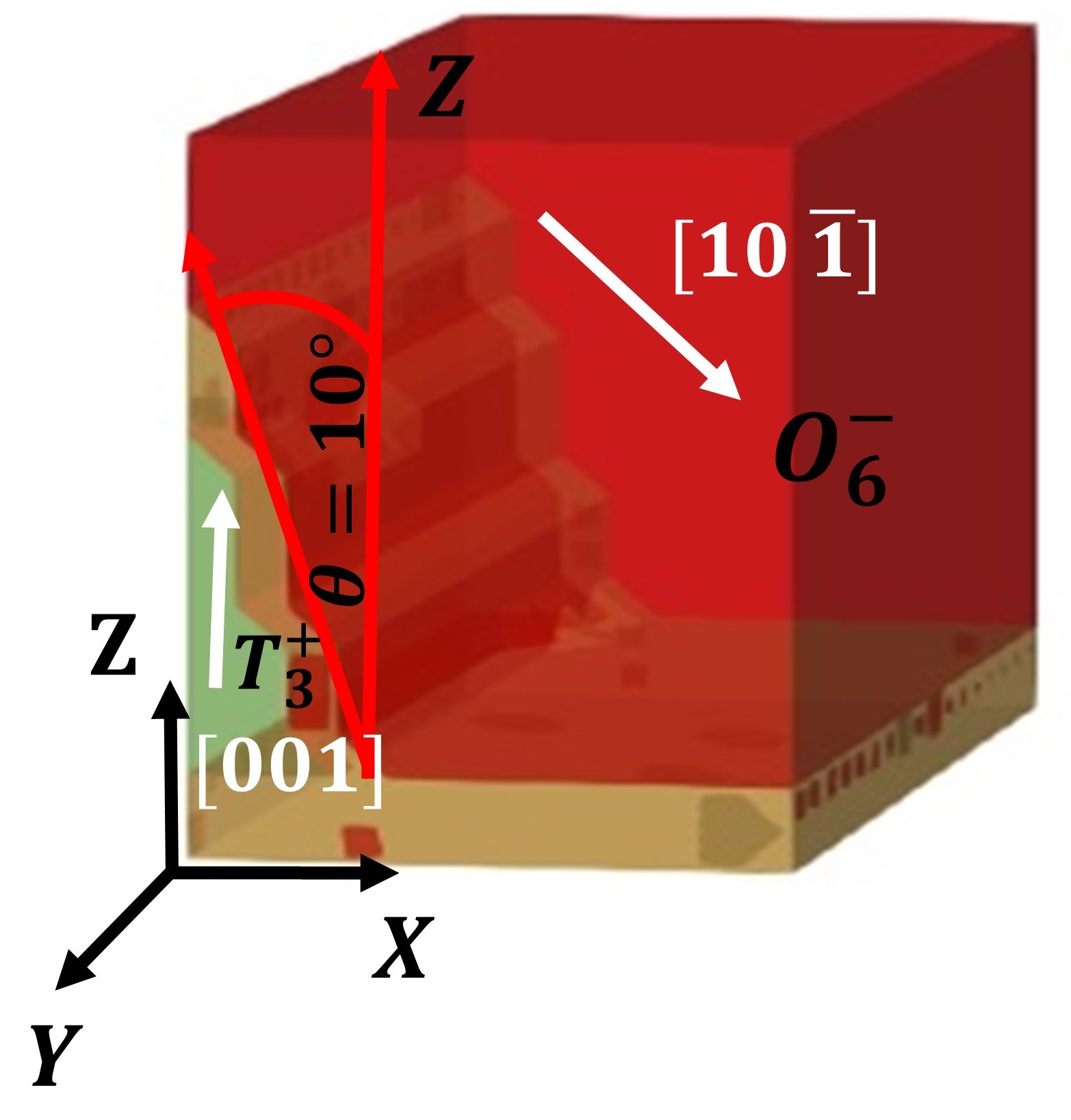}    
 \caption{\textbf{}}
  \label{fig:analysis4}
\end{subfigure}

\begin{subfigure}{0.3\textwidth}
  \centering
  \includegraphics[width=\linewidth]{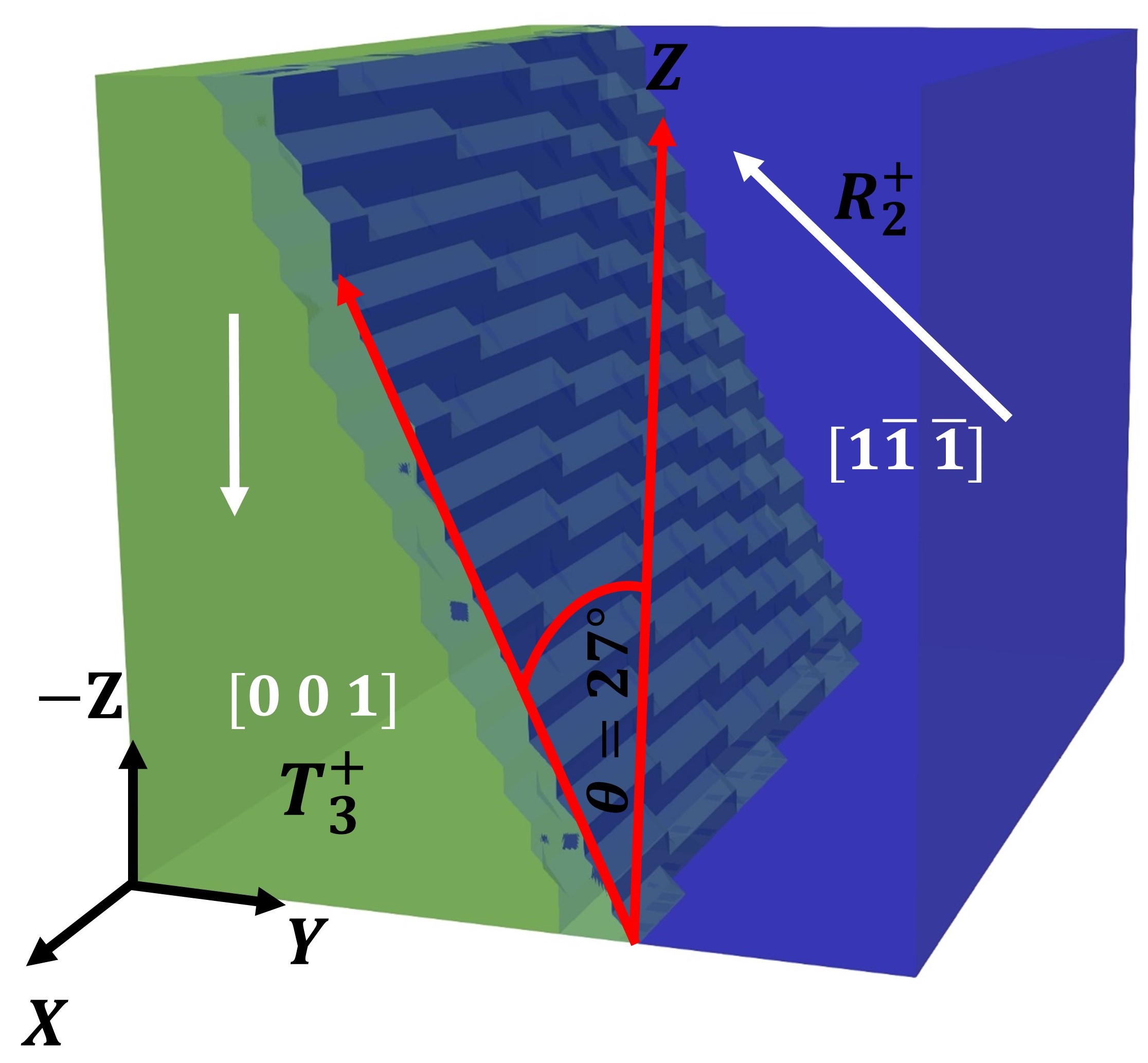}  
 \caption{\textbf{}}
  \label{fig:analysis1}
\end{subfigure}
\hspace{0.2 cm}
\begin{subfigure}{0.3\textwidth}
  \centering
  \includegraphics[width=\linewidth]{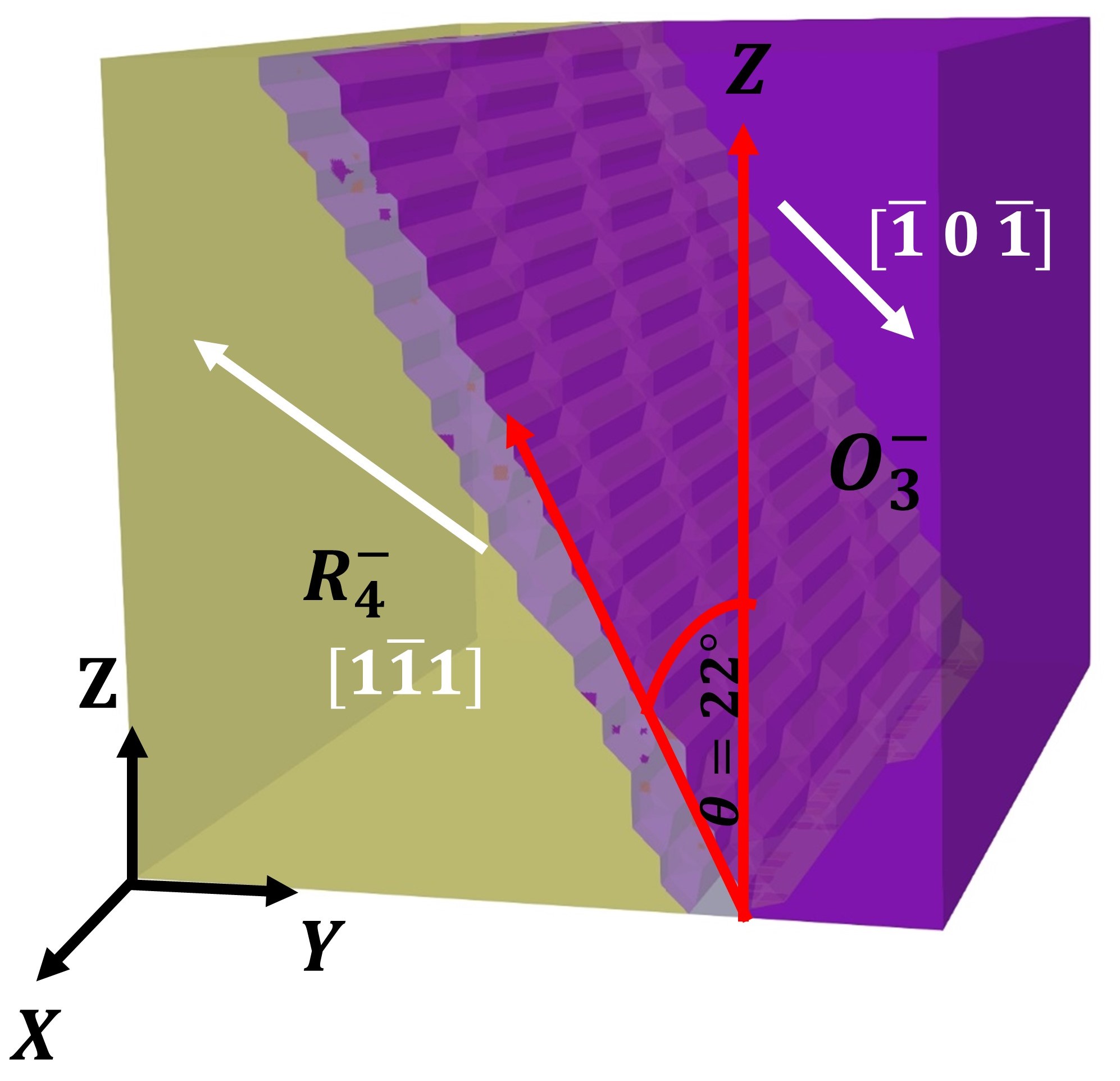}  
 \caption{\textbf{}}
  \label{fig:analysis2}
\end{subfigure}
\hspace{0.2 cm}
\begin{subfigure}{0.3\textwidth}
  \centering
  \includegraphics[width=1.12\linewidth]{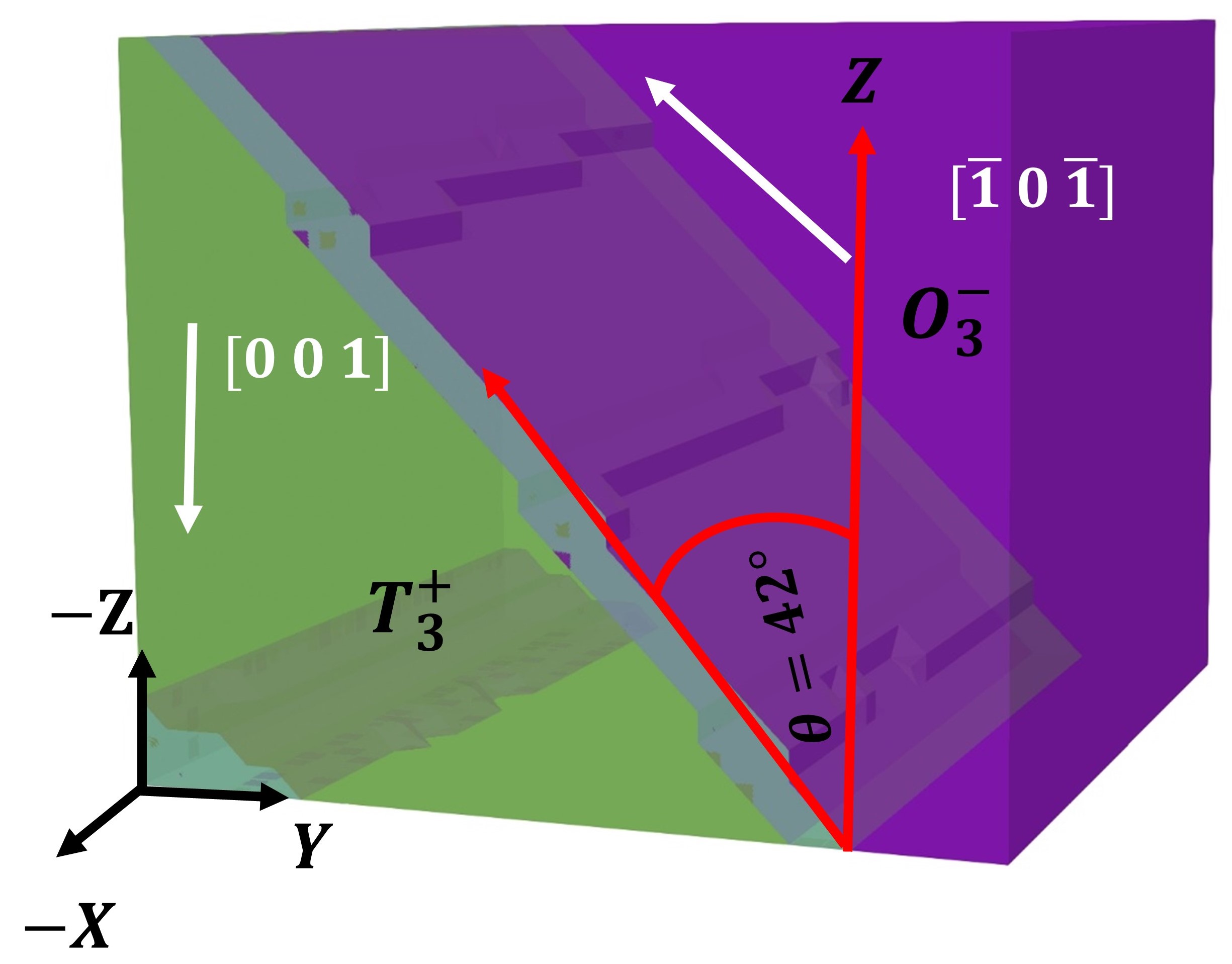}  
 \caption{\textbf{}}
  \label{fig:analysis3}
\end{subfigure}

\caption{Domain wall configurations in Case 1, Case 2, and Case 3 from our phase field simulations:
(a, b) Case 1: $120^{\circ}$ $O - O$ domain wall, $180^{\circ}$ $O - O$ domain wall 
(c) Case 2: $135^{\circ}$ $T - O$ domain wall, (d-f) Case 3: $125^{\circ}$ $T - R$ domain wall, $145^{\circ}$ $R - O$ domain wall, $135^{\circ}$ $T - O$ domain wall.}
\label{fig:analysis}
\end{figure}

The $O$ domain walls in Case 1, $T/O$ phase boundary in Case 2, and $R/O$, 
$O/T$ and $T/R$ phase boundaries in Case 3 have a common feature - they show a regular step-terrace structure where the steps are nearly perpendicular to domain wall orientations (Fig.~\ref{fig:analysis}), although step sizes vary for different 
combinations of orientation. In all cases, such a step-terrace structure suggests multistep switching via successive $90^{\circ}$ ferroelastic steps instead of single-step $180^{\circ}$ switching~\cite{xu2015ferroelectric}.

\begin{figure}[htbp]
\begin{subfigure}{0.5\textwidth}
  \centering
  \includegraphics[width=\linewidth]{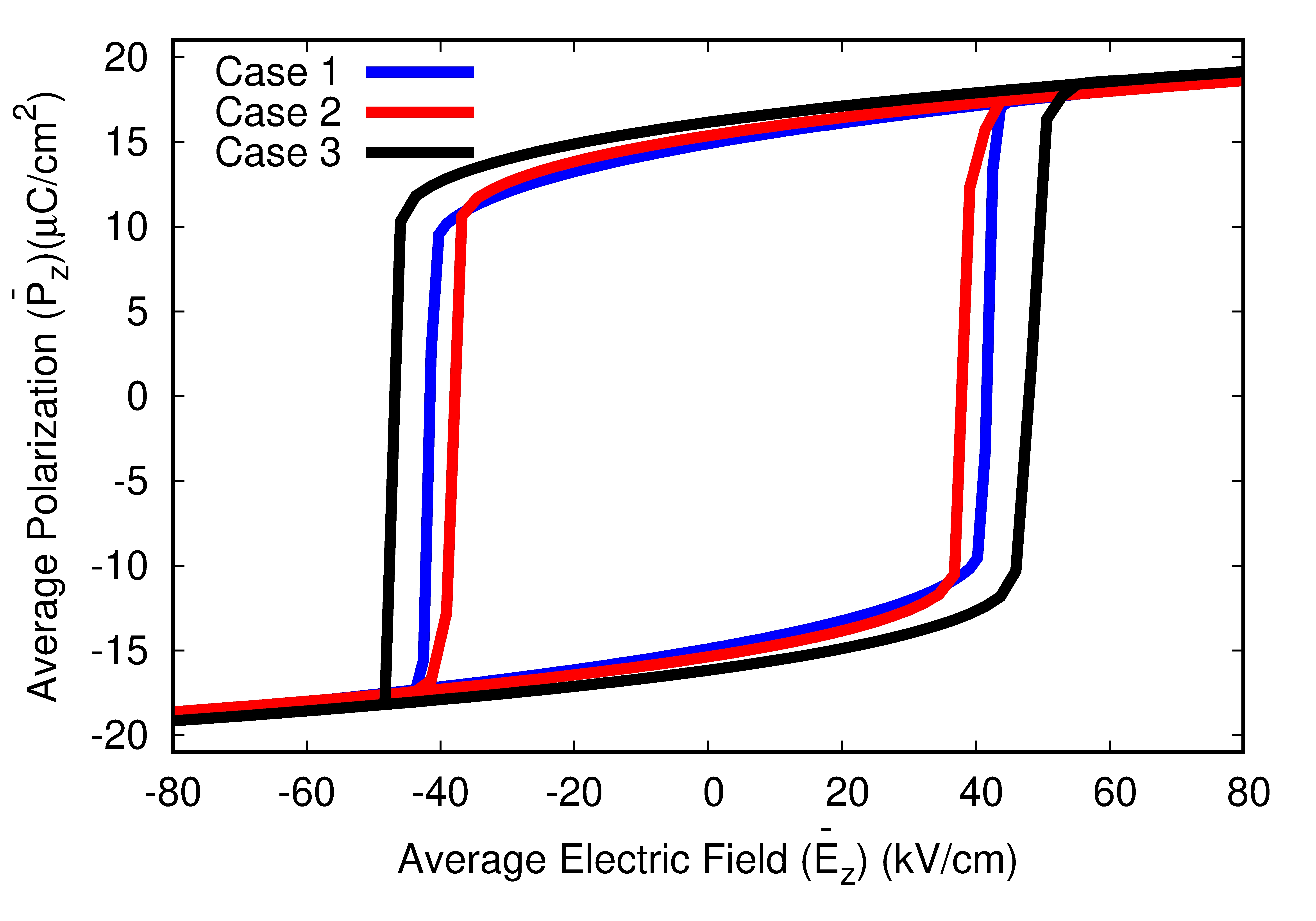}  
 \caption{\textbf{}}
  \label{fig:d33pe}
\end{subfigure}
\begin{subfigure}{0.5\textwidth}
  \centering
  \includegraphics[width=\linewidth]{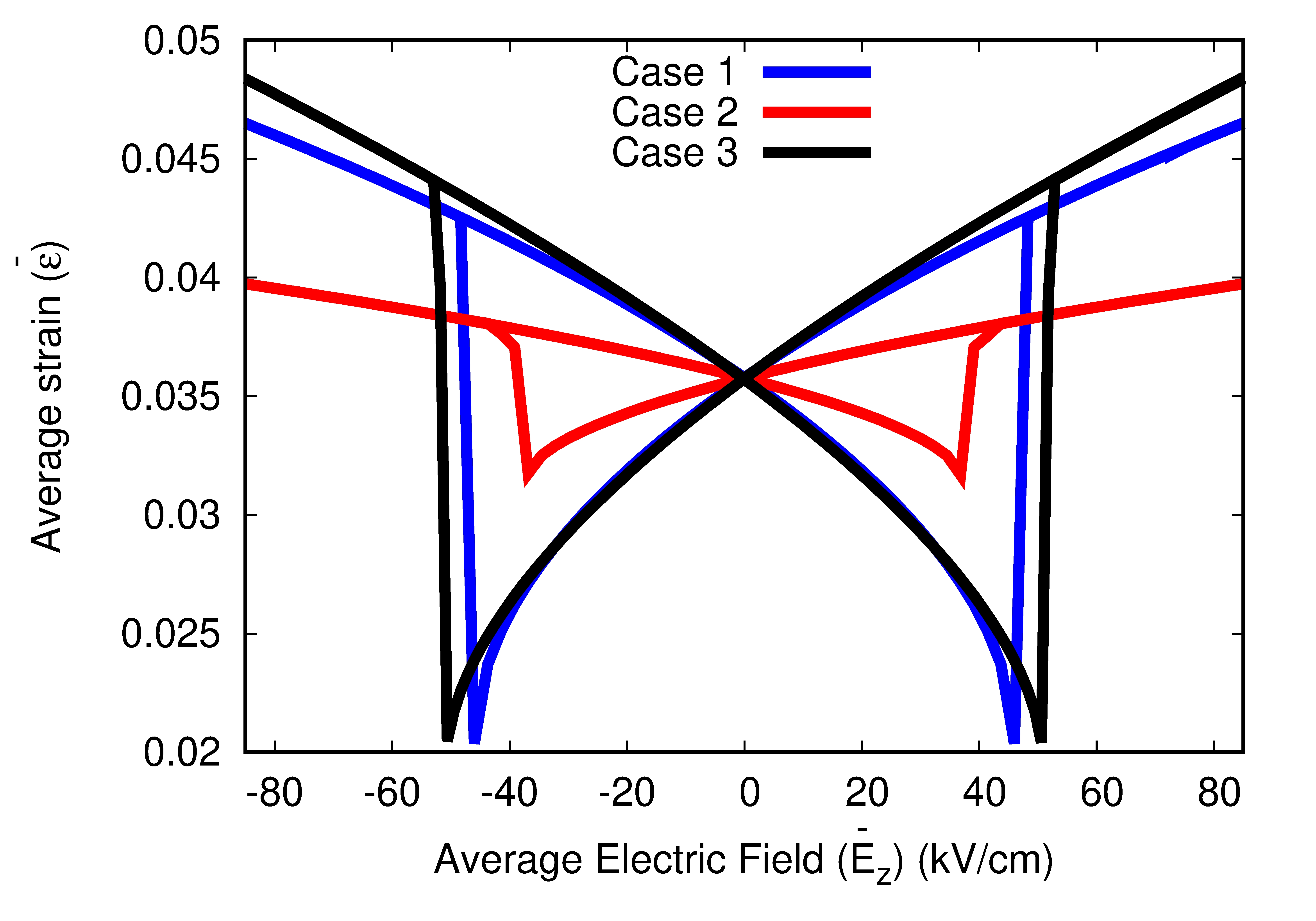}  
 \caption{\textbf{}}
  \label{fig:d33butter}
\end{subfigure} 
\begin{subfigure}{0.5\textwidth}
  \centering
  \includegraphics[width=\linewidth]{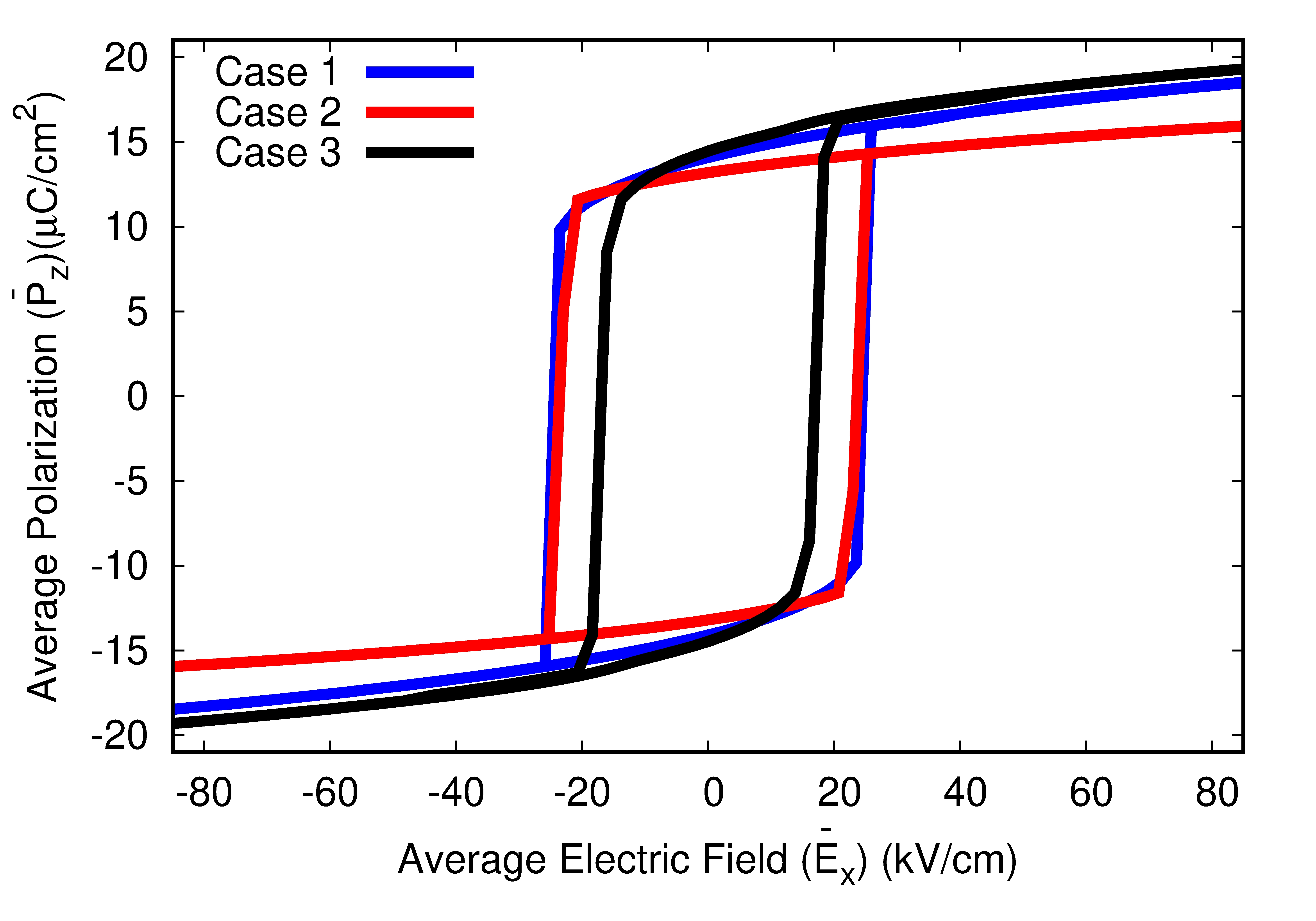}  
 \caption{\textbf{}}
  \label{fig:d31pe}
\end{subfigure} 
\begin{subfigure}{0.5\textwidth}
  \centering
  \includegraphics[width=\linewidth]{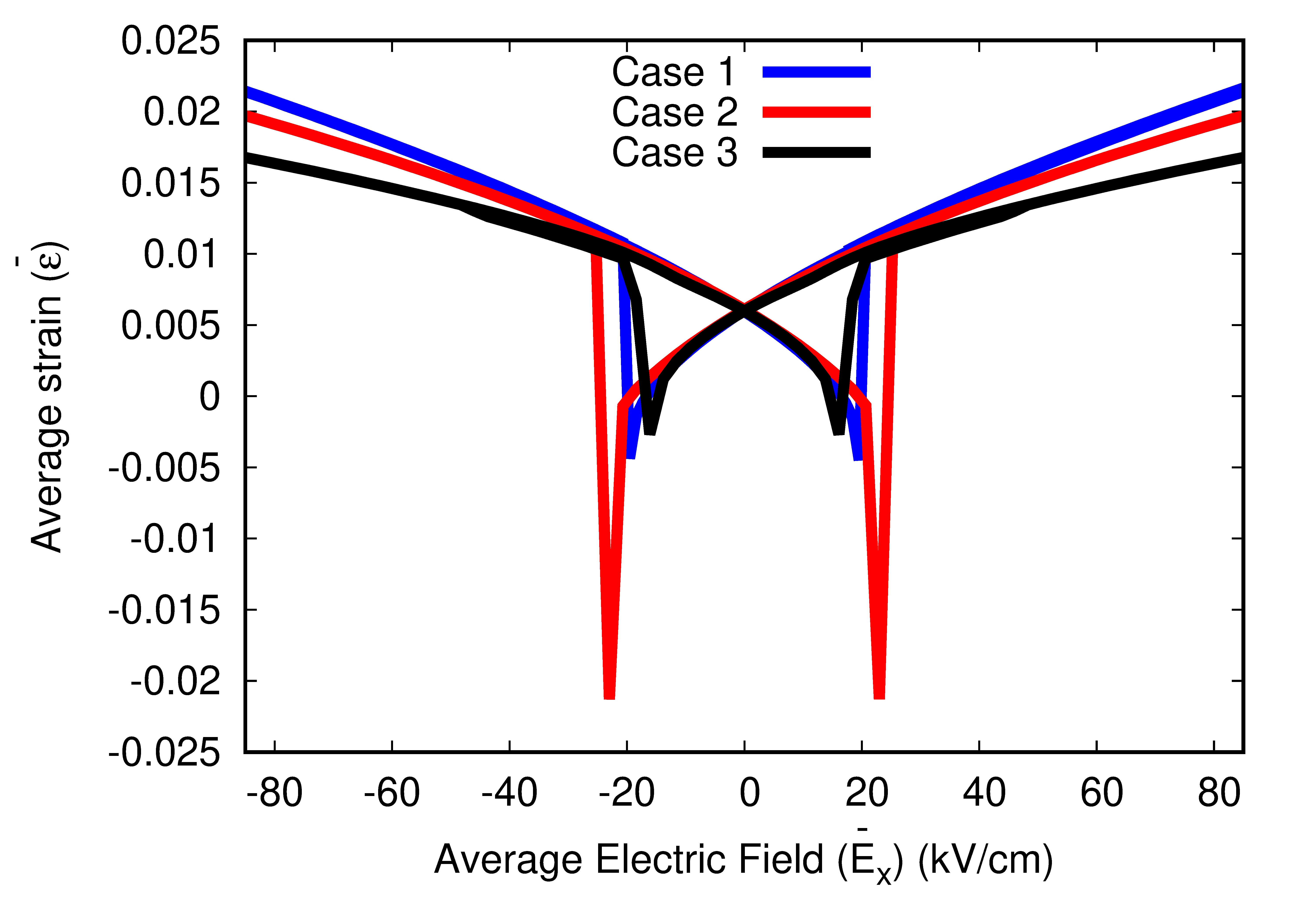}  
\caption{\textbf{}}
  \label{fig:d31butter}
\end{subfigure} 
 \caption{Corresponding hysteresis and the butterfly loops for Case 1 ($Q_z = 1$),
 Case 2 ($Q_z = 2$), and Case 3 ($Q_z = 2.5$). 
 (a), (b) Polarization and longitudinal strain hysteresis loops when applied electric field is along $[0 0 1]$ direction. 
 (c), (d) Polarization and transverse strain hysteresis loops when applied electric field is along $[1 0 0]$ direction. }
\label{fig:loops}
\end{figure}

Change in $Q_z$ can change the easy polarization directions in ferroelectric system 
(measured by the ease of switching under an applied field) leading to variations in switching characteristics. To quantify such a variation we measure the difference between effective $d_{31}$ and $d_{33}$ as a function of $Q_z$. Therefore, in each case, we subject the steady-state domain structure (obtained at zero electric field) to an applied field 
varying between $-300\si{\kilo\volt\centi\meter\tothe{-1}}$
and $300\si{\kilo\volt\centi\meter\tothe{-1}}$ with a step size of 
$\pm 5\si{\kilo\volt\centi\meter\tothe{-1}}$
along $[001]$ ($E_3$) and $[100]$ ($E_1$) directions 
to compute the electromechanical
switching properties given by polarization hysteresis loop ($\bar{P}_z - E_3$), 
longitudinal strain hysteresis loop ($\epsilon_{33} - E_3$) and 
transverse strain hysteresis loop ($\epsilon_{33} - E_1$).
The computed $\epsilon_{33} - E_3$ and  $\epsilon_{33} - E_1$ (Figs.~\ref{fig:d33butter} and~\ref{fig:d31butter}) 
loops for all cases show a typical butterfly shape that is symmetric about zero
applied field. We use the longitudinal and transverse strain hysteresis loops to 
determine the coercive field $E_c$ (defined as the field required for complete reversal of polarization) along $[001]$ and $[100]$ directions. 
When electric field is applied along $[100]$ direction, 
$E_c^{\textrm{Case1}}\approx E_c^{\textrm{Case2}}>E_c^{\textrm{Case3}}$. On the other hand,
when electric field is applied along $[001]$ direction, $E_c^{\textrm{Case3}}>E_c^{\textrm{Case1}}>E_c^{\textrm{Case2}}$. 
Moreover, for the isotropic case (Case 1, $Q_z=1$), the difference in longitudinal and transverse $E_c$ values is the lowest. The difference in $E_c$ values is the largest for Case 3 (showing three-phase coexistence) followed by Case 2 (showing two-phase coexistence). Thus, anisotropy in switching behaviour increases with increasing electrostrictive anisotropy and decreasing polar anisotropy. 

The effective piezoelectric coefficients $d_{33}$ and $d_{31}$ for each case are obtained from the slopes of the linear portions of the corresponding butterfly loops. The computed values of effective $d_{33}$, $d_{31}$ and their ratio
($d_{33}$/$d_{31}$) for all cases are listed in Table~\ref{table:d33}.
The ratio increases with increasing anisotropy in electrostriction 
$Q_z$ ($Q_z>1)$. 
\begin{table}[htbp]
\caption{$d_{33}$, $d_{31}$ and their ratio for the three cases} 
\centering 
\begin{tabular}{| m{3cm}| | m{3cm}| m{3cm} | m{3cm} |}
\hline\hline 
 & Case $1$ & Case $2$ & Case $3$\\ [2ex] 
\hline 
 $d_{33}(\si{\pico\coulomb\per\newton})$& $640$ & $585$ & $656$\\ %
 \hline
 $d_{31}(\si{\pico\coulomb\per\newton})$& $302$ & $230$ & $202$\\
 \hline
 $d_{33}/d_{31}$& $2.1$ & $2.5$ & $3.2$\\
\hline 
\end{tabular}
\label{table:d33} 
\end{table}
The table shows increase $d_{33}/d_{31}$ with increasing $Q_z$ indicating an increase in piezoelectric anisotropy.

As we have already noted, number of phases increase with increasing $Q_z$. When the increase in the number of phases induces an increase in the number of distinct crystallographic variants, we generally expect enhancement in electromechanical response due to consequent increase in energetically favourable switching pathways. Therefore, the $\bar{P}_z - E_3$ and  $\bar{\varepsilon}_{33}-E_3$ loops corresponding to $Q_z=2$ have the least width compared to $Q_z=1, 2.5$ when the field is along $[001]$ direction because the domain structure in the former contains the lowest number of distinct crystallographic variants ($n_{\textrm{var}}=3$). However, electromechanical response also depends on the orientation of domain walls relative to the direction of applied switching field. Thus, $\bar{P}_z - E_1$ and  $\bar{\varepsilon}_{33}-E_1$  for $Q_z=2$ shows the least width when the field is along $[100]$.

In general, Fig.~\ref{fig:loops}a, b show fatter $\bar{P}_z - E_3$, $\bar{\varepsilon}_{33}-E_3$ loops when electric field is applied along $[0 0 1]$ direction. On the contrary, $\bar{P}_z - E_1$ and $\bar{\varepsilon}_{33}-E_1$ loops are narrower for all cases of $Q_z$ when subjected to an electric field along 
$[1 0 0]$ direction. Moreover, $d_{33}$ is always larger than the corresponding $d_{31}$ with a monotonic increase in the $d_{33}/d_{31}$ ratio with increasing $Q_z$. 

The anisotropic nature of predicted hysteresis loops are commensurate with the underlying domain pattern and the domain wall structure for all cases of $Q_z$. 
Since the domain boundaries in equimolar BZCT have a strong ferroelastic nature, as evidenced by their step-terrace structure for all cases of $Q_z$, polarization reversal in each case happens via successive $90^{\circ}$ steps 
where the easy polarization rotation axes are determined by the orientation of step or terrace relative to the direction of applied field (Fig.~\ref{fig:analysis}).
Moreover, we find the aspect ratio between terrace (perpendicular to $[0 0 1]$)
and step (parallel to [0 0 1]) to be greater than unity in all cases. 
This indicates a larger polarization component
associated with the terrace that is normal to $[0 0 1]$ than a step that is parallel to $[1 0 0]$. Therefore, the energy required for polarization reversal is higher when the switching field is along $[0 0 1]$ direction than it is along 
$[1 0 0]$ direction. For a similar reason, strain associated with longitudinal strain hysteresis loop ($\bar{\varepsilon}_{33}-E_3$) is always larger than that associated with the transverse strain hysteresis loop ($\bar{\varepsilon}_{33}-E_1$).

\begin{figure}[htbp]
    \centering
    \includegraphics[width=0.75\linewidth]{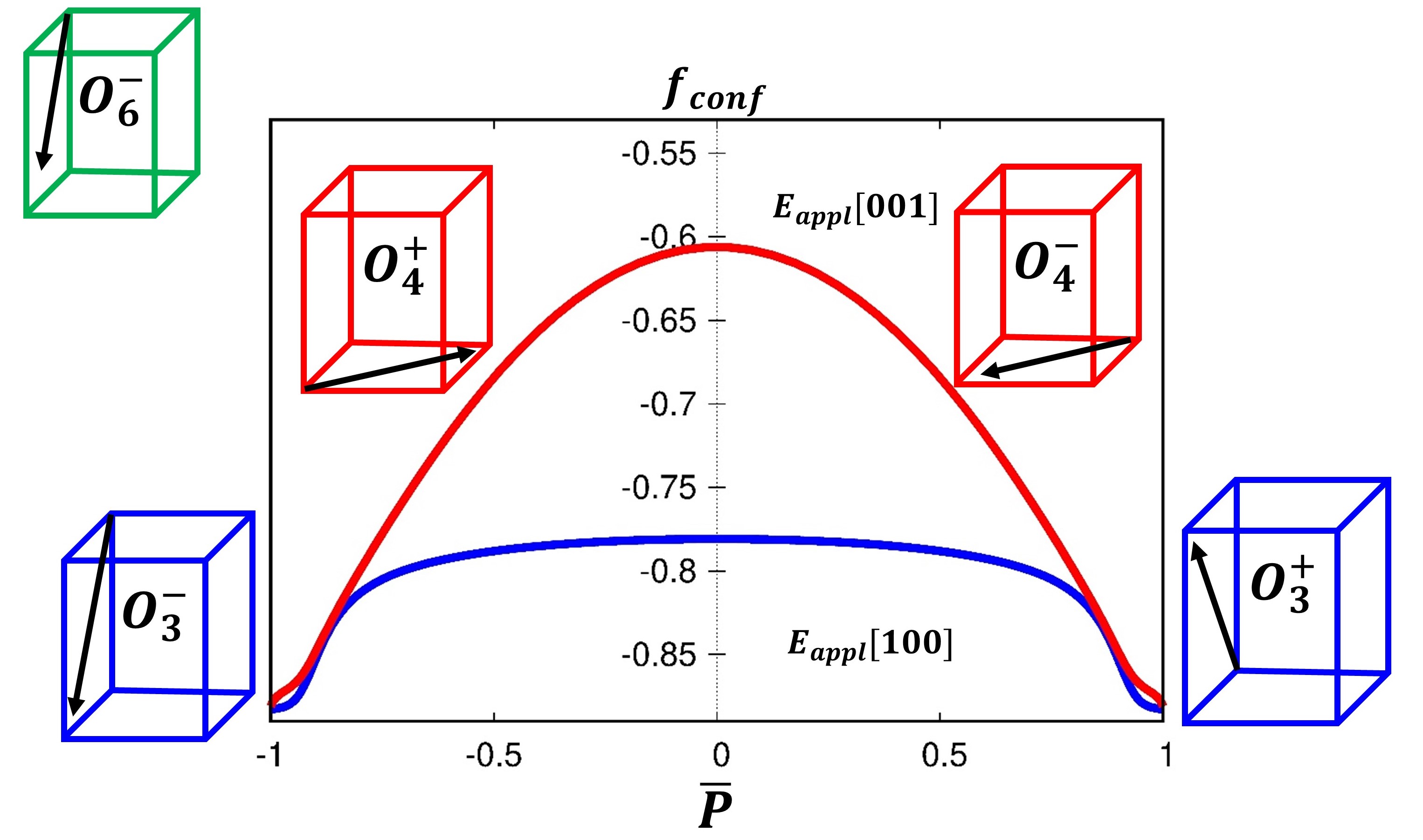}
    \caption{Case 2: Configuration energy (f$_{\textrm{conf}}$) as a function of average polarization.
    The green unit cell defines the initial $O_{6}^{-}$ state before applying electric field.
    Application of electric field along $[001]$ direction switches the system to $O_{4}^{-}$ state
    as shown by the red unit cell.
    When the applied electric field is along $[100]$ direction the system transforms to
    $O_{3}^{+}$ (shown by blue unit cell). The difference between the energy clearly indicates 
    that the switching of $O_{3}^{+}$ from $O_{6}^{-}$ is easier compared to $O_{4}^{-}$. 
    }
    \label{fig:configuration}
\end{figure}

Alternatively, one can explain the difference between $d_{33}$ and $d_{31}$ based on the energy barrier
associated with switching.
For example, in Case 2,
when we apply electric field along $[001]$ direction, the system transforms to $O_{4}^{-}$ from 
$O_{6}^{-}$. 
On the other hand, application of electric field along $[100]$ direction 
switches the system to $O_{3}^{+}$. In Fig.~\ref{fig:configuration} we show configuration energy (f$_{\textrm{conf}}$)
(i.e., the energy required to switch the polarization variants 
with applied electric field) as a
function of average polarization. The lower
energy barrier for $O_{3}^{+}$ (blue line) clearly indicates
that the switching from $O_{6}^{-}$ to $O_{3}^{+}$ with applied electric field along $[1 0 0]$ direction
requires lower energy (easy polarization switching) compared to the
other one (red line).
 
We also calculate the effective elastic moduli for all three cases as shown 
in Table~\ref{table:effective_mod}. Here we compare the 
elastic softening associated with phase coexistence.
The effective elastic stiffness tensor $C_{ijkl}^{eff}$
is obtained by measuring the stress response when the
system is subjected to applied strain~\cite{bhattacharyya2012spectral}.
For a given applied
strain $\bar{\varepsilon}_{ij}$ and an eigenstrain distribution $\varepsilon_{ij}^{0}(\mathbf{r})$, 
the stress field is given by
\begin{equation}
    \sigma_{ij}(\mathbf{r}) = C_{ijkl}[\bar{\varepsilon}_{kl} + \delta\varepsilon_{kl}(\mathbf{r}) - \varepsilon_{kl}^{0}(\mathbf{r})].
\end{equation}
The average stress $\sigma_{ij}^{ave}$ in a material is calculated as
\begin{equation}
    \sigma_{ij}^{ave} = \frac{1}{V}\int_{V}\sigma_{ij}(\mathbf{r})dV.
\end{equation}
Thus, the effective elastic stiffness tensor $C_{ijkl}^{eff}$ is written as
\begin{equation}
    \sigma_{ij}^{ave} = C_{ijkl}^{eff}\bar{\varepsilon}_{kl}.
\end{equation}
The effective elastic moduli corresponding to Case 3 ($Q_z = 2.5$) show the lowest
values among all cases indicating an increase in elastic softening with the increase
in the number of crystallographically distinct variants.

\begin{table}[ht]
\caption{Effective elastic moduli as a function of $Q_z$} 
\centering 
\begin{tabular}{| m{3cm}| | m{3cm}| m{3cm} | m{3cm} |}
\hline\hline 
 Effective modulus ($\si{\giga\pascal}$) & Case $1$ & Case $2$ & Case $3$\\ [2ex] 
\hline 
 $C_{11}^{eff}$& $242$ & $246$ & $239$\\ %
 \hline
 $C_{12}^{eff}$& $132$ & $137$ & $129$\\
 \hline
 $C_{44}^{eff}$& $55$ & $57$ & $51$\\
\hline 
\end{tabular}
\label{table:effective_mod} 
\end{table}

\begin{figure}[htpb]
\begin{subfigure}{0.4\textwidth}
  \centering
  \includegraphics[width=0.75\linewidth]{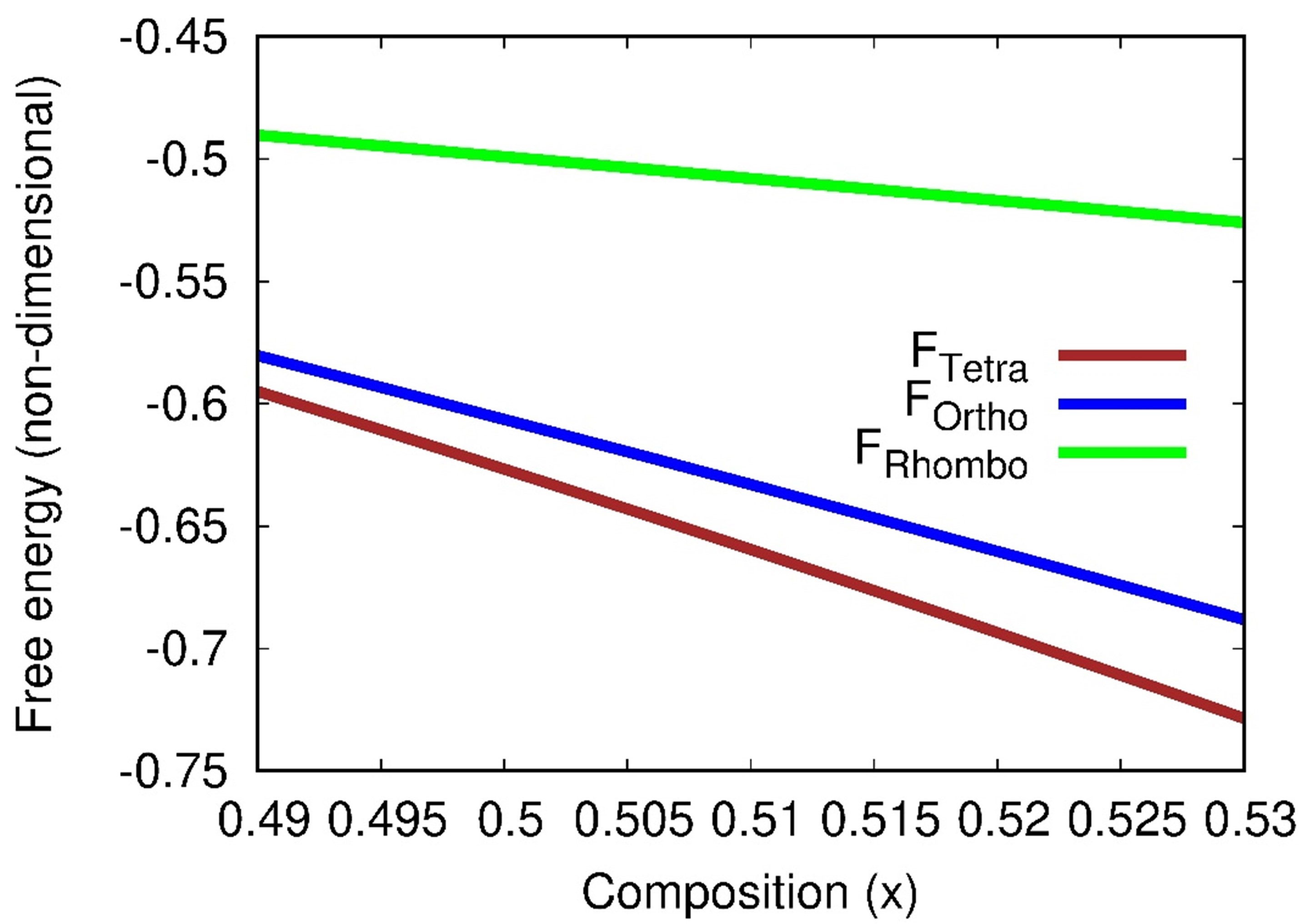}  
 \caption{\textbf{}}
  \label{fig:free_cons_zoom}
\end{subfigure}
\begin{subfigure}{0.4\textwidth}
  \centering
  \includegraphics[width=0.75\linewidth]{modi_fig/free_comp_normzoomed.jpg}  
 \caption{\textbf{}}
  \label{fig:free_norm_zoom}
\end{subfigure}
\begin{subfigure}{0.4\textwidth}
  \centering
  \includegraphics[width=0.72\linewidth]{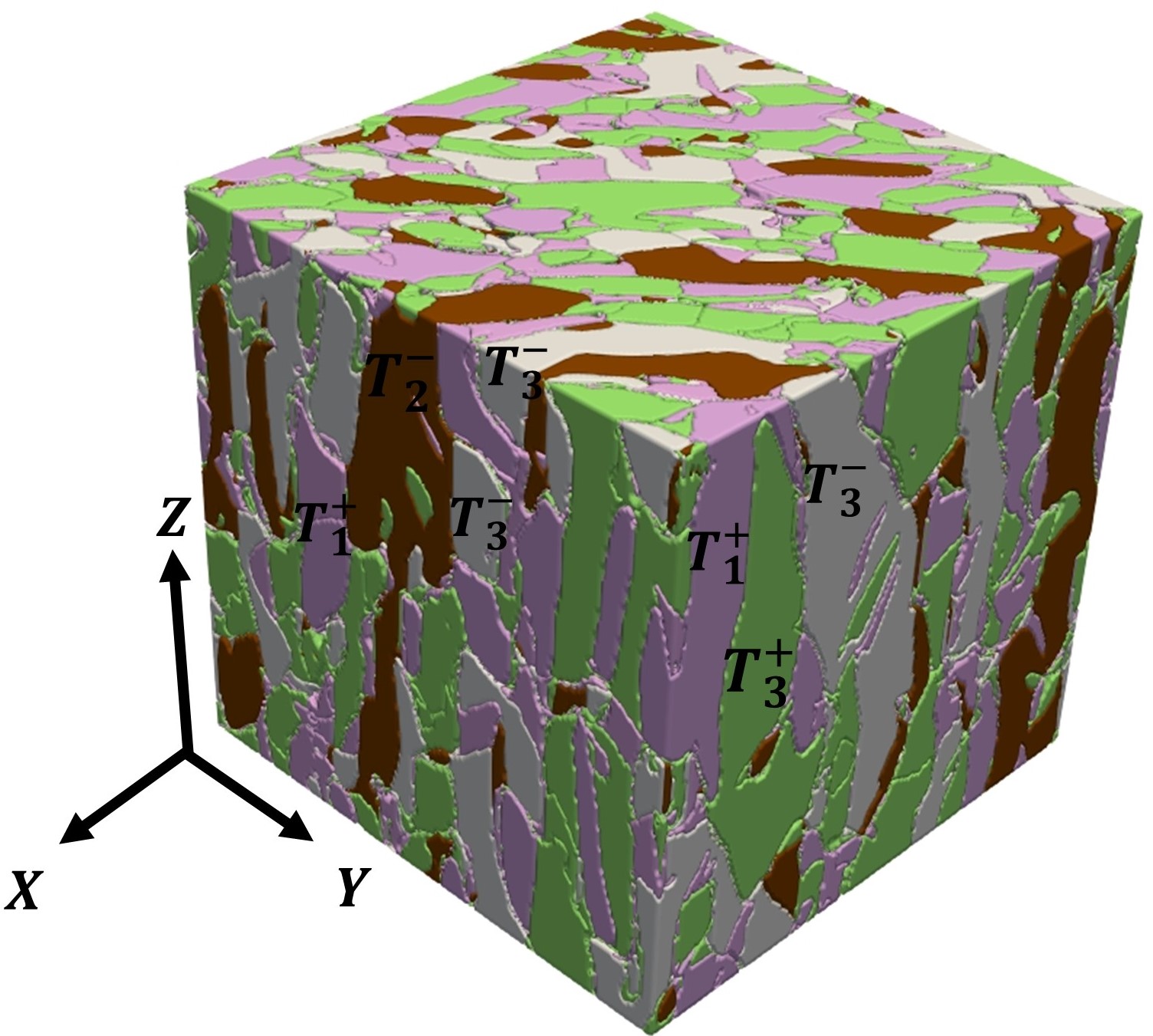}  
 \caption{\textbf{}}
  \label{fig:constrained_new}
\end{subfigure}
\begin{subfigure}{0.4\textwidth}
  \centering
  \includegraphics[width=0.7\linewidth]{modi_fig/domain1.jpeg.jpg}  
 \caption{\textbf{}}
  \label{fig:cons_com_domain}
\end{subfigure} 
\begin{subfigure}{0.4\textwidth}
  \centering
  \includegraphics[width=0.8\linewidth]{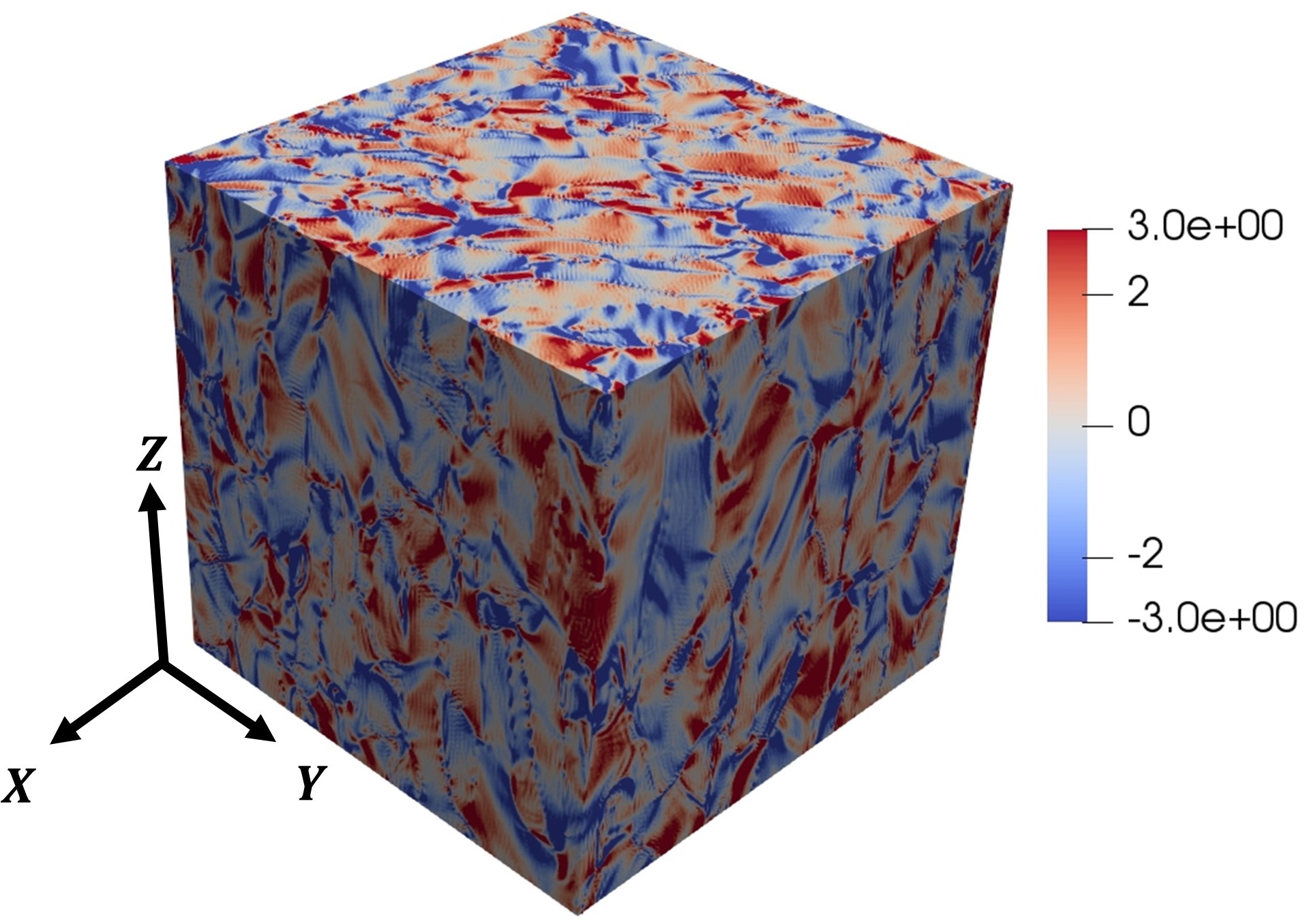}  
 \caption{\textbf{}}
  \label{fig:cons_elec}
\end{subfigure} 
\begin{subfigure}{0.4\textwidth}
  \centering
  \includegraphics[width=0.8\linewidth]{modi_fig/electric_Q.jpeg.jpg}  
 \caption{\textbf{}}
  \label{fig:stress_elec}
\end{subfigure} 
\begin{subfigure}{0.4\textwidth}
  \centering
  \includegraphics[width=0.8\linewidth]{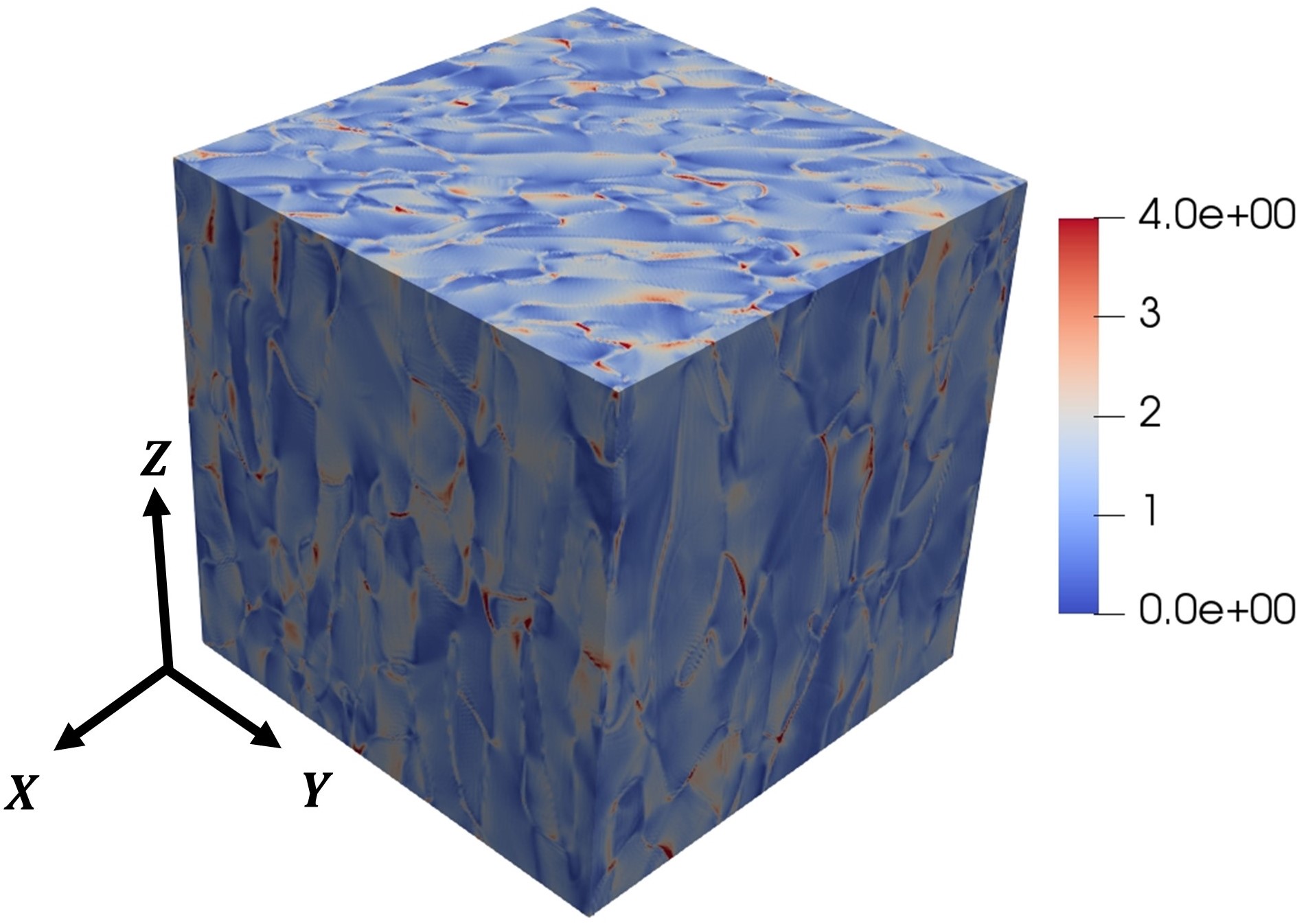}  
 \caption{\textbf{}}
  \label{fig:cons_elas}
\end{subfigure} 
\begin{subfigure}{0.4\textwidth}
  \centering
  \includegraphics[width=0.8\linewidth]{modi_fig/elasatic_Q.jpeg.jpg}  
 \caption{\textbf{}}
  \label{fig:stress_elas}
\end{subfigure} 
\centering
\begin{subfigure}{0.3\textwidth}
  \includegraphics[width=\linewidth]{modi_fig/colormap2.jpg}  
\end{subfigure} 
 \caption{Free energy - composition diagram of clamped BZCT system at room temperature indicating minimum free energy of the $T$  phase between $0.49\leq
 \leq 0.53$. (b) Corresponding free energy-composition diagram of stress-free BZCT system at room temperature. Simulated steady-state domain structures of equimolar BZCT at room temperature for (c) mechanically constrained and (d) stress-free conditions. Here $Q_z=1$.  
 (e, f) Electric energy distribution (nondimensional) corresponding to constrained
  and stress-free systems; (g, h) Elastic energy distribution (nondimensional) corresponding to constrained and stress-free systems.\\
    $O_{1}^{+}:[1 1 0]$, $O_{1}^{-}:[\bar{1} \bar{1} 0]$, $O_{2}^{+}:[0 1 1]$, $O_{2}^{-}:[0 \bar{1} \bar {1}]$,
    $O_{3}^{+}:[1 0 1]$, $O_{3}^{-}:[\bar{1} 0 \bar{1}]$, $O_{4}^{+}:[\bar{1} 1 0]$, $O_{4}^{-}:[1 \bar{1} 0]$, 
     $O_{5}^{+}:[0 \bar{1} 1]$, $O_{5}^{-}:[0 1 \bar{1}]$, $O_{6}^{+}:[\bar{1} 0 1]$, $O_{6}^{-}:[1 0 \bar{1}]$, 
     $T_{3}^{+}:[0 0 1]$, $T_{3}^{-}:[0 0 \bar{1}]$, $T_{1}^{+}:[1 0 0]$, $T_{2}^{-}:[0 \bar{1} 0]$.
     }
     \label{fig:constrained}
\end{figure}

Since our thermodynamic model includes elastic interactions, we can use the model 
to analyse phase stability in stress-free as well as 
mechanically constrained systems. When BZCT system with isotropic electrostrictive coefficients ($Q_z=1$) is clamped in all directions (imposed by setting all components of homogeneous/macroscopic strain  to 
be zero in the entire system), 
our model predicts change in room-temperature phase stability from orthorhombic (stress-free) to 
tetragonal (clamped) when the composition ranges between $0.49\leq x \leq 0.53$, 
as shown in the free energy-composition 
diagrams of stress-free and constrained BZCT (Figs.~\ref{fig:free_norm_zoom},~\ref{fig:free_cons_zoom}).
The simulated steady-state domain structure of constrained equimolar BZCT at room temperature contains only tetragonal variants confirming our thermodynamic stability analysis for the mechanically constrained system (Fig.~\ref{fig:constrained_new}). The domain structure of constrained BZCT consists of thin stripes of $T$ variants with curved domain walls, while the stress-free system possesses thicker and wider plates of $O$ variants with straight boundaries. 
Although elastic interactions show a marked increase for the constrained system, electric interactions associated with both systems remain nearly the same
(Figs.~\ref{fig:cons_elec},~\ref{fig:cons_elas}). The clamped system shows internal stress buildup given by $\bar{\sigma}_{ij}=q_{ijkl}\langle P_kP_l \rangle$.  
As a result, domain walls in this system show an increase in curvature.

\begin{figure}[htpb]
\begin{subfigure}{0.5\textwidth}
  \centering
  \includegraphics[width=\linewidth]{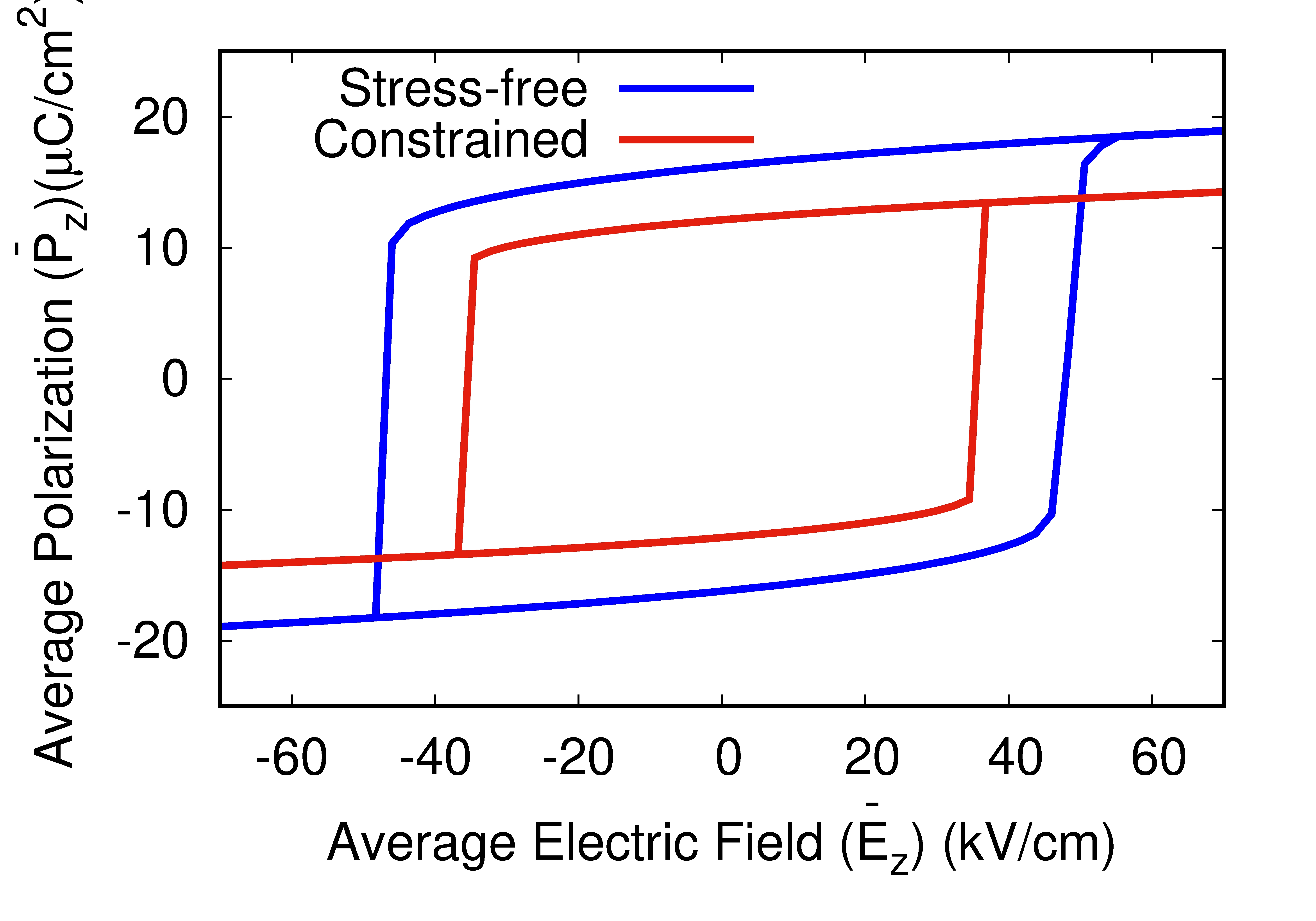}  
 \caption{\textbf{}}
  \label{fig:comparisonpe}
\end{subfigure}
\begin{subfigure}{0.5\textwidth}
  \centering
  \includegraphics[width=\linewidth]{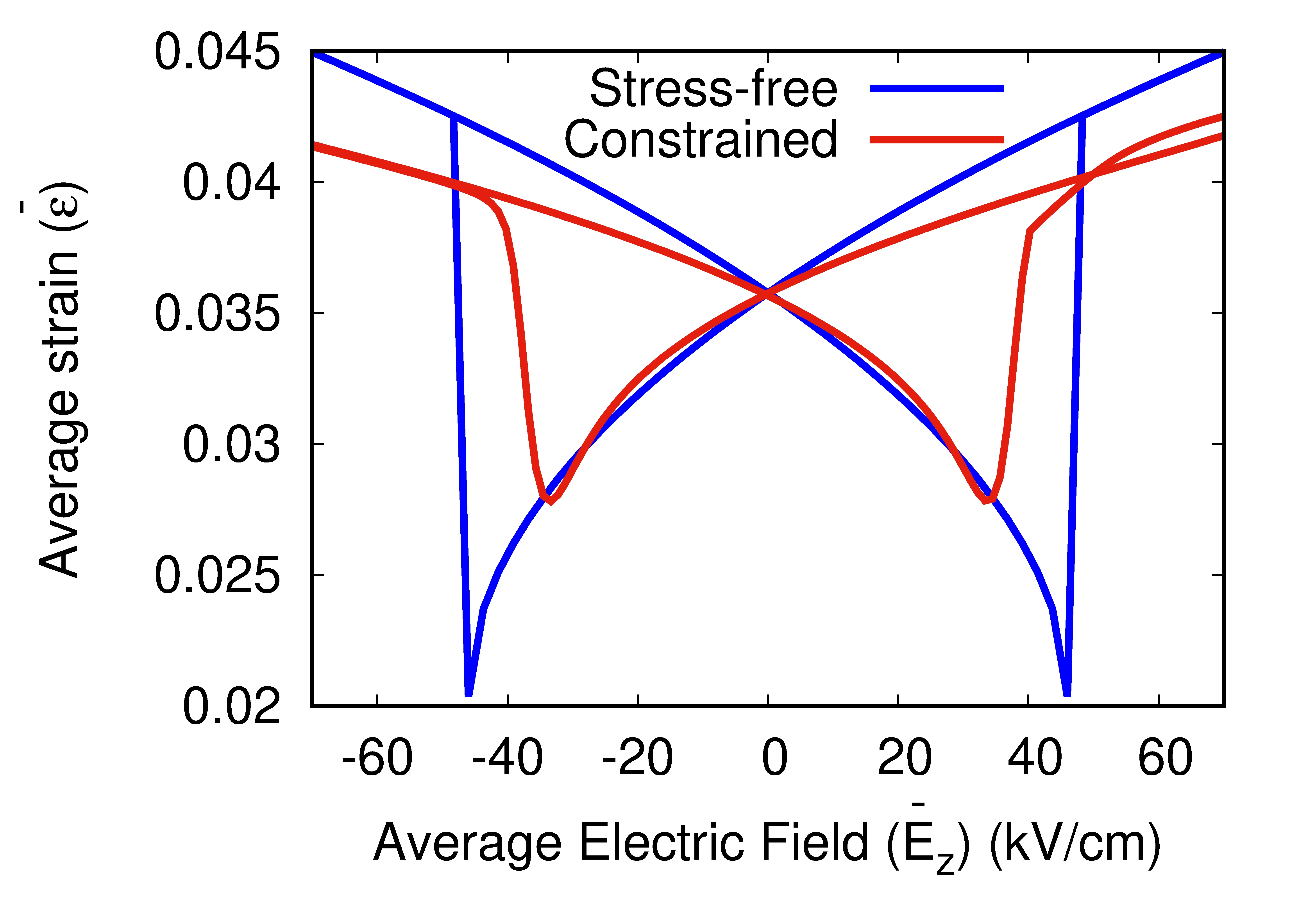}  
 \caption{\textbf{}}
  \label{fig:comparisonbutter}
\end{subfigure} 
 \caption{(a) Polarization hysteresis and (b) strain hysteresis loops for mechanically constrained and stress-free BZCT. Constrained system shows narrower and thinner loops.}
\label{fig:loopsnew}
\end{figure}

\begin{figure}[htpb]
\begin{subfigure}{0.5\textwidth}
  \centering
  \includegraphics[width=\linewidth]{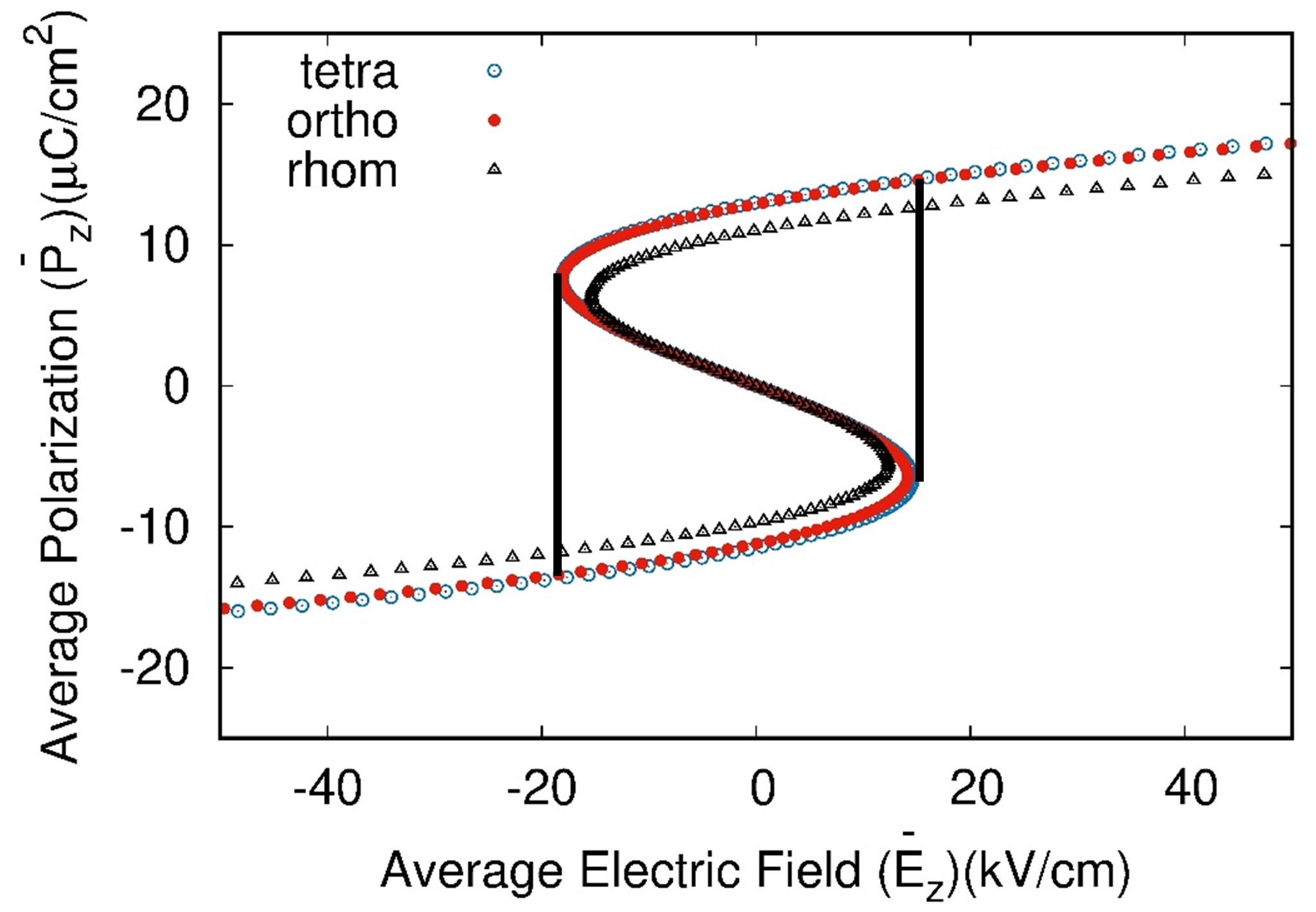}  
 \caption{\textbf{}}
  \label{fig:anapecons}
\end{subfigure}
\begin{subfigure}{0.5\textwidth}
  \centering
  \includegraphics[width=\linewidth]{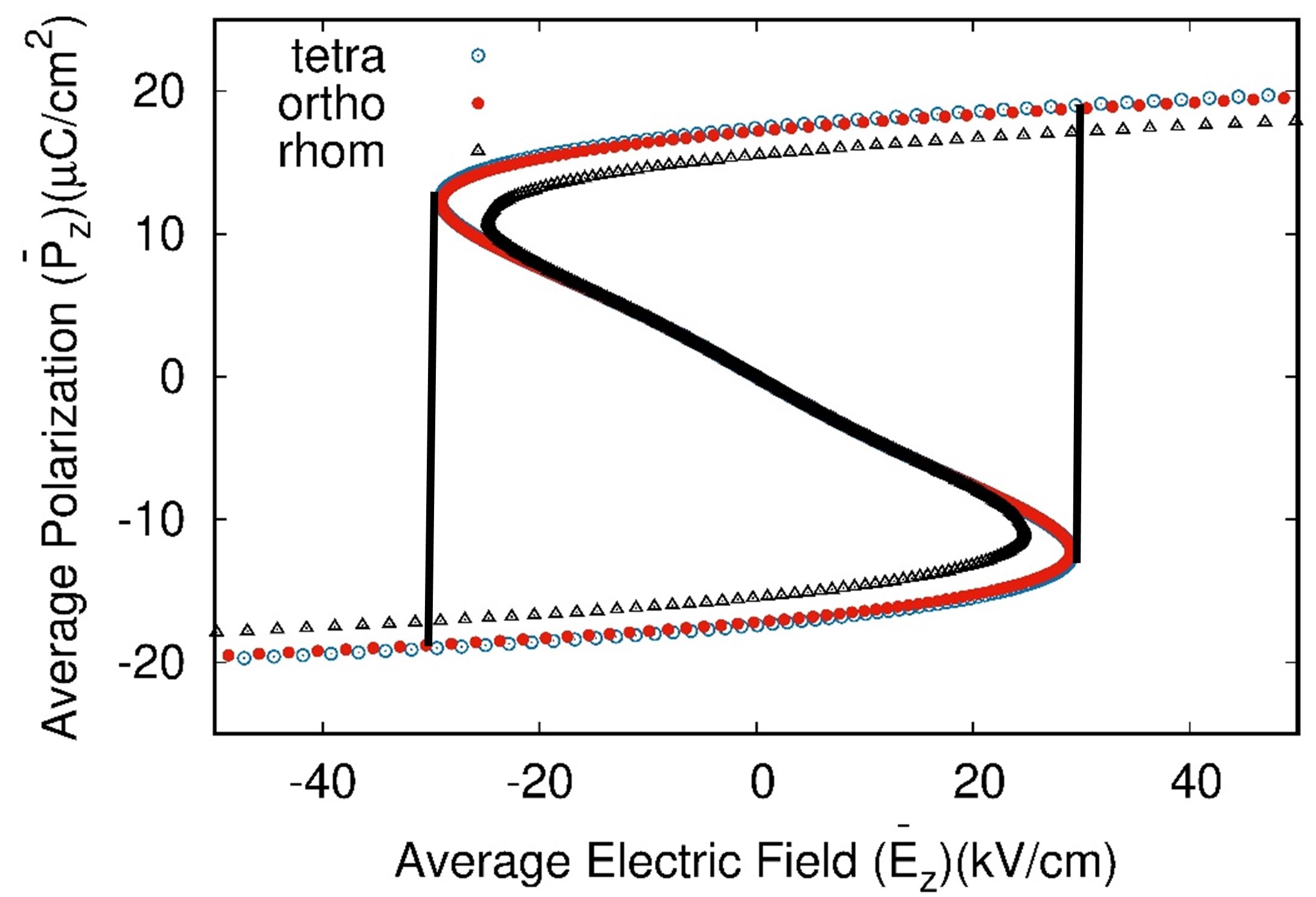}  
 \caption{\textbf{}}
  \label{fig:anapestress}
\end{subfigure} 
 \caption{Comparison between the analytical hysteresis loops for (a) Constrained system
 (b) Stress-free system.}
\label{fig:analytical hyteresis}
\end{figure}

The calculated polarization hysteresis loops ($\bar{P}_z - E_3$) and longitudinal
strain hysteresis loops ($\epsilon_{33} - E_3$) for constrained and stress-free systems are shown in Fig.~\ref{fig:loopsnew}.
The loops corresponding to the constrained system show lower $E_c$ value. 
The calculated effective $d_{33}$ of $634$ $\si{\pico\coulomb\per\newton}$ for the mechanically constrained system shows the closest match 
with the experimentally measured value ($d_{33}^{exp}= 620\si{\pico\coulomb\newton\tothe{-1}}$~\cite{keeble2013revised}).
The close match between the constrained value and the experimental value points to the fact that all measurements are carried out in mechanically constrained conditions. Indeed it is difficult to maintain ideal stress-free conditions in any experimental setup that requires measurement of strain. 
Moreover, all simulated loops from our phase-field model show good agreement with 
with the analytically obtained hysteresis loops (Fig.~\ref{fig:analytical hyteresis})
indicating thermodynamic consistency of our model.

\section{Conclusion}
We presented a thermodynamic model coupled with phase-field simulations to analyze 
phase stability, domain structure evolution and polarization switching properties
of bulk ferroelectric solid solution (BZCT) containing a morphotropic phase coexistence region. 
Since BZCT solid solution shows stress induced
phase transition preceding the paraelectric $\to$ ferroelectric transition, 
change in electromechanical processing conditions may induce structural changes in oxygen octahedra 
manifested by change in anisotropy in electrostriction in the paraelectric state.
Using our model we studied changes in phase stability as a function of electrostrictive
anisotropy. Our predictions of morphotropic phase boundaries 
and tricritical points show excellent agreement with experimental data obtained using high resolution X-ray diffraction studies and Rietveld analysis. The predicted diffusionless phase diagrams show change in phase stability within the morphotropic phase region from single-phase orthorhombic to a multi-phase mixture of tetragonal, rhombohedral and orthorhombic polar phases at room temperature with the increase in the anisotropy of electrostriction. Our predictions are in good agreement with recent experiments which show change in phase stability of ferroelectric phases at room temperature as a result of change in processing conditions in the paraelectric state~\cite{brajesh2015relaxor, brajesh2016structural}. 

The steady state domain structures predicted from our three-dimensional 
phase-field simulations of stress-free equimolar BZCT 
show orthorhombic variants when electrostriction is isotropic ($Q_z = 1$), coexistence of tetragonal and orthorhombic
variants at moderate anisotropy ($Q_z = 2$), and a mixture of tetragonal, orthorhombic and rhombohedral variants
at higher anisotropy ($Q_z = 2.5$). In all cases, the ferroelectric twin domains are in the form 
of plates where the domain boundaries are oriented along specific crystallographic directions 
given by mechanical compatibility condition $\Delta_{ij}s_is_j=0$ where 
$\Delta_{ij}=\varepsilon_{ij}^I-\varepsilon_{ij}^{II}$ is the difference in 
spontaneous strains of variants I and II. 
Moreover,  mobile domain walls for all cases of electrostrictive anisotropy show a step-terrace 
structure facilitating polarization reversal via $90^{\circ}$ ferroelastic steps. 
In all cases, we have 
applied external electric along $[001]$ and $[100]$ directions to study 
polarization switching 
characteristics in terms of polarization hysteresis, longitudinal strain hysteresis 
and transverse strain hysteresis loops. 
The loops are fatter when electric is applied along $[0 0 1]$
direction while they are thinner when the applied field in along $[1 0 0]$ direction.
We show a correlation between the step-terrace domain wall structure and the anisotropy in 
switching behavior. The ratio between $d_{33}/d_{31}$ increases with increasing electrostrictive 
anisotropy. However,
increase in electrostrictive anisotropy leads to reduction in polar anisotropy resulting in 
multi-phase coexistence. Application of mechanical constraint (clamping) can significantly modify 
phase stability in addition to changes in domain morphology in bulk BZCT system. For the 
constrained system (when $Q_z = 1$) we obtain 
clusters of $T$ variants at the equimolar composition and room temperature.
However, the $d_{33}$ value obtained for the constrained 
system ($634$ $\si{\pico\coulomb\per\newton}$) shows the closest match with the piezoresponse 
obtained experimentally.
In summary, our study establishes a framework to predict process-structure-property relations in BZCT ceramics which can be utilized to optimize the design of efficient electromechanical devices.

\section{Acknowledgements}
S.B., T.J, and S.B. gratefully acknowledge the support of DST (Grant No. EMR/2016/006007) for funding the computational research.


\bibliographystyle{elsarticle-num}

\bibliography{reference}








\end{document}